\documentclass[preprint,aps,nofootinbib]{revtex4}
\usepackage[utf8]{inputenc}

\usepackage{amsmath, amsthm, amssymb}
\usepackage{latexsym}
\usepackage{epsfig}
\usepackage{epstopdf}
\usepackage{graphicx}
\usepackage{axodraw2}
\usepackage{mathtools}
\usepackage{color}
\usepackage{hyperref}

\begin{document}

\title{Spectral functions from the real-time functional renormalization group}

\author{S.Huelsmann, S.~Schlichting, P.~Scior}
\email{scior@physik.uni-bielefeld.de}

\affiliation{Fakult\"at für Physik, Universit\"at Bielefeld, D-33615 Bielefeld, Germany}

\date{\today}

\begin{abstract}
We employ the functional renormalization group approach formulated on the Schwinger-Keldysh contour to calculate real-time correlation functions in scalar field theories. We provide a detailed description of the formalism, discuss suitable truncation schemes for real-time calculations as well as the numerical procedure to self-consistently solve the flow equations for the spectral function. Subsequently, we discuss the relations to other perturbative and non-perturbative approaches to calculate spectral functions, and present a detailed comparison and benchmark in $d=0+1$ dimensions.
\end{abstract}

\maketitle

\tableofcontents

\section{Introduction}
\label{sec:Intro}
Spectral functions of quarks, gluons and the gauge invariant states of QCD are important ingredients in the theoretical description of Heavy Ion Collisions performed at RHIC and LHC. The spectral function encodes important information about the real-time dynamics of the system, as well as thermal and in-medium modifications of quarks, gluons and hadrons. Thus, the knowledge of spectral functions of the various strongly interacting particles is highly desirable when trying to investigate e.g. dilepton production, transport coefficients or the melting of quarkonium states in the quark-gluon plasma (QGP).\newline
Unfortunately, extracting real-time information of strongly coupled systems is a difficult problem. The non-perturbative nature of QCD at energies below and around the phase transition prohibits the use of perturbative methods. Recently, there has been progress concerning the spectral functions of quarkonia and some transport coefficients coming from euclidean lattice simulations~\cite{Meyer:2007ic,Brandt:2015aqk,Ding:2018uhl,Aarts:2011sm,Aarts:2014nba}. However, the analytic continuation of the numerical data to Minkoswki space and other problems make these investigations quite challenging and so far there are no lattice results for spectral functions of lighter hadrons. \newline

So far our knowledge about spectral properties of thermal QCD matter comes primarily from calculations in low energy effective theories of QCD, based on a variety of different techniques including (re-summed) perturbative calculations \cite{Rapp:1997fs,Urban:1998eg,Roder:2005vt} as well as non-perturbative functional approaches \cite{Liu:2017qah,Mueller:2010ah,Fischer:2017kbq,Shen:2019jhl,Shen:2020jya,Patkos:2002xb}.  Recently, there has been great success in applying the analytically continued functional renormalization group (FRG) \cite{Floerchinger:2011sc,Kamikado:2013sia} to low energy effective models of QCD \cite{Tripolt:2013jra,Pawlowski:2015mia,Tripolt:2014wra,Jung:2016yxl,Jung:2019nnr,Pawlowski:2017gxj,Strodthoff:2016pxx,Wang:2017vis,Yokota:2016tip}. While many of the results from analytically continued FRG calculations have been impressive, it still is desirable to pursue non-perturbative functional calculations directly in Minkowski space. In this paper we adopt a real-time FRG approach on the Schwinger-Keldysh (SK) contour \cite{Gasenzer:2007za,Berges:2008sr,Berges:2012ty,Pietroni:2008jx,PhysRevLett.84.3686,Canet:2009vz,Delamotte:2004zg,Canet_2007,Karrasch_2010,Andergassen_2011,PhysRevB.89.134310} to extract spectral functions in the $O(N)$ model without the need for analytical continuation. By performing a careful perturbative analysis we show that -- in the absence of spontaneous symmetry breaking -- local potential approximations (LPA) are not able to generate a broadening of the spectral function. We therefore develop a truncation, based on a vertex expansion that includes momentum dependent four-point functions, which is able to capture the broadening of the spectral function as the propagators in this truncation are two-loop complete. One important feature of our method is that it is applicable for both quantum and classical-statistical field theories, such that we can compare and evaluate our results from the real-time FRG approach against non-perturbative classical-statistical real-time lattice simulations \cite{Aarts:2001yx,Berges:2009jz,Schlichting:2019tbr,Schweitzer:2020noq}.

This paper is organized as follows: We start in section~\ref{sec:FRG} with an introduction to dissipative classical- and quantum field theories on the SK contour and the formulation of the real-time FRG approach. After defining a $d+1$ dimensional regulator scheme that respects time-ordering on the SK
contour we introduce a diagrammatic notation simplifying the derivation of flow-equations for n-point functions. In section~\ref{sec:PT} we compare the RG flow to perturbative results, indicating the need for truncation schemes that go beyond the frequently used local potential approximation. Suitable truncation schemes are then developed in section~\ref{sec:Trunc}, and we explain our numerical implementation of the resulting flow-equation in section~\ref{sec:Numerics}. After presenting detailed comparisons and benchmarks in $d=0+1$ dimension in sec.~\ref{sec:Bench} we conclude our findings in section~\ref{sec:Conclusion}. Several appendices contain additional details intended for the non-expert reader.

\section{Real-time FRG on the Schwinger-Keldysh contour}
\label{sec:FRG}
\subsection{Schwinger Keldysh formulation of quantum and classical-statistical field theories}
We consider a $N$-component scalar quantum field theory in $d$ spatial dimensions, who's real-time correlation functions in thermal equilibrium can be obtained from the generating functional \cite{Berges:2004yj,Berges:2012ty}
\begin{eqnarray}
\label{eq:ZPart}
Z[J,\tilde{J}]=\int D \varphi D \tilde{\varphi} \exp \left\{ i S_{\mathcal{C}}[\varphi,\tilde{\varphi}] + \int_{x} \Big \lbrace \tilde J_a(x) \varphi_a(x) + J_a(x) \tilde \varphi_a(x) \Big \rbrace  \right\}\;,
\end{eqnarray}
where $S_{\mathcal{C}}[\varphi,\tilde{\varphi}]$ is the contour action on the Schwinger-Keldysh contour. Denoting the thermal distribution function of a bosonic quantum system as 
\begin{eqnarray}
n_{\rm eff}^{\rm qu}(\omega)=\hbar \left( n_{\rm BE}(\omega) +\frac{1}{2} \right)=\frac{\hbar}{2} \coth\left(\frac{\hbar\beta\omega}{2}\right)\;,
\end{eqnarray}
where $n_{\rm BE}(\omega)=\frac{1}{e^{\hbar\beta\omega}-1}$ is the Bose-Einstein distribution, the contour action $S_{\mathcal{C}}[\varphi,\tilde{\varphi}]$ for a dissipative quantum system coupled to an external heat-bath at inverse temperature $\beta=\frac{1}{k_B T}$ and with the rest-frame $u^{\mu}=(1,0,0,0)$ is explicitly given by
\begin{eqnarray}
\label{eq:ContourAction}
S_{\mathcal{C}}[\varphi,\tilde{\varphi}]&=& \frac{1}{2} \int_{x} 
\begin{pmatrix} \varphi_{a}(x), & \tilde \varphi_{a}(x) \end{pmatrix}
\begin{pmatrix}  
0 &  -\partial_{\mu}\partial^{\mu} + \frac{\gamma}{\beta} u^{\mu}\partial_{\mu} - m^2  \\
-\partial_{\mu}\partial^{\mu} - \frac{\gamma}{\beta} u^{\mu}\partial_{\mu} - m^2  &  2\frac{\gamma}{\beta}u^{\mu}\partial_{\mu}~n_{\rm eff}\left(-i u^{\mu}\partial_{\mu} \right) 
\end{pmatrix}
\begin{pmatrix} \varphi_{a}(x)  \\  \tilde \varphi_{a}(x) \end{pmatrix} \nonumber \\
&& - \frac{\lambda}{6N} \int_{x} \tilde \varphi_{a}(x) \varphi_{a}(x) \varphi_{b}(x) \varphi_{b}(x) - \frac{\lambda \hbar^2}{24N} \int_{x} \tilde \varphi_{a}(x) \tilde \varphi_{a}(x) \tilde \varphi_{b}(x) \varphi_{b}(x)\;,
\end{eqnarray}
where $\int_{x}=\int_{-\infty}^{\infty} dx^{0} \int d^{d}\mathbf{x}$ such that the real-time axis extends from $x^{0}=-\infty$ to $x^{0}=+\infty$ describing a time translation invariant system in thermal equilibrium \cite{Sieberer:2015hba}. While the contour action in Eq.~(\ref{eq:ContourAction}) describes a dissipative quantum system with Model A type dynamics \cite{Hohenberg:1977ym}, the case of a non-dissipative quantum system with conservative Model C/G type dynamics\footnote{Single component scalar theories $(N=1)$ classify as Model C, whereas multi-component scalar theories $(N\geq2)$ feature an additional conserved current, e.g. for $N=4$ one has $j^{\mu}_{ab}(x)=\epsilon^{abcd} \varphi_{c}(x) \partial^{\mu} \varphi_{d}(x)$, and therefore classify as Model G \cite{Rajagopal:1992qz}.} is obtained in the limit $\gamma\to0^{+}$, where the coupling to the external heat bath ultimately vanishes, but as usual in the $i \epsilon$ prescription is required at intermediate steps of the calculation to ensure the correct time ordering of the propagators and convergence of the functional integral. Specifically, in the absence of interactions $(\lambda=0)$ the free propagators of the theory in momentum space
\begin{eqnarray}
G_{0}^{ab}(p)=
\begin{pmatrix}
iF_{0}^{ab}(p) & G^{A,ab}_{0}(p) \\
G^{R,ab}_{0}(p) & i\tilde{F}_{0}^{ab}(p) 
\end{pmatrix}
= i \int_{x-y} 
\begin{pmatrix}
\langle \varphi_{a}(x) \varphi_{b}(y) \rangle_{c} & \langle \tilde{\varphi}_{a}(x) \varphi_{b}(y) \rangle_{c} \\
\langle \varphi_{a}(x) \tilde{\varphi}_{b}(y) \rangle_{c} & \langle \tilde{\varphi}_{a}(x) \tilde{\varphi}_{b}(y) \rangle_{c}
\end{pmatrix} e^{+ip(x-y)}
\end{eqnarray}
 are explicitly given by
\begin{eqnarray}
iF_{0}^{ab}(\omega,\mathbf{p})&=&\frac{2 i \frac{\gamma}{\beta} \omega~n_{\rm eff}(w)}{(\omega^2-E_p^2)^2+ \frac{\gamma^2}{\beta^2} \omega^2}  \delta^{ab} \;, \qquad  G^{A,ab}_{0}(\omega,\mathbf{p})=\frac{-1}{\omega^2 -E_p^2 - i \frac{\gamma}{\beta} \omega} \delta^{ab}\;, \\
G^{R,ab}_{0}(\omega,\mathbf{p})&=&\frac{-1}{\omega^2 -E_p^2 + i \frac{\gamma}{\beta} \omega} \delta^{ab}  \;, \quad \qquad~ i\tilde{F}_{0,ab}(\omega,\mathbf{p})=0\;,  \nonumber
\end{eqnarray}
with $E_{p}=\sqrt{\mathbf{p}^2+m^2}$ such that in the limit $\gamma\to 0^{+}$, the above expressions reduce to the familiar expressions for the retarded/advanced $(G^{R/A})$ and symmetric $(iF)$ two-point functions, who's operator definitions and basic properties are recalled in Appendix A.

Expressing the contour action in Fourier space
\begin{eqnarray}
\label{eq:ContourActionP}
S_{\mathcal{C}}[\varphi,\tilde{\varphi}]&=& \frac{1}{2} \int_{p} 
\begin{pmatrix} \varphi_{a}(-p), & \tilde \varphi_{a}(-p) \end{pmatrix}
\begin{pmatrix}  
0 &  p_{\mu}p^{\mu} +i\frac{\gamma}{\beta} u^{\mu}p_{\mu} - m^2  \\
p_{\mu}p^{\mu} -i\frac{\gamma}{\beta} u^{\mu}p_{\mu} - m^2  &  2i\frac{\gamma}{\beta}u^{\mu}p_{\mu}~n_{\rm eff}\left(u^{\mu}p_{\mu} \right) 
\end{pmatrix}
\begin{pmatrix} \varphi_{a}(p)  \\  \tilde \varphi_{a}(p) \end{pmatrix} \nonumber \\
&& - \frac{\lambda}{6N} \int_{pqkl} (2\pi)^{(d+1)}\delta(p+k+q+l)~\tilde \varphi_{a}(p) \varphi_{a}(q) \varphi_{b}(k) \varphi_{b}(l) \\
&&- \frac{\lambda \hbar^2}{24N} \int_{pqkl} (2\pi)^{(d+1)}\delta(p+k+q+l)~\tilde \varphi_{a}(p) \tilde \varphi_{a}(q) \tilde \varphi_{b}(k) \varphi_{b}(l)\;, \nonumber
\end{eqnarray}
where we denote $\int_{p}=\int \frac{d\omega}{2\pi} \int \frac{d^d\mathbf{p}}{(2\pi)^d}$ such that $\varphi(x)=\int_{p} \varphi(p) e^{+ipx}$, it becomes evident that the contour action in Eq.~(\ref{eq:ContourAction}) is invariant under the symmetry transformation \cite{Sieberer:2015hba}
\begin{eqnarray}
\mathcal{T}_{\beta} \varphi_{a}(\omega,\mathbf{p}) &=& ~~\cosh\Big(\frac{\hbar\beta \omega}{2}\Big) \varphi_{a}(-\omega,\mathbf{p}) + \frac{\hbar}{2} \sinh\Big(\frac{\hbar\beta \omega}{2}\Big) \tilde \varphi_{a}(-\omega,\mathbf{p})\;, \label{eq:sym-trafo-q} \\
\mathcal{T}_{\beta} \tilde \varphi_{a}(\omega,\mathbf{p})&=& \frac{2}{\hbar} \sinh\Big(\frac{\hbar\beta \omega}{2}\Big) \varphi_{a}(-\omega,\mathbf{p}) + ~~\cosh\Big(\frac{\hbar\beta \omega}{2}\Big) \tilde \varphi_{a}(-\omega,\mathbf{p})\;,  \notag
\end{eqnarray}
in the sense that $S_{\mathcal{C}}[\mathcal{T}_{\beta} \varphi, \mathcal{T}_{\beta} \tilde \varphi]=S_{\mathcal{C}}[\varphi,\tilde \varphi]$, which as discussed in \cite{Sieberer:2015hba} guarantees the validity of the fluctuation-dissipation relations for $n$-point correlation functions. Specifically for two point correlation functions, the fluctuation-dissipation relation takes the form
\begin{eqnarray}
iF_{ab}(\omega,\mathbf{p})&=& n_{\rm eff}(\omega) \left( G^{R}_{ab}(\omega,\mathbf{p})- G^{A}_{ab}(\omega,\mathbf{p}) \right)\;,
\end{eqnarray}
which along with the symmetry property of retarded/advanced propagators $G^{R,ab}(p)=G^{A,ba}(-p)$ implies that in thermal equilibrium there is only one independent two-point correlation function. When presenting explicit numerical results, we will therefore focus our attention on the investigation of the spectral function $\rho^{ab}(\omega,\mathbf{p})$, given by
\begin{eqnarray}
\rho^{ab}(\omega,\mathbf{p}) = \left( G^{R}_{ab}(\omega,\mathbf{p})- G^{A}_{ab}(\omega,\mathbf{p}) \right)\;.
\end{eqnarray}

Besides $N$-component scalar quantum field theory in $d$ spatial dimensions, we will also be interested in the corresponding classical-statistical field theories, who's dynamics can be formulated in terms of classical Langevin type field equations of motion
\begin{eqnarray}
\left[ \partial_{\mu}\partial^{\mu} + \frac{\gamma}{\beta} u^{\mu}\partial_{\mu} + m^2 + \frac{\lambda}{6 N} \Big( \varphi_{b}(x)\varphi_{b}(x) \Big) \right] \varphi_{a}(x)= \eta_{a}(x)\;,
\end{eqnarray}
where $\eta_{a}(x)$ represents a stochastic Gaussian white noise, with auto-correlation functions
\begin{eqnarray}
\langle \eta_{a}(x) \rangle = 0\;, \qquad \langle \eta_{a}(x) \eta_{b}(y) \rangle = \sqrt{\frac{2 \gamma}{\beta^2}}~\delta(x^{0}-y^{0})~\delta^{(d)}(\mathbf{x}-\mathbf{y})~\delta^{ab}\;.
\end{eqnarray}
By performing the usual Martin-Siggia-Rose-Janssen-de Dominicis path-integral re-formulation \cite{Martin:1973zz,Hertz:2016vpy}, the problem of calculating real-time observables in classical-statistical field theory can be formulated in an analogous fashion as a path integral in Eq.~(\ref{eq:ZPart}), where instead of Eq.~(\ref{eq:ContourAction}) the classical contour action $S^{\rm cl}_{\mathcal{C}}[\varphi,\tilde{\varphi}]$ is now  given by (see e.g.~\cite{Berges:2004yj,Berges:2012ty})
\begin{eqnarray}
\label{eq:ClassicalContourAction}
S^{\rm cl}_{\mathcal{C}}[\varphi,\tilde{\varphi}]&=& \frac{1}{2} \int_{x} 
\begin{pmatrix} \varphi_{a}(x), & \tilde \varphi_{a}(x) \end{pmatrix}
\begin{pmatrix}  
0 &  -\partial_{\mu}\partial^{\mu} + \frac{\gamma}{\beta} u^{\mu}\partial_{\mu} - m^2  \\
-\partial_{\mu}\partial^{\mu} - \frac{\gamma}{\beta} u^{\mu}\partial_{\mu} - m^2  &  2\frac{\gamma}{\beta}u^{\mu}\partial_{\mu}~n_{\rm eff}^{\rm cl}\left(-i u^{\mu}\partial_{\mu} \right) 
\end{pmatrix}
\begin{pmatrix} \varphi_{a}(x)  \\  \tilde \varphi_{a}(x) \end{pmatrix} \nonumber \\
&& - \frac{\lambda}{6N} \int_{x} \tilde \varphi_{a}(x) \varphi_{a}(x) \varphi_{b}(x) \varphi_{b}(x) \;,
\end{eqnarray}
where
\begin{eqnarray}
n_{\rm eff}^{\rm cl}(\omega)= \frac{1}{\beta \omega}\;,
\end{eqnarray}
is the Rayleigh-Jeans distribution. By explicit comparison with Eq.~(\ref{eq:ContourAction}) one finds that the classical contour action $S^{\rm cl}_{\mathcal{C}}[\varphi,\tilde{\varphi}]$ only contains the leading $\mathcal{O}(\hbar^{0})$ contributions, which as discussed extensively in the literature \cite{Aarts:1996qi,Aarts:1997kp,Aarts:2001yn,Berges:2004yj,Epelbaum:2014yja} amounts to a change of the statistical factor between Eqns.~(\ref{eq:ContourAction}) and (\ref{eq:ClassicalContourAction}), as well as the absence of the "quantum" $\tilde{\varphi} \tilde{\varphi} \tilde{\varphi} \varphi$ interaction term in the classical-statistical field theory. We also note for completeness that the classical-statistical theory in Eq.~(\ref{eq:ClassicalContourAction}) is invariant under the symmetry transformation \cite{Sieberer:2015hba}
\begin{eqnarray}
\mathcal{T}_{\beta}^{\rm cl} \varphi_{a}(\omega,\mathbf{p}) &=& \quad~~ \varphi_{a}(-\omega,\mathbf{p}) \;, \label{eq:sym-trafo-cl} \\
\mathcal{T}_{\beta}^{\rm cl} \tilde \varphi_{a}(\omega,\mathbf{p})&=& \beta\omega~\varphi_{a}(-\omega,\mathbf{p}) + \tilde\varphi_{a}(-\omega,\mathbf{p})\;,  \notag
\end{eqnarray}
which again guarantees the validity of the classical fluctuation-dissipation (Kubo-Martin-Schwinger) relations for $n$-point correlation functions.

Due to the fact that the quantum and classical-statistical theories only differ by the presence/absence of the quantum vertex and the change of statistical factors, the real-time functional renormalization group framework allows for an efficient simultaneous discussion of both classical-statistical and quantum field theories. Since in contrast to the quantum field theory, the classical-statistical field theory can be simulated in real-time from first principles by performing real-time lattice simulations \cite{Aarts:2001yx,Berges:2009jz,Schlichting:2019tbr,Schweitzer:2020noq}, the functional renormalization group results obtained in the classical-statistical regime can therefore be directly compared to exact numerical calculations, thus allowing for an important test of the methodology and benchmark of the quality of the underlying approximations.

\subsection{Effective action and flow equation}
Starting from the generating functional $Z[J,\tilde{J}]$ for quantum and classical-statistical field theories, the generating functional for connected correlation functions $W[J,\tilde{J}]$ is given by
\begin{eqnarray}
W[J,\tilde{J}]= -i \log Z[J,\tilde{J}] \label{Schwinger}
\end{eqnarray}
such that connected one- and two-point correlation functions are determined by
\begin{eqnarray}
 \frac{\delta W[J, \tilde J]}{\delta \tilde J_a (x)}=\phi_{a}(x) \;, \quad   \frac{\delta W[J, \tilde J]}{\delta J_a (x)} = \tilde\phi_{a}(x) 
\end{eqnarray}
and
\begin{eqnarray}
\label{eq:WPropagators}
 \frac{\delta^2 W[J, \tilde J]}{\delta \tilde J_a (x) \delta J_b(y)} &=& G^R_{k,ab}(x,y) \; , \quad \frac{\delta^2 W[J, \tilde J]}{\delta J_a (x) \delta \tilde J_b(y)} = G^A_{k,ab}(x,y)\;, \\
 \frac{\delta^2 W[J, \tilde J]}{\delta \tilde J_a (x) \delta \tilde J_b(y)} &=& i F_{k,ab}(x,y) \; , \quad \frac{\delta^2 W[J, J]}{\delta J_a (x) \delta \tilde J_b(y)} = i\tilde F_{k,ab}(x,y) \; ,
\end{eqnarray}
The one-particle irreducible (1PI) effective action is obtained by a Legendre transformation of Eq.~(\ref{Schwinger}) with respect to the sources $J$ and $\tilde J$, for fixed values of the field expectation values $\phi_{a}$, $\tilde \phi_{a}$, i.e.
\begin{equation}
    \Gamma[\phi, \tilde \phi] = W[J, \tilde J] - \int_x \Big \lbrace \tilde J_a(x) \phi_a(x) + J_a(x) \tilde \phi_a(x) \Big \rbrace \; .
\end{equation}
Even though the effective action contains the full information content about the dynamics of the theory, it is notoriously hard to compute due to the functional integrations in the generating functional. The basic idea of the functional renormalization group approach is therefore to construct the effective action step-by-step, by solving a set of functional differential flow equations which successively integrate out fluctuations at different scales. In order to construct the functional flow equations we follow standard procedure \cite{Wetterich:1992yh} and introduce a regulator term depending on the flow scale $k$, so that we replace the original action $S[\varphi, \tilde \varphi]$ in the generating functional by a scale dependent action
\begin{eqnarray}
S_k[\varphi, \tilde \varphi] &=& S[\varphi, \tilde \varphi] + \Delta_k S[\varphi, \tilde \varphi] \, ,
\end{eqnarray}
which includes a generic regulator term of the form
\begin{eqnarray}
\label{eq:regulator_term}
\Delta_k S[\varphi, \tilde \varphi] &=&  \frac{1}{2} \int_{xy} \begin{pmatrix}
\varphi_a(x), & \tilde \varphi_a(x)
\end{pmatrix}  \begin{pmatrix} 
    R^{\tilde F}_{k,ab}(x,y) & R^A_{k,ab}(x,y) \\ R^R_{k,ab}(x,y) & R^F_{k,ab}(x,y) 
\end{pmatrix}  \begin{pmatrix}
\varphi_b(y) \\ \tilde \varphi_b(y)
\end{pmatrix} \; .
\end{eqnarray}
Based on these modifications, the effective action $\Gamma_k[\phi, \tilde \phi]$ now depends on the scale $k$ and is explicitly given by 
\begin{equation}
  \label{eq:Gammak}
  \Gamma_k[\phi, \tilde \phi] = W_k[J, \tilde J] - \Delta_k S[\phi, \tilde \phi] - \int_x \Big \lbrace \tilde J_a(x) \phi_a(x) + J_a(x) \tilde \phi_a(x) \Big \rbrace \; .
\end{equation}  
Based on a suitable choice of regulator functions $R^{X}_{k,ab}(x,y)$, such that in the limit $k\to\Lambda$ the regulator suppresses all fluctuations, whereas in the limit $k\to0$ the all regulators vanish identically
\begin{eqnarray}
\lim_{k\to0} R^{X}_{k,ab}(x,y)=0\;,
\end{eqnarray}
and all fluctuations are included, the renormalization group flow interpolates between the classical action $S[\phi, \tilde \phi]$ at some ultra-violet (UV) cutoff scale $k\to\Lambda$ and the full effective action in the infrared, i.e.
\begin{equation}
    \lim_{k\to\Lambda} \Gamma_k[\phi, \tilde \phi] = S[\phi, \tilde \phi] \,, \quad \quad \lim_{k\to 0} \Gamma_k[\phi, \tilde \phi] = \Gamma[\phi, \tilde \phi] \, .
\end{equation}
We also note that on the Schwinger-Keldysh contour, the various regulators have to satisfy additional constraints to comply with the symmetries of the action of an equilibrium system, as will be discussed in more detail below. \\
By taking a renormalization group scale ($k$) derivative of the effective action $\Gamma_k[\phi, \tilde \phi]$ in Eq.~\eqref{eq:Gammak}, we obtain the flow equation for the effective action 
\begin{eqnarray}
\partial_k \Gamma_k[\phi, \tilde \phi] &=& \partial_k W_k[J, \tilde J] - \partial_k \Delta_k S[\phi, \tilde \phi] - \int_x \Big \lbrace \big(\partial_k \tilde J_a(x) \big) \phi_a(x) + \big(\partial_k J_a(x) \big ) \tilde \phi_a(x) \Big \rbrace \; , 
\end{eqnarray}
which upon performing a straightforward set of manipulations can be expressed as
\begin{eqnarray}
\partial_k \Gamma_k[\phi, \tilde \phi] &=& \frac{1}{2} \int_{xy} \text{Tr} \begin{pmatrix} 
   \dot R^{\tilde F}_{k,ab}(x,y) & \dot R^A_{k,ab}(x,y) \\ \dot R^R_{k,ab}(x,y) & \dot R^F_{k,ab}(x,y) 
\end{pmatrix}
\begin{pmatrix}
 \langle \varphi_b(y) \varphi_a(x) \rangle_{c} && \langle \varphi_b(y) \tilde \varphi_a(x) \rangle_{c} \\
 \langle \tilde \varphi_b(y) \varphi_a(x) \rangle_{c} && \langle \tilde \varphi_b(y) \tilde \varphi_a(x) \rangle_{c} \, .
\end{pmatrix}
\end{eqnarray}
where all two-point functions in the last line are understood to be connected. By use of the relations in Eq.~(\ref{eq:WPropagators}) we then arrive at the most general form for the flow equation \cite{Berges:2012ty}
\begin{eqnarray}
   \partial_k \Gamma_k[\phi, \tilde \phi] = - \frac{i}{2} \int_{xy} \Big[& \dot R^R_{k,ab}(x,y) G^R_{k,ba}(y,x) + \dot R^A_{k,ab}(x,y) G^A_{k,ba}(y,x) \label{eq:FlowEqnGamma} \\ 
   &+ \dot R^F_{k,ab}(x,y) i \tilde F_{k,ba}(y,x)  +\dot R^{\tilde F}_{k,ab}(x,y) iF_{k,ba} (y,x) \Big]  \; . \notag
\end{eqnarray}

\subsection{Propagators and two-point functions}
The flow equation for the effective action~\eqref{eq:FlowEqnGamma} is given in terms of scale dependent propagators, which are related to the derivatives of the effective action. Denoting the second functional derivatives of the effective action as
\begin{eqnarray}
 \Gamma^{\tilde \phi \phi}_{k,ab}(x,y) = \frac{\delta^2 \Gamma_k[\phi, \tilde \phi]}{\delta \tilde \phi_a(x) \delta \phi_b(y)}\, , \quad \Gamma^{\phi \phi}_{k,ab}(x,y) = \frac{\delta^2 \Gamma_k[\phi, \tilde \phi]}{\delta \phi_a(x) \delta \phi_b(y)} \, ,
\end{eqnarray}
the expressions for the various propagators are then given by \cite{Berges:2008sr}
\begin{eqnarray}
 G^R_k &=& - \Big \lbrace (\Gamma^{\tilde \phi \phi}_k +R^R_k) - (\Gamma^{\tilde \phi \tilde \phi}_k + R^F_k) (\Gamma^{\phi \tilde \phi}_k + R^A_k)^{-1} (\Gamma^{\phi \phi}_k  + R^{\tilde{F}}_k ) \Big \rbrace^{-1} \, , \notag \\
G^A_k &=& - \Big \lbrace (\Gamma^{\phi \tilde \phi}_k +R^A_k) - (\Gamma^{\phi \phi}_k + R^{\tilde F}_k) (\Gamma^{\tilde\phi \phi}_k + R^R_k)^{-1} (\Gamma^{\tilde\phi \tilde \phi}_k  + R^{F}_k ) \Big \rbrace^{-1} \, , \label{eq:propagators_full} \\
iF_k &=& - \Big \lbrace (\Gamma^{\phi \phi}_k  + R^{\tilde{F}}_k ) - (\Gamma^{\phi \tilde \phi}_k + R^A_k) (\Gamma^{\tilde\phi \tilde \phi}_k  + R^{F}_k )^{-1} (\Gamma^{\tilde \phi \phi}_k +R^R_k) \Big \rbrace^{-1} \notag \, ,\\
i \tilde F_k &=& - \Big \lbrace (\Gamma^{\tilde \phi \tilde \phi}_k  + R^{{F}}_k ) - (\Gamma^{\tilde \phi \phi}_k + R^R_k) (\Gamma^{\phi \phi}_k  + R^{ \tilde F}_k )^{-1} (\Gamma^{\phi \tilde \phi}_k +R^A_k) \Big \rbrace^{-1} \notag \, .
\end{eqnarray}
Similarly, by taking functional derivatives of the propagators in Eq.~(\ref{eq:propagators_full}) one obtains the flow equations for $n$-point correlation functions, which in the end have to be evaluated at the minimum of the effective action. Since $\tilde \phi=0$ and $\Gamma^{\phi \phi} =0$ vanish due to discrete symmetries of the effective action, the propagators evaluated at the minimum of the effective action then simplify to\footnote{Note that also the regulator needs to be chosen in accordance with the symmetry requirements, and we further chose $R_k^{\tilde F}=0$, as any other choice would violate causality.}
\begin{eqnarray}
 G^R_k =& - (\Gamma^{\tilde \phi \phi}_k +R^R_k)^{-1} \, , \quad \quad G^A_k &= - (\Gamma^{\phi \tilde \phi}_k +R^A_k)^{-1} \, , \\
 iF_k =& G^R_k (\Gamma^{\tilde\phi \tilde \phi}_k  + R^{F}_{k} ) G_k^A \, ,\quad \quad i \tilde F_k &=0 \, .
\end{eqnarray}
Using the fluctuation dissipation relation in Eq.~\eqref{eq:fluc_dis} for scale dependent propagators, then implies the following relations between the different two-point functions appearing in the effective action
\begin{eqnarray}
\Gamma_{k}^{\tilde{\phi}\tilde{\phi}}(p) = n_{\rm eff}(p_0) \left( \Gamma^{\tilde{\phi}\phi}_{k}(p) - \Gamma^{\phi\tilde{\phi}}_{k}(p) \right)\;,
\end{eqnarray}
which needs to be satisfied at any scale $k$.

\subsection{Regulator functions}

Even though the detailed choice of regulators is irrelevant if the functional differential flow equation is solved exactly, in practice the hierarchy of flow equations for $n$-point correlation functions has to be truncated at a finite order making the solution sensitive to the regulator choice. Since finding suitable regulators for real-time calculations turns out to be a rather subtle issue, we will now comment in more detail on the general conditions for the regulator functions in the real-time FRG framework and specify explicit choices below.

Clearly, the most essential property of the regulator is that it suppresses the effect of fluctuations in the real-time path integral. Expressing the regulator matrix $R_{k,ab}(\omega,\mathbf{p})= R_{k}(\omega,\mathbf{p}) \delta_{ab}$ for a space-time translation invariant system in Fourier space, as 
\begin{eqnarray}
R_{k}(\omega,\mathbf{p}) = \begin{pmatrix} 
    R^{\tilde F}_{k}(\omega,\mathbf{p}) & R^A_{k}(\omega,\mathbf{p}) \\ R^R_{k}(\omega,\mathbf{p}) & R^F_{k}(\omega,\mathbf{p}) 
\end{pmatrix}\; ,
\end{eqnarray}
this can e.g. be achieved if the imaginary part of the bi-linear form 
\begin{eqnarray}
\Delta S_{k}[\varphi,\tilde \varphi]=\frac{1}{2}\int_{p}(\varphi^{*}_a(\omega,\mathbf{p}),\tilde \varphi^{*}_a(\omega,\mathbf{p})) R_{k,ab}(\omega,\mathbf{p}) \begin{pmatrix}
  \varphi_b(\omega,\mathbf{p})\\ \tilde \varphi_b(\omega,\mathbf{p})
\end{pmatrix} \;,
\end{eqnarray}
is positive semi-definite, such that the associated term in the path integral $e^{i \Delta S_{k}[\varphi,\tilde \varphi]}$ gives rise to an exponential suppression of fluctuations below the renormalization group scale.

Besides its regulating properties, it is also desirable that the introduction of the regulator does not explicitly break the symmetries of the system. Specifically, in our context of real-time dynamics in equilibrium systems, this boils down to the invariance of the regulator term $\Delta S_{k}[\varphi,\tilde \varphi]$ under the symmetry transformation in eqn.~(\ref{eq:sym-trafo-q}) for quantum and eqn.~(\ref{eq:sym-trafo-cl}) for classical system, which can be satisfied with
\begin{eqnarray}
R_{k,ab}^{F}(\omega,\mathbf{p})=n_{\rm eff}(\omega) \left[R_{k,ab}^{R}(\omega,\mathbf{p})-R_{k,ab}^{A}(\omega,\mathbf{p})\right]\;,  \qquad R_{k}^{\tilde{F}}=0\;.
\end{eqnarray}
Vice versa, if the regulator functions $R_{k}(\omega,\mathbf{p})$ are chosen to comply with the above symmetry condition, this also guarantees the validity of the fluctuation dissipation relation for the scale $(k)$ dependent $n$-point correlation functions,   such that for example the fluctuation-dissipation relation
\begin{eqnarray}
 \label{eq:fluc_dis}
iF_{k}(\omega,\mathbf{p})= \, n_{\rm eff}(\omega) \left( G_{k}^{R}(\omega,\mathbf{p}) - G_{k}^{A}(\omega,\mathbf{p}) \right) \; ,
\end{eqnarray}
will automatically be satisfied at all scales.

Specifically for the real-time FRG approach, it is also highly desirable that the introduction of the regulator $R_{k}$ respects the time ordering properties of the retarded/advanced and symmetric propagators in coordinate space, such that e.g. the scale dependent propagator $G^{R}_{k}(x,y)$ remains retarded, i.e. vanishes for space-like separations $(x-y)^2<0$, throughout the entire renormalization group evolution. Vice versa, in momentum space, this condition dictates, that the regulator term does not introduce spurious complex poles of the advanced/retarded propagators, which would result in a violation of causality. Note that there is no analogue of such a causality constraint for Euclidean FRG calculation, indicating the additional difficulties that appear in real-time QFT calculations.

Clearly the simplest possible way to comply with causality, is to employ a frequency independent (purely spatial) regulator acting as an effective mass term, such that following \cite{Mesterhazy:2015uja}
\begin{eqnarray}
\label{eq:MassRegulator}
\text{Re}R_{k}^{R}(\omega,\mathbf{p})=\text{Re}R_{k}^{A}(\omega,\mathbf{p})=r_k(\mathbf{p}) \qquad \text{Im}~R_k^{R}(\omega,\mathbf{p})=\text{Im}~R_k^{A}(\omega,\mathbf{p})=0\;, 
\end{eqnarray}
whereas the symmetric regulator functions $R_{k}^{F}$ and $R_{k}^{\tilde F}$ vanish identically in this scheme. One particular choice of the regulator function, which has been frequently employed in the literature \cite{Litim:2001up} is
\begin{equation}
    r_k(\mathbf{p}) = (k^2-\mathbf{p}^2) \theta(k^2-\mathbf{p}^2) \; . \label{eq:simple_reg}
\end{equation}
However, a purely spatial regulator scheme has the obvious disadvantage that it can not be applied in $0+1$ dimensions, and moreover is also not particularly suitable for higher dimensional lattice models which feature a discrete set of spatial momenta. We will therefore explore a different possibility, where inspired by the free inverse propagator the regulator takes the form \cite{Canet:2009vz}
\begin{eqnarray}
\label{eq:complicated_reg}
R_{k}(\omega,\mathbf{p})= 
\begin{pmatrix} 
0 & -\mu_{k}(\omega,\mathbf{p})+i \omega\gamma_{k}(\omega,\mathbf{p}) \\
-\mu_{k}(\omega,\mathbf{p})-i \omega\gamma_{k}(\omega,\mathbf{p}) & 2 i \omega \gamma_{k}(\omega,\mathbf{p})n_{\rm eff}(\omega)
\end{pmatrix}  r_{k}(\omega,\mathbf{p})\;.
\end{eqnarray}
We emphasize that in the above expression $\mu_{k}(\omega,\mathbf{p})$ and $\gamma_{k}(\omega,\mathbf{p})$ are real-valued even functions of the frequency $\omega$, such that in addition to the real part $\propto \mu_{k}(\omega,\mathbf{p})$ which corresponds to an effective mass term, the regulator also features a non-vanishing imaginary part $\propto \omega \gamma_{k}(\omega,\mathbf{p})$, which for $\gamma_{k}(\omega,\mathbf{p}) >0$ corresponds to an effective damping rate. Specifically, for our $0+1$ dimensional case study, we will choose
\begin{eqnarray}
\mu_{k}(\omega)=k^2-\omega^2\;, \qquad \gamma_{k}(\omega)=\frac{\gamma}{\beta m^2} k\left(k-|\omega|\right)\;,
\label{eq:causal_reg}
\end{eqnarray}
such that both $\mu_{k}(\omega)$ and $\gamma_{k}(\omega)$ diverges as $k^2$ in the limit $k \to \Lambda$. Since the regulator diverges for sufficiently large $k$ all fluctuations are suppressed when the renormalization group scale $k$ approaches the UV cut-off scale $\Lambda$ such that the effective action $\Gamma_{k=\Lambda}[\phi,\tilde \phi]=S_{\mathcal{C}}[\phi,\tilde \phi]$ is given by the bare action. In analogy to Euclidean FRG calculations \cite{Wetterich:1992yh}, this can be easily demonstrated via a saddle point approximation of the path integral and we provide a short discussion in Appendix~B for completeness.

\subsection{Diagramatics}
While Eq.~(\ref{eq:FlowEqnGamma}) provides the flow equation for the effective action, it is more convenient  in practice to work directly with the flow equations for $n$-point correlation functions, which are obtained from Eq.~(\ref{eq:FlowEqnGamma}) by functional differentiation w.r.t to $\phi$ and $\tilde \phi$. Even though the differentiations can be carried out analytically, it is significantly more straightforward to employ graphical rules to perform the functional differentiations. We follow previous works in this context and start with the following
diagramatic representations of the propagators and regulators 
\begin{eqnarray}
\begin{picture}(349,60) (-23,-90)
      \SetWidth{1.0}
      \SetColor{Black}
      \Text(0,-45)[lb]{$\displaystyle G^R_k(x,y) =$}
      
      \Line[color=Blue](80,-38.5)(100,-38.5)
      \Line[color=Red](100,-38.5)(120,-38.5)

      \Text(70,-42)[lb]{$\displaystyle x$}
      \Text(123,-44)[lb]{$\displaystyle y,$}

      \Text(190,-45)[lb]{$\displaystyle \dot R_k^R(x,y) =$}

      \Line[color=Red](270,-38.5)(290,-38.5)
      \Line[color=Blue](290,-38.5)(310,-38.5)    
      \Line(287,-35.5)(293,-41.5)\Line(287,-41.5)(293,-35.5)  

      \Text(260,-42)[lb]{$\displaystyle x$}
      \Text(313,-44)[lb]{$\displaystyle y,$}

      \Text(0,-70)[lb]{$\displaystyle G^A_k(x,y) =$}

      \Line[color=Red](80,-63.5)(100,-63.5)
      \Line[color=Blue](100,-63.5)(120,-63.5)

      \Text(70,-67)[lb]{$\displaystyle x$}
      \Text(123,-69)[lb]{$\displaystyle y,$}

      \Text(190,-70)[lb]{$\displaystyle \dot R_k^A(x,y) =$}

      \Line[color=Blue](270,-63.5)(290,-63.5)
      \Line[color=Red](290,-63.5)(310,-63.5)
      \Line(287,-60.5)(293,-66.5)\Line(287,-66.5)(293,-60.5)  

      \Text(260,-67)[lb]{$\displaystyle x$}
      \Text(313,-69)[lb]{$\displaystyle y,$}

      \Text(0,-95)[lb]{$\displaystyle iF_k(x,y) =$}

      \Line[color=Blue](80,-88.5)(100,-88.5)
      \Line[color=Blue](100,-88.5)(120,-88.5)

      \Text(70,-92)[lb]{$\displaystyle x$}
      \Text(123,-94)[lb]{$\displaystyle y,$}

      \Text(190,-95)[lb]{$\displaystyle \dot R_A^F(x,y) =$}

      \Line[color=Red](270,-88.5)(290,-88.5)
      \Line[color=Red](290,-88.5)(310,-88.5)
      \Line(287,-85.5)(293,-91.5)\Line(287,-91.5)(293,-85.5)  

      \Text(260,-92)[lb]{$\displaystyle x$}
      \Text(313,-94)[lb]{$\displaystyle y.$}

\end{picture}
\end{eqnarray}
With these, the diagramatic representation of the flow equation~\eqref{eq:FlowEqnGamma} takes the compact form

\begin{eqnarray}
\partial_k \Gamma_k = - \frac{i}{2}
\begin{picture}(100,10) (160,-27)
      \SetWidth{1.0}
      \SetColor{Black}
      \Arc[color=Green](180,-23)(15,0,360)
      \Line(177,-5)(183,-11)\Line(177,-11)(183,-5)
\end{picture}
\hspace*{-2cm} ,
\end{eqnarray}
where -- as a novelty of our notation -- a green line is shorthand notation for either blue or red and the flow equation is a sum of all allowed color permutations. Notably the introduction of this compact matrix notation is particularly useful when deriving flow equations for higher $n$-point functions. Since the functional differentiation of the various propagators gives rise to all possible insertions of intermediate propagators, e.g.
\begin{eqnarray}
\frac{\delta}{\delta \chi_{a}(z)} G^R_k(x,y) &=& G^R_k(x,a) \Gamma^{\tilde \phi \phi}_{\chi,k}(a,z,b) G^R_k(b,y) + G^R_k(x,a) \Gamma^{\tilde \phi \tilde \phi}_{\chi,k}(a,z,b) i\tilde F_k(b,y)\\ &&+ iF_k(x,a) \Gamma^{\phi \tilde \phi}_{\chi,k}(a,z,b) i \tilde F_k(b,y) + iF_k(x,a) \Gamma^{\phi \phi}_{\chi,k}(a,z,b) G^R_k(b,y) \notag \, ,
\end{eqnarray}
the short hand notation 
\begin{eqnarray}
\frac{\delta}{\delta \chi_{a}(z)} G^R_k(x,y) =
\begin{picture}(100,20) (180,-35)
      \SetWidth{1.0}
      \SetColor{Black}
      \Line[color=Blue](205,-20)(225,-20)
      \Line[color=Green](225,-20)(265,-20)
      \Line[color=Red](265,-20)(285,-20)
      \Line(245,-20)(245,-40)

      \Text(195,-30)[lb]{$\displaystyle x$}
      \Text(290,-30)[lb]{$\displaystyle y$}
      \Text(245,-50)[lb]{$\displaystyle z$}
      \Text(300,-25)[lb]{$\displaystyle ,$}
\end{picture}
\end{eqnarray}
allows for an efficient bookkeeping with a drastically reduced number of the diagrams. Based on the diagrammatic shorthand notation, the flow equation of a generic two-point function can be compactly expressed as 
\begin{eqnarray}
\partial_{k} \Gamma_{k, ab}^{\alpha \bar \alpha}(x \bar x) = - \frac{i}{2}
\begin{picture}(100,30)(140,-30)
      \SetWidth{1.0}
      \SetColor{Black}
      \Arc[color=Green](180,-24)(15,0,360)
      \Line(177,-6)(183,-12)\Line(177,-12)(183,-6)
      \Text(210,-30)[lb]{$\displaystyle ,$}
      \Line(165,-39)(195,-39)
      \Text(145,-45)[lb]{$x,a$}
      \Text(200,-45)[lb]{$\bar x,b$}
\end{picture}
\end{eqnarray}
where also the black lines on the external legs can be either blue or red, depending on the particular two-point function under consideration.

\section{Explicit comparison to perturbation theory}
\label{sec:PT}
Before we proceed with our discussion of the real-time FRG approach, it proves insightful to analyze which set of perturbative contributions are included in the real-time functional renormalization group calculation.  Generally, our strategy for this purpose will be to expand the effective action into terms proportional to powers of $\lambda^n$
\begin{equation}
\Gamma_k[\phi, \tilde{\phi}] = S[\phi, \tilde{\phi}] + \sum_n \Delta^{(n)} \Gamma_k[\phi, \tilde{\phi}] \; .
\end{equation}
and then write down separate flow equations for all terms $\Delta^{(n)} \Gamma[\phi, \tilde{\phi}]$ to bring them into the form of a total differential such that the integration w.r.t to the scale parameter $k$ becomes trivial.
\subsection{One loop contributions to propagators and vertices}
Starting from the FRG flow equation for the effective action in Eq.~(\ref{eq:FlowEqnGamma}) it is evident, that to one loop order only bare propagators and vertices can appear on the RHS of the flow equation. Explicit evaluation of the (scale dependent) bare propagators yields the following expressions
\begin{align}
&G^{R(0)}_{k}(x\bar{x}) = \frac{-1}{S^{\tilde{\phi}\phi} + R_{k}^{R}} (x\bar{x}) \;,  \qquad
G^{A(0)}_{k}(x\bar{x}) = \frac{-1}{S^{\phi \tilde{\phi}} + R_{k}^{A}} (x\bar{x})\;, \qquad  \\
& \qquad\qquad iF^{(0)} _{k}(x\bar{x}) = \int_{v\bar{v}} G^{R(0)}_{k}(xv)~\left(S^{\tilde{\phi}\tilde{\phi}}(v\bar{v}) + R_{k}^{F} (v\bar{v})    \right)~G^{A(0)}_{k}(\bar{v}\bar{x})\;. \nonumber
\end{align}
Since the scale $(k)$ dependence only enters through the regulator itself, one finds the following explicit relations for the scale derivatives of the propagators
\begin{eqnarray}
\label{eq:G0Derivatives}
\partial_{k} G^{R(0)}_{k}(x\bar{x}) &=& \int_{v\bar{v}} G^{R(0)}_{k}(xv) \dot{R}_{k}^{R}(v\bar{v}) G^{R(0)}_{k}(\bar{v}\bar{x}) \; , \nonumber \\
\partial_{k} G^{A(0)}_{k}(x\bar{x}) &=& \int_{v\bar{v}} G^{A(0)}_{k}(xv) \dot{R}_{k}^{A}(v\bar{v}) G^{A(0)}_{k}(\bar{v}\bar{x}) \; ,  \\
\partial_{k} iF^{(0)}_{k}(x\bar{x}) &=& \int_{v\bar{v}}  \left[ G^{R(0)}_{k}(xv) \dot{R}_{k}^{R}(v\bar{v}) iF^{(0)}_{k}(\bar{v}\bar{x}) + iF^{(0)}_{k}(xv) \dot{R}_{k}^{A}(v\bar{v}) G^{A(0)}_{k}(\bar{v}\bar{x}) \right. \notag \\
&& \qquad \left.+ G^{R(0)}_k(xv)\dot{R}_{k}^{F}(v\bar{v}) G^{A(0)}_{k}(\bar{v}\bar{x})  \right] \; , \nonumber
\end{eqnarray}
which can be used to integrate the flow equations w.r.t. to $k$ as described below. Similarly, at one loop level all vertices appearing on the RHS of the flow equation are simply given in terms of the bare vertices and take the following explicit form
\begin{eqnarray}
S^{\phi\phi\phi\tilde{\phi}}_{ab\bar{b}\bar{a}}(xy\bar{y}\bar{x})
&=& \lambda_{cl} \left[\delta_{ab} \delta_{\bar{a}\bar{b}} +  \delta_{a\bar{a}} \delta_{b\bar{b}} + \delta_{a\bar{b}} \delta_{\bar{a}b} \right] 
\delta(x-y) \delta(\bar{x}-\bar{y}) \delta(x-\bar{x})\; ,  \\
S^{\phi\tilde{\phi}\tilde{\phi}\tilde{\phi}}_{ab\bar{b}\bar{a}}(xy\bar{y}\bar{x})
&=& \lambda_{qu} \left[\delta_{ab} \delta_{\bar{a}\bar{b}} +  \delta_{a\bar{a}} \delta_{b\bar{b}} + \delta_{a\bar{b}} \delta_{\bar{a}b} \right] 
\delta(x-y) \delta(\bar{x}-\bar{y}) \delta(x-\bar{x})\; ,
\end{eqnarray}
where we denote $\lambda_{cl}=-\frac{\lambda}{3N}$ and $\lambda_{qu}=-\frac{\lambda}{12N}$ in the following.

Specifically, for the two point functions the relevant flow equations then evaluate to
\begin{eqnarray}
\partial_{k} \Delta^{(1)}\Gamma_{k}^{\phi\tilde{\phi}}(x\bar{x}) &=& -\frac{i}{2} \delta(x-\bar{x}) (N+2) \lambda_{cl} \int_{u\bar{u}v\bar{v}} \delta(x-u)\delta(x-\bar{u})  \\
&& \times \left[ G^{R(0)}_{k}(uv) \dot{R}_{k}^{R}(v\bar{v}) iF^{(0)}_{k}(\bar{v}\bar{u}) + iF^{(0)}_{k}(uv) \dot{R}_{k}^{A}(v\bar{v}) G^{A(0)}_{k}(\bar{v}\bar{u})\right.\\
&& \quad \left. + G^{R(0)}_{k}(uv) \dot{R}_{k}^{F}(v\bar{v}) G^{A(0)}_{k}(\bar{v}\bar{u}) \right]\;, \nonumber \\
\partial_{k} \Delta^{(1)}\Gamma_{k}^{\tilde{\phi}\tilde{\phi}}(x\bar{x}) &=& -\frac{i}{2} \delta(x-\bar{x}) (N+2) \lambda_{qu}  \int_{u\bar{u}v\bar{v}} \delta(x-u)\delta(x-\bar{u}) \\
&& \times \left[ G^{R(0)}_{k}(uv) \dot{R}_{k}^{R}(v\bar{v}) G^{R(0)}_{k}(\bar{v}\bar{u}) + G^{A(0)}_{k}(uv) \dot{R}_{k}^{A}(v\bar{v}) G^{A(0)}_{k}(\bar{v}\bar{u}) \right]\;, \nonumber \\
\partial_{k} \Delta^{(1)}\Gamma_{k}^{\phi\phi}(x\bar{x}) &=& -\frac{i}{2} \delta(x-\bar{x}) (N+2) \lambda_{cl}  \int_{u\bar{u}v\bar{v}} \delta(x-u)\delta(x-\bar{u}) \\
&& \times \left[ G^{R(0)}_{k}(uv) \dot{R}_{k}^{R}(v\bar{v}) G^{R(0)}_{k}(\bar{v}\bar{u}) + G^{A(0)}_{k}(uv) \dot{R}_{k}^{A}(v\bar{v}) G^{A(0)}_{k}(\bar{v}\bar{u}) \right]\;, \nonumber 
\end{eqnarray}
where the 'flavor' factor $(N+2)$ comes from the following contraction of $O(N)$ indices
\begin{eqnarray}
(\delta_{a\bar{a}}\delta_{c\bar{c}} +\delta_{ac} \delta_{\bar{a}\bar{c}} +\delta_{ac} \delta_{\bar{a}\bar{c}}) \delta_{c\bar{c}} = (N+2) \delta_{a\bar{a}} \; .
\end{eqnarray}
By comparison with Eqns.~(\ref{eq:G0Derivatives}), one recognizes the RHS as total $k$-derivatives and the flow equations can be integrated w.r.t to the scale parameter $k$ yielding\footnote{Note that at $k=k_{UV}$ the action is given by the bare equation. Hence the corresponding boundary terms on the LHS vanish. Similarly, one the RHS the resulting propagators are suppressed for sufficiently large choice of the cut-off scale $k_{UV}$, again giving rise to vanishing boundary terms.} 
\begin{eqnarray}
\label{eq:DeltaGammaOneTwoPoint}
 \Delta^{(1)}\Gamma_{k}^{\phi\tilde{\phi}}(x\bar{x}) &=& -\frac{i}{2} \delta(x-\bar{x}) (N+2) \lambda_{cl}~iF^{(0)}_{k}(x\bar{x})\;,\nonumber \\
 \Delta^{(1)}\Gamma_{k}^{\tilde{\phi}\tilde{\phi}}(x\bar{x}) &=&  -\frac{i}{2} \delta(x-\bar{x}) (N+2) \lambda_{qu} \left[ G^{R(0)}_{k}(x\bar{x}) +  G^{A(0)}_{k}(x\bar{x})\right]= 0\;,  \\
  \Delta^{(1)}\Gamma_{k}^{\phi\phi}(x\bar{x}) &=&  -\frac{i}{2} \delta(x-\bar{x}) (N+2) \lambda_{cl} \left[ G^{R(0)}_{k}(x\bar{x}) +  G^{A(0)}_{k}(x\bar{x})\right]=0\;,\nonumber
\end{eqnarray}
irrespective of the details of the regulator, as long as the latter ensures the suppression of UV boundary terms and does not introduce violations of causality such that the terms in the last two lines vanish.

Based on the expressions in Eq.~(\ref{eq:DeltaGammaOneTwoPoint}), one immediately realizes that the only contribution at the one loop level is a manifestly real and local correction, which physically amounts to the familiar one loop mass shift $\Delta^{(1)}m^2_{k} = - \frac{1}{V_{d+1}}\int_{x\bar{x}} \Delta^{(1)}\Gamma_{k}^{\phi\tilde{\phi}}(x\bar{x})$. Hence one concludes that, in the absence of spontaneous symmetry breaking, any non-trivial modifications of the spectral shape only occur starting at the two loop level, and it is therefore important to understand how these are generated within the real-time FRG approach.

Beside the one loop correction to the two point function, we will also need the one loop corrections to the four point functions, which enters the perturbative calculation of the spectral function at the two loop level. Evidently, the one loop corrections to the four point functions can be obtained in an analogous fashion from the flow equation of the four point function
\begin{align}
\partial_{k} \Delta^{(1)}\Gamma^{\alpha\beta\bar{\beta}\bar{\alpha}}_{ab\bar{b}\bar{a}}(xy\bar{y}\bar{x}) =  -\frac{i}{2}\, \Bigg \lbrace\;\; 
 &\begin{picture}(110,40) (170,-37)
      \SetWidth{1.0}
      \SetColor{Black}
      \Arc[color=Green](210,-33)(15,0,360)
      \Line(207,-15)(213,-21)\Line(207,-21)(213,-15)
      \Line(185,-25)(194,-33)
      \Line(185,-41)(194,-33)
      \Line(235,-25)(226,-33)
      \Line(235,-41)(226,-33)
      \Text(172,-50)[lb]{\scriptsize $x,a \alpha$}
      \Text(172,-23)[lb]{\scriptsize $y,b \beta$}
      \Text(235,-23)[lb]{\scriptsize $\bar y,\bar b \bar \beta$}
      \Text(235,-50)[lb]{\scriptsize $\bar x,\bar a \bar \alpha$}
\end{picture}\hspace*{-1cm}+
\begin{picture}(110,40) (170,-37)
      \SetWidth{1.0}
      \SetColor{Black}
      \Arc[color=Green](210,-33)(15,0,360)
      \Line(207,-15)(213,-21)\Line(207,-21)(213,-15)
      \Line(185,-25)(194,-33)
      \Line(185,-41)(194,-33)
      \Line(235,-25)(226,-33)
      \Line(235,-41)(226,-33)
      \Text(172,-50)[lb]{\scriptsize $x,a \alpha$}
      \Text(172,-23)[lb]{\scriptsize $y,b \beta$}
      \Text(235,-23)[lb]{\scriptsize $\bar x,\bar a \bar \alpha$}
      \Text(235,-50)[lb]{\scriptsize $\bar y,\bar b \bar \beta$}
\end{picture}\hspace*{-1cm}+
\begin{picture}(110,40) (170,-37)
      \SetWidth{1.0}
      \SetColor{Black}
      \Arc[color=Green](210,-33)(15,0,360)
      \Line(207,-15)(213,-21)\Line(207,-21)(213,-15)
      \Line(185,-25)(194,-33)
      \Line(185,-41)(194,-33)
      \Line(235,-25)(226,-33)
      \Line(235,-41)(226,-33)
      \Text(172,-50)[lb]{\scriptsize $x,a \alpha$}
      \Text(172,-23)[lb]{\scriptsize $\bar x,\bar a \bar \alpha$}
      \Text(235,-23)[lb]{\scriptsize $y, b \beta$}
      \Text(235,-50)[lb]{\scriptsize $\bar y,\bar b \bar \beta$}
\end{picture}\\
+&\begin{picture}(110,40) (170,-37)
      \SetWidth{1.0}
      \SetColor{Black}
      \Arc[color=Green](210,-33)(15,0,360)
      \Line(207,-15)(213,-21)\Line(207,-21)(213,-15)
      \Line(185,-25)(194,-33)
      \Line(185,-41)(194,-33)
      \Line(235,-25)(226,-33)
      \Line(235,-41)(226,-33)
      \Text(172,-50)[lb]{\scriptsize $x,a \alpha$}
      \Text(172,-23)[lb]{\scriptsize $\bar x,\bar a \bar \alpha$}
      \Text(235,-23)[lb]{\scriptsize $\bar y,\bar b \bar \beta$}
      \Text(235,-50)[lb]{\scriptsize $y,b \beta$}
\end{picture}\hspace*{-1cm}+
\begin{picture}(110,40) (170,-37)
      \SetWidth{1.0}
      \SetColor{Black}
      \Arc[color=Green](210,-33)(15,0,360)
      \Line(207,-15)(213,-21)\Line(207,-21)(213,-15)
      \Line(185,-25)(194,-33)
      \Line(185,-41)(194,-33)
      \Line(235,-25)(226,-33)
      \Line(235,-41)(226,-33)
      \Text(172,-50)[lb]{\scriptsize $x,a \alpha$}
      \Text(172,-23)[lb]{\scriptsize $\bar y,\bar b \bar \beta$}
      \Text(235,-23)[lb]{\scriptsize $\bar x,\bar a \bar \alpha$}
      \Text(235,-50)[lb]{\scriptsize $y,b \beta$}
\end{picture}\hspace*{-1cm}+
\begin{picture}(110,40) (170,-37)
      \SetWidth{1.0}
      \SetColor{Black}
      \Arc[color=Green](210,-33)(15,0,360)
      \Line(207,-15)(213,-21)\Line(207,-21)(213,-15)
      \Line(185,-25)(194,-33)
      \Line(185,-41)(194,-33)
      \Line(235,-25)(226,-33)
      \Line(235,-41)(226,-33)
      \Text(172,-50)[lb]{\scriptsize $x,a \alpha$}
      \Text(172,-23)[lb]{\scriptsize $\bar y,\bar b \bar \beta$}
      \Text(235,-23)[lb]{\scriptsize $y, b \beta$}
      \Text(235,-50)[lb]{\scriptsize $\bar x,\bar a \bar \alpha$}
\end{picture}\hspace*{-0.7cm} \Bigg \rbrace \; . \notag
\end{align}
Based on the apparent symmetries of the corresponding diagrams, we can decompose the one-loop corrections to the four point functions according to
\begin{eqnarray}
\hspace*{-0.7cm} \Delta^{(1)}\Gamma^{\alpha\beta\bar{\beta}\bar{\alpha}}_{k,~ab\bar{b}\bar{a}}(xy\bar{y}\bar{x})
&=& \label{1l-4pt}
\left[(N+4) \delta_{ab} \delta_{\bar{a}\bar{b}} + 2 \delta_{a\bar{a}} \delta_{b\bar{b}} +2 \delta_{a\bar{b}} \delta_{\bar{a}b} \right] 
\delta(x-y) \delta(\bar{x}-\bar{y}) \Delta^{(1)}\Gamma_{k}^{\alpha\beta,\bar{\beta}\bar{\alpha}}(x\bar{x}) \nonumber \\
&+&
\left[2\delta_{ab} \delta_{\bar{a}\bar{b}} + (N+4)  \delta_{a\bar{a}} \delta_{b\bar{b}} +2 \delta_{a\bar{b}} \delta_{\bar{a}b} \right] 
\delta(x-\bar{x}) \delta(y-\bar{y}) \Delta^{(1)}\Gamma_{k}^{\alpha\bar{\alpha},\beta\bar{\beta}}(xy)  \\
&+&
\left[2 \delta_{ab} \delta_{\bar{a}\bar{b}} + 2 \delta_{a\bar{a}} \delta_{b\bar{b}} + (N+4) \delta_{a\bar{b}} \delta_{\bar{a}b} \right] 
\delta(x-\bar{y}) \delta(\bar{x}-y) \Delta^{(1)}\Gamma_{k}^{\alpha\bar{\beta},\beta\bar{\alpha}}(x\bar{x}) \nonumber \; ,
\end{eqnarray}
where the $O(N)$ index structure of the expression is obtained by evaluating the index contraction of bare propagators and vertices according to
\begin{eqnarray}
(\delta_{ab}\delta_{ef} + \delta_{af} \delta_{be} + \delta_{ae} \delta_{bf}) 
\delta_{e\bar{e}} \delta_{f\bar{f}} 
(\delta_{\bar{a}\bar{b}}\delta_{\bar{e}\bar{f}} + \delta_{\bar{a}\bar{f}} \delta_{\bar{b}\bar{e}} + \delta_{\bar{a}\bar{e}} \delta_{\bar{b}\bar{f}}) = (N+4) \delta_{ab} \delta_{\bar{a}}\delta_{\bar{b}} + 2 \delta_{a\bar{a}} \delta_{b\bar{b}} +2 \delta_{a\bar{b}} \delta_{\bar{a}b}\;. \notag
\end{eqnarray}
One is then left with the calculation of the one-loop vertex functions $\Delta^{(1)}\Gamma_{k}^{\phi\phi,\phi\tilde{\phi}}(x\bar{x})$ of the classical ($\phi\phi\phi\tilde{\phi}$) vertex, $\Delta^{(1)}\Gamma_{k}^{\phi\tilde{\phi},\tilde{\phi}\tilde{\phi}}(x\bar{x})$ of the quantum ($\phi \tilde{\phi}\tilde{\phi}\tilde{\phi}$) vertex, as well as the two vertex functions $\Delta^{(1)}\Gamma_{k}^{\phi\tilde{\phi},\phi\tilde{\phi}}(x\bar{x})$ and $\Delta^{(1)}\Gamma_{k}^{\phi\phi,\tilde{\phi}\tilde{\phi}}(x\bar{x})$ of the anomalous ($\phi\phi \tilde{\phi}\tilde{\phi}$) vertex. By combining the individual terms in an appropriate fashion, we can compactly express the result in the form
\begin{eqnarray}
\partial_{k} \Delta^{(1)}\Gamma_{k}^{\phi\phi,\phi\tilde{\phi}}(x\bar{x})&=&
-\frac{i}{2}\lambda_{cl}^2 \left[ \left( \partial_{k} iF^{(0)}_{k}(x\bar{x})\right) G_{k}^{A(0)}(x\bar{x}) +  iF^{(0)}_{k}(x\bar{x})  \left( \partial_{k} G_{k}^{A(0)}(x\bar{x}) \right) \right.  \\
&& \quad \quad ~ +  \left .\left( \partial_{k} iF^{(0)}_{k}(\bar{x}x)\right) G_{k}^{R(0)}(\bar{x}x) +  iF^{(0)}_{k}(\bar{x}x)  \left( \partial_{k} G_{k}^{R(0)}(\bar{x}x) \right) \right]\;, \nonumber \\
\partial_{k} \Delta^{(1)}\Gamma_{k}^{\tilde{\phi}\tilde{\phi},\tilde{\phi}\phi}(x\bar{x})&=& 
- \frac{i}{2} \lambda_{cl} \lambda_{qu}  \left[ \left( \partial_{k} iF^{(0)}_{k}(x\bar{x})\right) G_{k}^{A(0)}(x\bar{x}) +  iF^{(0)}_{k}(x\bar{x})  \left( \partial_{k} G_{k}^{A(0)}(x\bar{x}) \right) \right.  \\
&& \quad \qquad ~~ +  \left .\left( \partial_{k} iF^{(0)}_{k}(\bar{x}x)\right) G_{k}^{R(0)}(\bar{x}x) +  iF^{(0)}_{k}(\bar{x}x)  \left( \partial_{k} G_{k}^{R(0)}(\bar{x}x) \right) \right]\;, \nonumber \\
\partial_{k} \Delta^{(1)}\Gamma_{k}^{\phi\tilde{\phi},\phi\tilde{\phi}}(x\bar{x})&=&
-\frac{i}{2}\lambda_{cl}^2  \left[ \left( \partial_{k} iF^{(0)}_{k}(x\bar{x})\right) iF_{k}^{(0)}(x\bar{x}) +  iF^{(0)}_{k}(x\bar{x})  \left( \partial_{k} iF_{k}^{(0)}(x\bar{x}) \right) \right] \\
&&-\frac{i}{2}\lambda_{cl} \lambda_{qu} \left[ \left( \partial_{k} G_{k}^{A(0)}(x\bar{x})\right) G_{k}^{A(0)}(x\bar{x}) +  G_{k}^{A(0)}(\bar{x}x)  \left( \partial_{k} G_{k}^{A(0)}(\bar{x}x) \right) \right. \nonumber \\
&& \quad \qquad ~~ +  \left . \left( \partial_{k} G_{k}^{R(0)}(x\bar{x})\right) G_{k}^{R(0)}(x\bar{x}) +  G_{k}^{R(0)}(\bar{x}x)  \left( \partial_{k} G_{k}^{R(0)}(\bar{x}x) \right) \right]\;, \nonumber \\
\partial_{k} \Delta^{(1)}\Gamma_{k}^{\phi\phi,\tilde{\phi}\tilde{\phi}}(x\bar{x})&=&
-\frac{i}{2}\lambda_{cl} \lambda_{qu}  \left[ \left( \partial_{k}G_{k}^{A(0)}(x\bar{x})\right) G_{k}^{R(0)}(x\bar{x}) +  G_{k}^{A(0)}(x\bar{x})  \left( \partial_{k} G_{k}^{R(0)}(x\bar{x}) \right) \right. \\
&& \quad \qquad ~~ +  \left .\left( \partial_{k}G_{k}^{A(0)}(\bar{x}x)\right) G_{k}^{R(0)}(\bar{x}x) +  G_{k}^{A(0)}(\bar{x}x)  \left( \partial_{k} G_{k}^{R(0)}(\bar{x}x) \right) \right]\;. \nonumber
\end{eqnarray}
Since the RHS represents a total derivative w.r.t to the scale $k$, the above flow equations can be integrated yielding the following result for the (scale dependent) one-loop vertex functions
\begin{eqnarray}
\Delta^{(1)}\Gamma_{k}^{\phi\phi,\phi\tilde{\phi}}(x\bar{x})&=& -\frac{i}{2}\lambda_{cl}^2~2 iF^{(0)}_{k}(x\bar{x}) G_{k}^{A(0)}(x\bar{x})\;, \\
\Delta^{(1)}\Gamma_{k}^{\tilde{\phi}\tilde{\phi},\tilde{\phi}\phi}(x\bar{x})&=& -\frac{i}{2} \lambda_{cl} \lambda_{qu}~2 iF^{(0)}_{k}(x\bar{x}) G_{k}^{A(0)}(x\bar{x})\;,  \\
\Delta^{(1)}\Gamma_{k}^{\phi\tilde{\phi},\phi\tilde{\phi}}(x\bar{x})&=& -\frac{i}{2}\lambda_{cl}^2  \left( iF_{k}^{(0)}(x\bar{x}) \right)^2 -\frac{i}{2}\lambda_{cl} \lambda_{qu} \left[ \left( G_{k}^{R(0)}(x\bar{x}) \right)^2  + \left( G_{k}^{A(0)}(x\bar{x}) \right)^2 \right]\;,  \\
\Delta^{(1)}\Gamma_{k}^{\phi\phi,\tilde{\phi}\tilde{\phi}}(x\bar{x})&=& -\frac{i}{2}\lambda_{cl} \lambda_{qu}~2 G_{k}^{A(0)}(x\bar{x}) G_{k}^{R(0)}(x\bar{x})=0\;, 
\end{eqnarray}
where we exploited the symmetries
\begin{eqnarray}
G_{k}^{A(0)}(x\bar{x})= G_{k}^{R(0)}(\bar{x}x)\;, \qquad iF^{(0)}_{k}(x\bar{x})=iF^{(0)}_{k}(\bar{x}x)\;,
\end{eqnarray}
to further compactify the expressions. We note in passing that for the quantum theory, the (tree-level) symmetry relation $\lambda_{cl}=4\lambda_{qu}$ between the local classical and quantum vertices also holds for the non-local vertex functions at the one loop level, i.e.
$\Delta^{(1)}\Gamma_{k}^{\phi\phi,\phi\tilde{\phi}}(x\bar{x})= 4 \Delta^{(1)}\Gamma_{k}^{\phi\tilde{\phi},\tilde{\phi}\tilde{\phi}}(x\bar{x})$. Nevertheless, there is also a non-local $\Delta^{(1)}\Gamma_{k}^{\phi\tilde{\phi},\phi\tilde{\phi}}(x\bar{x})$ vertex generated at one loop level, in both classical and quantum theories.

\subsection{Two loop contributions to propagators}
Since the flow equation for the propagators is of one-loop form, we can obtain the two loop contribution in a similar fashion, by using one propagator or respectively one vertex at one-loop order and use bare versions for all other quantities, i.e. 
   \begin{equation}
   \Delta^{(2)} \partial_k {\Gamma}_{k,ab}^{\tilde{\phi}\phi} 
      \, =\, -\frac{i}{2} \, \Big\{
    \begin{picture}(100,20) (100,-40)
      \SetWidth{0.5}
      \SetColor{Black}
      \Text(140,-39)[lb]{$\displaystyle +$}
      \Text(190,-39)[lb]{$\displaystyle +$}
      \Text(5,-48)[lb]{$\displaystyle $}
    \Text(110,-55)[lb]{$a$}
    \Text(126,-55)[lb]{$b$}
    \Text(160,-55)[lb]{$a$}
    \Text(176,-55)[lb]{$b$}
    \Text(210,-55)[lb]{$a$}
    \Text(226,-55)[lb]{$b$}
      \SetWidth{1.0}
      \DoubleArc[color=Green](120,-34)(10,100,180){2}
      \DoubleArc[color=Green](120,-34)(10,180,-80){2}
      \Arc[color=Green](120,-34)(10,0,100)
      \Arc[color=Green](120,-34)(10,-80,0)
      \SetWidth{0.5}
      \SetWidth{1.0}
      \Line(117,-21)(123,-27)\Line(117,-27)(123,-21)
      \Line[color=Red](110,-45)(120,-45)
      \Line[color=Blue](120,-45)(130,-45)
      \Arc[color=Green](170,-34)(10,90,270)
      \DoubleArc[color=Green](170,-34)(10,0,90){2}
      \DoubleArc[color=Green](170,-34)(10,-90,0){2}
     \SetWidth{0.5}
      \SetWidth{1.0}
      \Line(167,-21)(173,-27)\Line(167,-27)(173,-21)
      \Line[color=Red](160,-45)(170,-45)
      \Line[color=Blue](170,-45)(180,-45)
     \SetWidth{1.0}
      \Arc[color=Green](220,-34)(10,90,180)
      \Arc[color=Green](220,-34)(10,180,270)
      \Arc[color=Green](220,-34)(10,0,90)
      \Arc[color=Green](220,-34)(10,-90,0)
      \SetWidth{0.5}
      \SetWidth{1.0}
      \Line(217,-21)(223,-27)\Line(217,-27)(223,-21)
      \Line[color=Red](210,-45)(220,-45)
      \Line[color=Blue](220,-45)(230,-45)
      \Vertex(220,-44){3.4}
    \end{picture}
    \hspace*{1.3cm} \Big\}\,
  \end{equation}
where the black dot denotes the perturbative one-loop vertex and double lines denote the perturbative one-loop propagators given by 
  \begin{align}
      &\Delta^{(1)}G^R_k = G^{R(0)}_k \Delta^{(1)} \Gamma_k^{\tilde{\phi}\phi} G^{R(0)}_k \; , \qquad
      \Delta^{(1)}G^A_k = G^{A(0)}_k \Delta^{(1)} \Gamma_k^{\phi \tilde{\phi}} G^{A(0)}_k \; , \\
      &\qquad \Delta^{(1)}iF_k = G^{R(0)}_k \Delta^{(1)} \Gamma_k^{\tilde{\phi}\phi} iF^{(0)}_k + iF^{(0)}_k \Delta^{(1)}\Gamma_k^{\phi \tilde{\phi}} G^{A(0)}_k\;. \nonumber 
  \end{align}
  By inserting the corresponding expressions into the above flow equation, one finds that the contributions to $\partial_k \Delta^{(2)} \Gamma_k^{\phi \tilde{\phi}}(x\bar{x})$ fall in two topologically different categories, given by "double bubble" and "sunset" diagrams respectively.  By performing a straightforward but cumbersome set of manipulations, the contributions from diagrams with a one-loop propagator, can be expressed as
    \begin{align}
      &\left.\partial_k \Delta^{(2)} \Gamma_k^{\phi \tilde{\phi}}(x\bar{x})\right|_{\text{propagator}}^{\text{double bubble}}= - \frac{i}{2} \delta(x-\bar x) (N+2) \lambda_{cl} \delta_{a \bar a}  \notag \\
      & \int_{w \bar w} \Big \lbrace \partial_k \Big[ G^{R(0)}_k(xw) \Delta^{(1)} \Gamma^{\tilde \phi \phi}_k(w \bar w) iF^{(0)}(\bar w x)
          + iF^{(0)}(xw) \Delta^{(1)} \Gamma^{\phi \tilde \phi}_k(w \bar w) G^{A(0)}_k(\bar w x) \Big] \\
          &- G^{R(0)}_k(xw) (\partial_k \Delta^{(1)} \Gamma^{\tilde \phi \phi}_k(w \bar w)) iF^{(0)_k}(\bar w x)
          -iF^{(0)}_k(xw) (\partial_k \Delta^{(1)} \Gamma^{ \phi \tilde \phi}_k(w \bar w)) G^{A(0)}_k(\bar w x) \Big\rbrace\;. \notag
   \end{align}
   By combining this contribution with a corresponding set of double-bubble diagrams with a one-loop vertex, which upon further manipulations and dropping off vanishing terms can be compactly expressed in the form
   \begin{align}
    &\left.\partial_k \Delta^{(2)} \Gamma_k^{\phi \tilde{\phi}}(x\bar{x})\right|_{\text{vertex}}^{\text{double bubble}}= -\frac{i}{2} \delta(x- \bar x) (N+2)  \lambda_{cl} \delta_{a\bar a} \\
     &\int_{w \bar{w}} \Big(  G^{R(0)}_k(xw) iF^{(0)}_k(\bar{uw}\bar{x})  +   iF^{(0)}_k(xw) G^{A(0)}_k(\bar{w}\bar{x}) \Big) \partial_k \Delta^{(1)} \Gamma^{\phi \tilde \phi}_k(w \bar{w}) \notag \;,
    \end{align}
    one finds that the sum of two contributions yields a total derivative w.r.t $k$, such that
     \begin{align}
          &\left.\Delta^{(2)} \Gamma_k^{\phi \tilde{\phi}}(x\bar{x})\right|^{\text{double bubble}}= - \frac{i}{2} \delta(x-\bar x) (N+2) \lambda_{cl} \delta_{a \bar a} \notag \\
      & \int_{w \bar w}   \Big[ G^{R(0)}_k(xw) \Delta^{(1)} \Gamma^{\tilde \phi \phi}_k(w \bar w) iF^{(0)}(\bar w x)
          + iF^{(0)}(xw) \Delta^{(1)} \Gamma^{\phi \tilde \phi}_k(w \bar w) G^{A(0)}_k(\bar w x) \Big]\;,  
             \end{align}
    yielding
             \begin{align}
          &\left.\Delta^{(2)} \Gamma_k^{\phi \tilde{\phi}}(x\bar{x})\right|^{\text{double bubble}}=\left( \frac{i}{2} \right)^2 \delta(x-\bar x) (N+2)^2 \lambda_{cl}^2 \delta_{a \bar a} \label{eq:Delta2Gamma2TwoLoopBubble} \\
      &\int_{w} \Big[ G^{R(0)}_k(xw) iF^{(0)}_k(w w) iF^{(0)}(w x) + iF^{(0)}(xw) iF^{(0)}(w x) G^{A(0)}_k(\bar w x) \Big]\;. \notag
     \end{align}
     Similar to the one loop correction $\Delta^{(1)} \Gamma_k^{\phi \tilde{\phi}}(x\bar{x})$, this term is manifestly real and local providing the two loop correction to the mass shift. However, there is also the contribution from the sunset diagrams which can be compactly expressed as
     \begin{align}\label{eq:Delta2Gamma2TwoLoopSunset}
    \Delta^{(2)} \Gamma_{k}^{\phi \tilde \phi}(x \bar x)\Big|^\text{sunset} = -\frac{3}{2} (N+2) \delta_{a \bar a} \Big \lbrace &
    \lambda_{cl}^2 i F_{k}^{(0)}(x\bar x)  iF_{k}^{(0)}(x\bar x) G^{A(0)}_k(x \bar x) + \lambda_{cl} \lambda_{qu} \frac{1}{3} (G^{A(0)}_k(x\bar x))^3 \Big \rbrace\; .
    \end{align}
 Clearly this contribution to the effective action is non-local and posses a non-vanishing imaginary part, which describes the collisional broadening of the spectral function. We further emphasize, that in the real-time FRG framework the sunset contribution arises entirely due to the one-loop vertex correction, indicating the importance of including non-local vertex structures into the truncation of the real-time FRG flow equations. By including these non-local vertex structures, as in Eq.~(\ref{1l-4pt}), one is then able to derive the two-loop perturbative contributions to the damping rate \cite{Wang:1995qf}, as discussed in Appendix C.
\section{Non-trivial truncations for real-time calculations}
\label{sec:Trunc}
Based on our perturbative analysis of the flow equations in the preceding section, we conclude that a two loop complete truncation scheme for the two-point function is necessary to describe the collisional broadening of the spectral function in the symmetric phase. We have also observed, that a two loop complete truncation scheme for the two point function necessarily has to include a non-local four field interaction (e.g. generated at the one loop level), indicating the the local potential approximation that is commonly used in Euclidean FRG calculations is insufficient for the purpose of real time calculations.

Now in order to devise a more suitable truncation scheme, we first note that we can generally express the scale dependent effective action in a vertex expansion as
\begin{eqnarray}
\Gamma_{k}[\phi,\tilde{\phi}]= \sum_{n=1}^{Q} \frac{1}{n!} \left(\prod_{i=1}^{n} \int_{x_i} \right) \sum_{j=0}^{n} \begin{pmatrix}
n \\ j
\end{pmatrix} \Gamma^{(j,n-j)}_k(\{x\}) \left(\prod_{l=1}^{j} \phi(x_{l}) \right)  \left(\prod_{m=j+1}^{n}  \tilde{\phi}(x_{m}) \right)\;,
\end{eqnarray}
Since in the symmetric phase only the $n-$even terms contribute, a two-loop complete expansion can be achieved by truncating the vertex expansion at the level of the four-point function ($Q=4$) keeping only two- and four-point functions. Hence, the simplest possible two-loop complete expansion scheme is given by
\begin{eqnarray}
\label{eq:EffActTrunc}
\Gamma_{k}[\phi,\tilde{\phi}]=\frac{1}{2} && \int_{x \bar{x}} \begin{pmatrix}\phi_a(x) & \tilde{\phi}_a(x) \end{pmatrix}  \begin{pmatrix} 0 & \Gamma_{k,a\bar{a}}^{\phi\tilde{\phi}}(x\bar{x})  \\   \Gamma_{k,a\bar{a}}^{\tilde{\phi}\phi}(x\bar{x}) & \Gamma_{k,a\bar{a}}^{\tilde{\phi}\tilde{\phi}}(x\bar{x}) \end{pmatrix} \begin{pmatrix} \phi_{\bar{a}}(\bar{x} ) \\ \tilde{\phi}_{\bar{a}}(\bar{x}) \end{pmatrix}  \notag \\
+ \frac{1}{3!} &&\int_{x\bar{x}y\bar{y}} \phi_a(x) \phi_{b}(y)  \Gamma_{k,ab\bar{b}\bar{a}}^{\phi\phi\phi\tilde{\phi}}(xy\bar{y}\bar{x}) \phi_{\bar{b}}(\bar{y}) \tilde{\phi}_{\bar{a}}(\bar{x}) \\
+ \frac{1}{2!2!} &&\int_{x\bar{x}y\bar{y}} \phi_a(x) \tilde{\phi}_{b}(y) \Gamma_{k,ab\bar{b}\bar{a}}^{\phi\tilde{\phi}\phi\tilde{\phi}}(xy\bar{y}\bar{x})  \phi_{\bar{b}}(\bar{y}) \tilde{\phi}_{\bar{a}}(\bar{x}) \nonumber \\
+ \frac{1}{3!} &&\int_{x\bar{x}y\bar{y}}  \phi_a(x) \tilde{\phi}_{b}(y)  \Gamma_{k,ab\bar{b}\bar{a}}^{\phi\tilde{\phi}\tilde{\phi}\tilde{\phi}}(xy\bar{y}\bar{x})\tilde{\phi}_{\bar{b}}(\bar{y}) \tilde{\phi}_{\bar{a}}(\bar{x})\;, \notag
\end{eqnarray}
where the above truncation only takes vertices into account that can be generated at one-loop level, i.e. the $(\phi \phi \phi \phi)$, $(\tilde \phi \tilde \phi \tilde \phi \tilde \phi)$ and $(\phi \phi, \tilde \phi \tilde \phi)$ vertices vanish. With regards to the non-vanishing vertex functions, we employ a generalization of the one-loop result in Eq.~(\ref{1l-4pt}) as our ansatz
\begin{eqnarray}
\label{eq:VertexAnsatzGen}
\hspace*{-1cm} \Gamma^{\phi\phi\phi\tilde{\phi}}_{k,ab\bar{b}\bar{a}}(xy\bar{y}\bar{x}) 
&=& 
\left[v_{cl,A,k}^{diag}(x\bar{x}) \delta_{ab} \delta_{\bar{a}\bar{b}} + v_{cl,A,k}^{off}(x\bar{x}) \delta_{a\bar{a}} \delta_{b\bar{b}} + v_{cl,A,k}^{off}(x\bar{x})  \delta_{a\bar{b}} \delta_{\bar{a}b} \right] 
\delta(x-y) \delta(\bar{x}-\bar{y}) \notag \\
&+&
\left[v_{cl,A,k}^{off}(y\bar{x})\delta_{ab} \delta_{\bar{a}\bar{b}} + v_{cl,A,k}^{diag}(y\bar{x})  \delta_{a\bar{a}} \delta_{b\bar{b}} +v_{cl,A,k}^{off}(y\bar{x}) \delta_{a\bar{b}} \delta_{\bar{a}b} \right] 
\delta(x-\bar{x}) \delta(y-\bar{y})  \label{four-pt1} \\
&+&
\left[v_{cl,A,k}^{off}(\bar{y}\bar{x}) \delta_{ab} \delta_{\bar{a}\bar{b}} + v_{cl,A,k}^{off}(\bar{y}\bar{x}) \delta_{a\bar{a}} \delta_{b\bar{b}} + v_{cl,A,k}^{diag}(\bar{y}\bar{x}) \delta_{a\bar{b}} \delta_{\bar{a}b} \right] 
\delta(x-\bar{y}) \delta(\bar{x}-y)  \; , \nonumber
\end{eqnarray}
\begin{eqnarray}
\hspace*{-1cm} \Gamma^{\phi\tilde{\phi}\phi\tilde{\phi}}_{k,ab\bar{b}\bar{a}}(xy\bar{y}\bar{x})
&=& 
\left[v_{anom,k}^{diag}(x\bar{x}) \delta_{ab} \delta_{\bar{a}\bar{b}} + v_{anom,k}^{off}(x\bar{x}) \delta_{a\bar{a}} \delta_{b\bar{b}} +v_{anom,k}^{off}(x\bar{x}) \delta_{a\bar{b}} \delta_{\bar{a}b} \right] 
\delta(x-y) \delta(\bar{x}-\bar{y}) \label{four-pt2}   \\
&+&
\left[v_{anom,k}^{off}(y\bar{x})\delta_{ab} \delta_{\bar{a}\bar{b}} + v_{anom,k}^{diag}(y\bar{x})  \delta_{a\bar{a}} \delta_{b\bar{b}} +v_{anom,k}^{off}(y\bar{x}) \delta_{a\bar{b}} \delta_{\bar{a}b} \right] 
\delta(x-\bar{x}) \delta(y-\bar{y}) \; ,  \nonumber 
\end{eqnarray}
\begin{eqnarray}
\hspace*{-1cm} \Gamma^{\phi\tilde{\phi}\tilde{\phi}\tilde{\phi}}_{k,ab\bar{b}\bar{a}}(xy\bar{y}\bar{x})
&=&
\left[v_{qu,R,k}^{diag}(x\bar{x}) \delta_{ab} \delta_{\bar{a}\bar{b}} + v_{qu,R,k}^{off}(x\bar{x}) \delta_{a\bar{a}} \delta_{b\bar{b}} + v_{qu,R,k}^{off}(x\bar{x})  \delta_{a\bar{b}} \delta_{\bar{a}b} \right] 
\delta(x-y) \delta(\bar{x}-\bar{y})  \notag \\
&+&
\left[v_{qu,R,k}^{off}(y\bar{x})\delta_{ab} \delta_{\bar{a}\bar{b}} + v_{qu,R,k}^{diag}(y\bar{x})  \delta_{a\bar{a}} \delta_{b\bar{b}} +v_{qu,R,k}^{off}(y\bar{x}) \delta_{a\bar{b}} \delta_{\bar{a}b} \right] 
\delta(x-\bar{x}) \delta(y-\bar{y})  \label{four-pt4} \\
&+&
\left[v_{qu,R,k}^{off}(\bar{y}\bar{x}) \delta_{ab} \delta_{\bar{a}\bar{b}} + v_{qu,R,k}^{off}(\bar{y}\bar{x}) \delta_{a\bar{a}} \delta_{b\bar{b}} + v_{qu,R,k}^{diag}(\bar{y}\bar{x}) \delta_{a\bar{b}} \delta_{\bar{a}b} \right] 
\delta(x-\bar{y}) \delta(\bar{x}-y)  \nonumber \; ,
\end{eqnarray}
with scale dependent vertex functions $v^{X}_{Y,k}(x\bar{x})$. While at the one-loop level the diagonal $v^{diag}_{Y,k}(x\bar{x})$ and off-diagonal $v^{off}_{Y,k}(x\bar{x})$ vertex functions are simply related by a factor of $(N+4)/2$, this is not the case beyond one-loop  and we generally have to distinguish between diagonal and off-diagonal vertex functions. Based on the symmetries of the effective action for an equilibrium system, the above vertex functions satisfy the following symmetry relations
\begin{eqnarray}
v_{cl,R,k}^X(x\bar{x})&=& v_{cl,A,k}^X(\bar{x}x)\;, \label{vertex-rel1} \\
v_{qu,R,k}^X(x\bar{x})&=& \frac{1}{4} v_{cl,R,k}^X(x\bar{x})\;, 
\end{eqnarray}
as well as the fluctuation dissipation relation
\begin{eqnarray}
\tilde{v}_{anom,k}^X(p)= n_{\rm eff}(p_0) \Big( \tilde{v}_{cl,R,k}^X(p) - \tilde{v}_{cl,A,k}^X(p) \Big)\;.  \label{vertex-rel2}
\end{eqnarray}

\subsection{Explicit form of flow equations for two-point functions}
Based on the truncation of the effective action in Eq.~(\ref{eq:EffActTrunc}), the two-point equations obey the flow equation
\begin{eqnarray}
\partial_{k} \Gamma_{k}^{\phi\tilde{\phi}}(x\bar{x}) &=& \frac{-i}{2} 
\begin{picture}(35,20)(200,-40) 
  \SetWidth{1.0}
  \Arc[color=Green](220,-34)(10,90,180)
  \Arc[color=Green](220,-34)(10,180,270)
  \Arc[color=Green](220,-34)(10,0,90)
  \Arc[color=Green](220,-34)(10,-90,0)
  \SetWidth{0.5}
  \SetWidth{1.0}
  \Line(217,-21)(223,-27)\Line(217,-27)(223,-21)
  \Line[color=Blue](210,-45)(220,-45)
  \Line[color=Red](220,-45)(230,-45)
  \Vertex(220,-44){3.4}
\end{picture},
\\
\partial_{k} \Gamma_{k}^{\tilde{\phi}\tilde{\phi}}(x\bar{x})&=& \frac{-i}{2}
\begin{picture}(35,20)(200,-40) 
  \SetWidth{1.0}
  \Arc[color=Green](220,-34)(10,90,180)
  \Arc[color=Green](220,-34)(10,180,270)
  \Arc[color=Green](220,-34)(10,0,90)
  \Arc[color=Green](220,-34)(10,-90,0)
  \SetWidth{0.5}
  \SetWidth{1.0}
  \Line(217,-21)(223,-27)\Line(217,-27)(223,-21)
  \Line[color=Red](210,-45)(220,-45)
  \Line[color=Red](220,-45)(230,-45)
  \Vertex(220,-44){3.4}
\end{picture},
\end{eqnarray}
which upon inserting the explicit expressions for the non-local vertex functions in Eq.~(\ref{four-pt1}-\ref{four-pt4}), gives rise to the following structure of the flow equations
\begin{align}
\partial_{k} \Gamma_{k}^{\phi\tilde{\phi}}(x\bar{x}) =& \frac{-i}{2} \label{flow-2pt1}
\begin{axopicture}(200,40)(0,7)
  \SetWidth{0.5}
  \SetTextScale{15}
  \Text(2,0)[lb]{$\displaystyle\Big ($}
  \SetTextScale{15}
  \Text(13,-2)[lb]{$x$}
  \Text(60,-2)[lb]{$\bar x$}
  \Text(57,7)[lb]{\quad+}
  \SetWidth{1}
  \Bezier[color=Green](28,8)(19,18)(19,28)(39,28)
  \Line[color=Blue](20,-2)(28,8)
  \Bezier[color=Green](50,8)(59,18)(59,28)(39,28)
  \Line[color=Red](50,8)(58,-2)
  \Line(36,25)(42,31)\Line(36,31)(42,25)
  \COval(39,8)(3,10)(0.0){Black}{Black}
  \Text(78,-2)[lb]{$x$}
  \Text(125,-2)[lb]{$\bar x$}
  \Bezier[color=Green](93,8)(93,18)(124,18)(104,28)
  \Line[color=Blue](85,-2)(93,8)
  \Bezier[color=Green](115,8)(115,18)(83,18)(104,28)
  \Line[color=Red](115,8)(123,-2)
  \COval(104,8)(3,10)(0.0){Black}{Black}
  \Line(101,25)(107,31)\Line(101,31)(107,25)
  \Text(122,7)[lb]{\quad+}
  \Text(143,-2)[lb]{$x$}
  \Text(167,-2)[lb]{$\bar x$}
  \Line[color=Blue](158,8)(150,-2)
  \Line[color=Red](158,8)(165,-2)
  \Arc[color=Green](158,36)(8,0,360)
  \COval(158,18)(10,3)(0.0){Black}{Black}
  \Line(155,41)(161,47)\Line(155,47)(161,41)
  \Text(175,-0)[lb]{$\displaystyle \Big ) \; ,$}
\end{axopicture}
\\
\partial_{k} \Gamma_{k}^{\tilde{\phi}\tilde{\phi}}(x\bar{x})=& \frac{-i}{2} \label{flow-2pt2}
\begin{axopicture}(200,50)(0,7)
  \SetWidth{0.5}
  \SetTextScale{15}
  \Text(2,0)[lb]{$\displaystyle\Big ($}
  \SetTextScale{15}
  \Text(13,-2)[lb]{$x$}
  \Text(60,-3)[lb]{$\bar x$}
  \Text(57,7)[lb]{\quad+}
  \SetWidth{1}
  \Bezier[color=Green](28,8)(19,18)(19,28)(39,28)
  \Line[color=Red](20,-2)(28,8)
  \Bezier[color=Green](50,8)(59,18)(59,28)(39,28)
  \Line[color=Red](50,8)(58,-2)
  \Line(36,25)(42,31)\Line(36,31)(42,25)
  \COval(39,8)(3,10)(0.0){Black}{Black}
  \Text(78,-2)[lb]{$x$}
  \Text(125,-3)[lb]{$\bar x$}
  \Bezier[color=Green](93,8)(93,18)(124,18)(104,28)
  \Line[color=Red](85,-2)(93,8)
  \Bezier[color=Green](115,8)(115,18)(83,18)(104,28)
  \Line[color=Red](115,8)(123,-2)
  \COval(104,8)(3,10)(0.0){Black}{Black}
  \Line(101,25)(107,31)\Line(101,31)(107,25)
  \Text(122,7)[lb]{\quad+}
  \Text(143,-2)[lb]{$x$}
  \Text(167,-3)[lb]{$\bar x$}
  \Line[color=Red](158,8)(150,-2)
  \Line[color=Red](158,8)(165,-2)
  \Arc[color=Green](158,36)(8,0,360)
  \COval(158,18)(10,3)(0.0){Black}{Black}
  \Line(155,41)(161,47)\Line(155,47)(161,41)
  \Text(175,-0)[lb]{$\displaystyle \Big ) \; .$}
\end{axopicture}
\end{align}
featuring a two-loop structure of sunset diagrams in the first and second column and double bubble diagrams in the thrid column. By introducing the following short-hand notation for the one-loop integrals
\begin{eqnarray}
B^{R/A}_{k}(x\bar{x})&=&\int_{z\bar{z}} G_{k}^{R/A}(xz) \dot{R}_{k}^{R/A}(z\bar{z}) G_{k}^{R/A}(\bar{z}\bar{x})\;, \label{eq:def-B}  \\
B^{F}_{k}(x\bar{x})&=&\int_{z\bar{z}} G_{k}^{R}(xz) \dot{R}_{k}^{R}(z\bar{z}) iF(\bar{z}\bar{x}) + iF(xz) \dot{R}_{k}^{A}(z\bar{z}) G_{k}^{A}(\bar{z}\bar{x}) + G_{k}^{R}(xz) \dot{R}_{k}^{F}(z\bar{z})G_{k}^{A}(\bar{z}\bar{x})\;, \notag
\end{eqnarray}
the flow-equations for the two-point functions then take the form
\begin{align}
&\partial_{k} \Gamma_{k}^{\phi\tilde{\phi}}(x\bar{x})= \frac{-i}{2} \Bigg[ \delta(x\bar{x}) \int_{y} \left[ N v_{cl,A,k}^{diag}(y) + 2 v_{cl,A,k}^{off}(y)  \right] B_{F}(0) \label{2pt-flow1} \\
&+  \left[ 2 v_{cl,A,k}^{diag}(x\bar{x})  + 2(N+1) v_{cl,A,k}^{off}(x\bar{x}) \right]  B_{F}(x\bar{x}) + \Big[ 2 v_{anom,k}^{diag}(x\bar{x})  + 2(N+1) v_{anom,k}^{off}(x\bar{x}) \Big] B_{A}(x\bar{x})\Bigg] \;,\notag \\
 \notag \\
& \partial_{k} \Gamma_{k}^{\tilde{\phi}\tilde{\phi}}(x\bar{x}) = 
\frac{-i}{2} \bigg\{\Big[ 2 v_{anom,k}^{diag}(x\bar{x})  + 2(N+1) v_{anom,k}^{off}(x\bar{x}) \Big] B_{F}(x\bar{x}) \label{2pt-flow2} \\
&+ \left[ 2 v_{qu,R,k}^{diag}(x\bar{x})  + 2(N+1) v_{qu,R,k}^{off}(x\bar{x}) \right]  B_{R}(x\bar{x})  + \left[ 2 v_{qu,A}^{diag}(x\bar{x})  + 2(N+1) v_{qu,A}^{off}(x\bar{x}) \right]  B_{A}(x\bar{x}) \bigg\} \; .\nonumber
\end{align}
where we dropped acausal contributions proportional to $B^{R/A}(0)$. 
\subsection{Vertex flow}
Evidently, to close the system of equations we still need expressions for the vertex functions. In the following we will compare two different truncations.
\subsubsection{One-loop vertex functions}
We start by using the one-loop expressions of the vertex functions with self-consistently determined propagators. Explicitly, for the four-point functions eqs.~(\ref{four-pt1}-\ref{four-pt4}) the pertubative one-loop expressions determined at each step of the renormalization group evolution take the form
\begin{align}
v_{cl,A,k}^{diag}(x\bar{x}) &=\lambda_{cl}\delta(x\bar{x})-(N+4)i  \lambda_{cl}^2~iF_{k}(x\bar{x}) G_{k}^{A}(x\bar{x})\;, \notag \\
v_{cl,A,k}^{off}(x\bar{x}) &= \qquad \qquad \qquad \quad -2 i  \lambda_{cl}^2~iF_{k}(x\bar{x}) G_{k}^{A}(x\bar{x})\;, \notag \\
v_{anom,k}^{diag}(x\bar{x}) &= (N+4) \left(-\frac{i}{2} \lambda_{cl}^2 \left( iF_{k}(x\bar{x})\right)^{2} -\frac{i}{2} \lambda_{cl} \lambda_{qu} \left[ \left( G^{R}_{k}(x\bar{x})\right)^{2} + \left(G^{A}_{k}(x\bar{x})\right)^{2} \right] \right)\;, \label{eq:one-loop-vertex} \\
v_{anom,k}^{off}(x\bar{x}) &=\hspace{1.2cm} 2 \left(-\frac{i}{2} \lambda_{cl}^2 \left( iF_{k}(x\bar{x})\right)^{2} -\frac{i}{2} \lambda_{cl} \lambda_{qu} \left[ \left( G^{R}_{k}(x\bar{x})\right)^{2} + \left(G^{A}_{k}(x\bar{x})\right)^{2} \right] \right)\;, \notag \\
v_{qu,R,k}^{diag}(x\bar{x}) &=\lambda_{qu}\delta(x\bar{x})-(N+4)i  \lambda_{cl}\lambda_{qu}~iF_{k}(x\bar{x}) G_{k}^{R}(x\bar{x})\;, \notag \\
v_{qu,R,k}^{off}(x\bar{x}) &= \qquad \qquad \qquad \quad -2 i  \lambda_{cl}\lambda_{qu}~iF_{k}(x\bar{x}) G_{k}^{R}(x\bar{x})\;, \notag
\end{align}
such that diagonal and off-diagonal vertex functions differ only by their corresponding flavor factors.

\subsubsection{Vertex flow equation}
\label{sec:Vertex}
While Eq.~(\ref{eq:one-loop-vertex}) represents the simplest possible two-loop complete truncation, there are many possible ways to improve upon this truncation; for instance the four point couplings $v_{k}^{cl}(x\bar{x}),v_{k}^{anom}(x\bar{x})$ and $v_{k}^{qu}(x\bar{x})$ could be determined self-consistently by projecting the corresponding flow equations for the four point functions. Clearly, the main advantage of this procedure is the automatic renormalization of the coupling that comes with solving the flow equations for the four point functions. While this amounts to a selective resummation of higher order contributions, it is also clear that this does not improve the perturbative completeness of the calculation, unless more complicated non-local structures and higher order vertices are to be included as well.

Since within our truncation, the classical, quantum and anomalous four-point functions are not independent (see, eqs.~(\ref{vertex-rel1}- \ref{vertex-rel2})) we will only solve the flow-equation for the classical vertex function and reconstruct the other two vertices from the classical vertex function. Based on our discussion in Sec.~\ref{sec:FRG}, the flow-equation for the classical vertex takes the form
\begin{align}
    \partial_k \Gamma^{\tilde{\phi}\phi\phi\phi}_{k,ab\bar{b}\bar{a}}(xy\bar{y}\bar{x}) = 
 -\frac{i}{2}\, \Bigg \lbrace\;\; 
 &\begin{picture}(110,40) (170,-37)
      \SetWidth{1.0}
      \SetColor{Black}
      \Arc[color=Green](210,-33)(15,0,360)
      \Line(207,-15)(213,-21)\Line(207,-21)(213,-15)
      \Line[color=Blue](185,-25)(194,-33)
      \Line[color=Red](185,-41)(194,-33)
      \Line[color=Blue](235,-25)(226,-33)
      \Line[color=Blue](235,-41)(226,-33)
      \Vertex(194,-33){3.4}
      \Vertex(226,-33){3.4}
      \Text(172,-50)[lb]{\scriptsize $x,a$}
      \Text(172,-23)[lb]{\scriptsize $y,b$}
      \Text(235,-23)[lb]{\scriptsize $\bar y,\bar b$}
      \Text(235,-50)[lb]{\scriptsize $\bar x,\bar a$}
\end{picture}\hspace*{-1cm}+
\begin{picture}(110,40) (170,-37)
      \SetWidth{1.0}
      \SetColor{Black}
      \Arc[color=Green](210,-33)(15,0,360)
      \Line(207,-15)(213,-21)\Line(207,-21)(213,-15)
      \Line[color=Blue](185,-25)(194,-33)
      \Line[color=Red](185,-41)(194,-33)
      \Line[color=Blue](235,-25)(226,-33)
      \Line[color=Blue](235,-41)(226,-33)
      \Vertex(194,-33){3.4}
      \Vertex(226,-33){3.4}
      \Text(172,-50)[lb]{\scriptsize $x,a$}
      \Text(172,-23)[lb]{\scriptsize $y,b$}
      \Text(235,-23)[lb]{\scriptsize $\bar x,\bar a$}
      \Text(235,-50)[lb]{\scriptsize $\bar y,\bar b$}
\end{picture}\hspace*{-1cm}+
\begin{picture}(110,40) (170,-37)
      \SetWidth{1.0}
      \SetColor{Black}
      \Arc[color=Green](210,-33)(15,0,360)
      \Line(207,-15)(213,-21)\Line(207,-21)(213,-15)
      \Line[color=Blue](185,-25)(194,-33)
      \Line[color=Red](185,-41)(194,-33)
      \Line[color=Blue](235,-25)(226,-33)
      \Line[color=Blue](235,-41)(226,-33)
      \Vertex(194,-33){3.4}
      \Vertex(226,-33){3.4}
      \Text(172,-50)[lb]{\scriptsize $x,a$}
      \Text(172,-23)[lb]{\scriptsize $\bar x,\bar a$}
      \Text(235,-23)[lb]{\scriptsize $y, b$}
      \Text(235,-50)[lb]{\scriptsize $\bar y,\bar b$}
\end{picture}\\
+&\begin{picture}(110,40) (170,-37)
      \SetWidth{1.0}
      \SetColor{Black}
      \Arc[color=Green](210,-33)(15,0,360)
      \Line(207,-15)(213,-21)\Line(207,-21)(213,-15)
      \Line[color=Blue](185,-25)(194,-33)
      \Line[color=Red](185,-41)(194,-33)
      \Line[color=Blue](235,-25)(226,-33)
      \Line[color=Blue](235,-41)(226,-33)
      \Vertex(194,-33){3.4}
      \Vertex(226,-33){3.4}
      \Text(172,-50)[lb]{\scriptsize $x,a$}
      \Text(172,-23)[lb]{\scriptsize $\bar x,\bar a$}
      \Text(235,-23)[lb]{\scriptsize $\bar y,\bar b$}
      \Text(235,-50)[lb]{\scriptsize $y,b$}
\end{picture}\hspace*{-1cm}+
\begin{picture}(110,40) (170,-37)
      \SetWidth{1.0}
      \SetColor{Black}
      \Arc[color=Green](210,-33)(15,0,360)
      \Line(207,-15)(213,-21)\Line(207,-21)(213,-15)
      \Line[color=Blue](185,-25)(194,-33)
      \Line[color=Red](185,-41)(194,-33)
      \Line[color=Blue](235,-25)(226,-33)
      \Line[color=Blue](235,-41)(226,-33)
      \Vertex(194,-33){3.4}
      \Vertex(226,-33){3.4}
      \Text(172,-50)[lb]{\scriptsize $x,a$}
      \Text(172,-23)[lb]{\scriptsize $\bar y,\bar b$}
      \Text(235,-23)[lb]{\scriptsize $\bar x,\bar a$}
      \Text(235,-50)[lb]{\scriptsize $y,b$}
\end{picture}\hspace*{-1cm}+
\begin{picture}(110,40) (170,-37)
      \SetWidth{1.0}
      \SetColor{Black}
      \Arc[color=Green](210,-33)(15,0,360)
      \Line(207,-15)(213,-21)\Line(207,-21)(213,-15)
      \Line[color=Blue](185,-25)(194,-33)
      \Line[color=Red](185,-41)(194,-33)
      \Line[color=Blue](235,-25)(226,-33)
      \Line[color=Blue](235,-41)(226,-33)
      \Vertex(194,-33){3.4}
      \Vertex(226,-33){3.4}
      \Text(172,-50)[lb]{\scriptsize $x,a$}
      \Text(172,-23)[lb]{\scriptsize $\bar y,\bar b$}
      \Text(235,-23)[lb]{\scriptsize $y, b$}
      \Text(235,-50)[lb]{\scriptsize $\bar x,\bar a$}
\end{picture}\hspace*{-1cm} \Bigg \rbrace \; , \notag
\end{align}
where the black dots correspond to insertions of the full four-point vertex functions. Solving the flow-equation for the full four-point function with all it's space-time arguments is prohibitively expensive. Hence, our strategy will be to project the flow-equation onto the vertex functions $v(x \bar x)$ and solve the corresponding flow-equations. We will now switch to momentum space as the projection to the vertex functions is simpler here. The classical four-point function in momentum space takes the form
\begin{eqnarray}
\hspace*{-1cm} &&\Gamma^{\phi\phi\phi\tilde{\phi}}_{k,ab\bar{b}\bar{a}}(p|q \bar q \bar p) = \frac{(2 \pi)^{2+1}}{V} \delta\left( \frac{p+q+\bar q +\bar p}{2} \right)  \Big \lbrace   \label{four-pt-mom} \\
&& 
v_{cl,R,k}^{diag}(\frac{p+q-\bar p - \bar q}{2}) \delta_{ab} \delta_{\bar{a}\bar{b}} + v_{cl,R,k}^{off}(\frac{p+q-\bar p - \bar q}{2}) \delta_{a\bar{a}} \delta_{b\bar{b}} + v_{cl,R,k}^{off}(\frac{p+q-\bar p - \bar q}{2})  \delta_{a\bar{b}} \delta_{\bar{a}b}  \notag  \\
&+&
v_{cl,R,k}^{off}(\frac{p+\bar p -q - \bar q}{2})\delta_{ab} \delta_{\bar{a}\bar{b}} + v_{cl,R,k}^{diag}(\frac{p+\bar p -q - \bar q}{2})  \delta_{a\bar{a}} \delta_{b\bar{b}} +v_{cl,R,k}^{off}(\frac{p+\bar p -q - \bar q}{2}) \delta_{a\bar{b}} \delta_{\bar{a}b} \nonumber \\
&+&
v_{cl,R,k}^{off}(\frac{p+ \bar q-\bar p -q}{2}) \delta_{ab} \delta_{\bar{a}\bar{b}} + v_{cl,R,k}^{off}(\frac{p+ \bar q-\bar p -q}{2}) \delta_{a\bar{a}} \delta_{b\bar{b}} + v_{cl,R,k}^{diag}(\frac{p+ \bar q-\bar p -q}{2}) \delta_{a\bar{b}} \delta_{\bar{a}b}  \Big \rbrace  \;, \nonumber
\end{eqnarray}
and we will use the following relation to project the flow equation onto the diagonal and off-diagonal vertex functions
\begin{eqnarray}
 & & \hspace*{-1cm} 3 \left(v_{cl,R,k}^{diag}(p) + 2 v_{cl,R,k}^{off}(0)\right)\delta_{ab} \delta_{\bar{a}\bar{b}} + 3 (v_{cl,R,k}^{off}(p) + v_{cl,R,k}^{diag}(0) + v_{cl,R,k}^{off}(0)) \left(\delta_{a\bar{a}} \delta_{b\bar{b}} + \delta_{a\bar{b}} \delta_{\bar{a}b}\right) = \label{projection} \\
& &\Gamma^{\phi\phi\phi\tilde{\phi}}_{k,ab\bar{b}\bar{a}}(-\frac{p}{2},+\frac{p}{2}, -\frac{p}{2}, -\frac{p}{2})
+\Gamma^{\phi\phi\phi\tilde{\phi}}_{k,a\bar{a}b\bar{b}}(-\frac{p}{2},-\frac{p}{2}, +\frac{p}{2}, -\frac{p}{2})
+\Gamma^{\phi\phi\phi\tilde{\phi}}_{k,a\bar{b}\bar{a}b}(-\frac{p}{2},-\frac{p}{2}, +\frac{p}{2}, -\frac{p}{2}) \; . \notag
\end{eqnarray}
By performing the projection of the flow equation according to Eq.~(\ref{projection}), the flow-equation for the projected vertex function then takes the following diagrammatic form\footnote{Since for $p=0$ the LHS of Eq.~(\ref{projection}) involves diagonal and off-diagonal vertex functions in exactly the same way, there is a hidden ambiguity of how to treat momentum independent contributions to the vertex function. However, as the vertex functions in Eq.~(\ref{four-pt-mom}) always appear in the combination $v^{diag}_{cl,R}+v^{off}_{cl,R}+v^{off}_{cl,R}$, any momentum independent contribution can be arbitrarily distributed between diagonal and off-diagonal vertex functions, and in practice we split the momentum independent contribution and absorb parts in both the diagonal and off-diagonal parts of the vertex function.}
\begin{align}
 &\partial_k \Big[ \left(v_{cl,R,k}^{diag}(p) + 2 v_{cl,R,k}^{off}(0)\right)\delta_{ab} \delta_{\bar{a}\bar{b}} +  (v_{cl,R,k}^{off}(p) + v_{cl,R,k}^{diag}(0) + v_{cl,R,k}^{off}(0)) \left(\delta_{a\bar{a}} \delta_{b\bar{b}} + \delta_{a\bar{b}} \delta_{\bar{a}b}\right) \Big] = \notag \\
 &\hspace{1.5cm}-i\, \Big\lbrace\;\; 
 \begin{picture}(110,40) (170,-37)
      \SetWidth{1.0}
      \SetColor{Black}
      \Arc[color=Green](210,-33)(15,0,360)
      \Line(207,-15)(213,-21)\Line(207,-21)(213,-15)
      \Line[color=Blue](185,-25)(194,-33)
      \Line[color=Red](185,-41)(194,-33)
      \Line[color=Blue](235,-25)(226,-33)
      \Line[color=Blue](235,-41)(226,-33)
      \Vertex(194,-33){3.4}
      \Vertex(226,-33){3.4}
      \Text(165,-50)[lb]{\scriptsize $+p/2,a$}
      \Text(165,-23)[lb]{\scriptsize $+p/2,b$}
      \Text(227,-23)[lb]{\scriptsize $-p/2,\bar b$}
      \Text(227,-50)[lb]{\scriptsize $-p/2,\bar a$}
\end{picture}\hspace*{-0.7cm}+
\begin{picture}(110,40) (170,-37)
      \SetWidth{1.0}
      \SetColor{Black}
      \Arc[color=Green](210,-33)(15,0,360)
      \Line(207,-15)(213,-21)\Line(207,-21)(213,-15)
      \Line[color=Blue](185,-25)(194,-33)
      \Line[color=Red](185,-41)(194,-33)
      \Line[color=Blue](235,-25)(226,-33)
      \Line[color=Blue](235,-41)(226,-33)
      \Vertex(194,-33){3.4}
      \Vertex(226,-33){3.4}
      \Text(165,-50)[lb]{\scriptsize $+p/2,a$}
      \Text(165,-23)[lb]{\scriptsize $-p/2,\bar a$}
      \Text(227,-23)[lb]{\scriptsize $+p/2,b$}
      \Text(227,-50)[lb]{\scriptsize $-p/2,\bar b$}
\end{picture}\hspace*{-0.7cm}+
\begin{picture}(110,40) (170,-37)
      \SetWidth{1.0}
      \SetColor{Black}
      \Arc[color=Green](210,-33)(15,0,360)
      \Line(207,-15)(213,-21)\Line(207,-21)(213,-15)
      \Line[color=Blue](185,-25)(194,-33)
      \Line[color=Red](185,-41)(194,-33)
      \Line[color=Blue](235,-25)(226,-33)
      \Line[color=Blue](235,-41)(226,-33)
      \Vertex(194,-33){3.4}
      \Vertex(226,-33){3.4}
      \Text(165,-50)[lb]{\scriptsize $+p/2,a$}
      \Text(165,-23)[lb]{\scriptsize $-p/2,\bar b$}
      \Text(227,-23)[lb]{\scriptsize $-p/2,\bar a$}
      \Text(227,-50)[lb]{\scriptsize $+p/2, b$}
\end{picture}
\hspace*{-0.7cm}\Big \rbrace \label{4pt-flow}  \; . \\ \notag
\end{align}
where it is important to state that Eq.~(\ref{projection}) is fully symmetrized, such that all six permutations of the outer indices on the right-hand side of the flow-equation enter in exactly the same way. Clearly, this flow-equation has a rather complicated structure, as there are three different vertex-propagator combinations for each diagram drawn, and additionally, every vertex comes with its substructure, see Eq.~(\ref{four-pt-mom}). 

Generally, the flow-equation for the vertex functions in Eq.~(\ref{4pt-flow}) contains 81 terms and can be found in Appendix E. Since the resulting expression is rather lengthy, we only state the explicit form for the single component $N=1$ theory
\begin{align}
    &\partial_k (v_k^{cl,R}(p) + 2 v_k^{cl,R}(0)) = -i \int_l \bigg \lbrace \label{eq:vert-flow-N1}  \\
    &\Big(v_k^{cl,R}(p) + v_k^{cl,R}(l) + v_k^{cl,R}(-l) \Big) G_k^R(\frac{p}{2}+l) B_k^F\left(\frac{p}{2}-l\right) \Big(v_k^{cl,R}(p) + v_k^{cl,R}(l) + v_k^{cl,R}(l) \Big) \notag \\
    +&\Big(v_k^{cl,R}(0) + v_k^{cl,R}(\frac{p}{2}+l) + v_k^{cl,R}(\frac{p}{2}-l) \Big) G_k^R(l) B_k^F(-l) \Big(v_k^{cl,R}(0) + v_k^{cl,R}(\frac{p}{2}+l) + v_k^{cl,R}(\frac{p}{2}-l) \Big) \notag \\
    +&\Big(v_k^{cl,R}(0) + v_k^{cl,R}(\frac{p}{2}+l) + v_k^{cl,R}(\frac{p}{2}-l) \Big) G_k^R(l) B_k^F(-l) \Big(v_k^{cl,R}(0) + v_k^{cl,R}(\frac{-p}{2}+l) + v_k^{cl,R}(\frac{p}{2}-l) \Big) \notag \\
    +&\Big(v_k^{cl,R}(p) + v_k^{cl,R}(l) + v_k^{cl,R}(-l) \Big) iF_k(\frac{p}{2}+l) B_k^R(\frac{p}{2}-l) \Big(v_k^{cl,R}(-l) + v_k^{cl,R}(-l) + v_k^{cl,R}(p) \Big) \notag \\
    +&\Big(v_k^{cl,R}(0) + v_k^{cl,R}(\frac{p}{2}+l) + v_k^{cl,R}(\frac{p}{2}-l) \Big) iF_k(l) B_k^R(-l) \Big(v_k^{cl,R}(\frac{p}{2}-l) + v_k^{cl,R}(\frac{-p}{2}-l) + v_k^{cl,R}(0) \Big) \notag \\
    +&\Big(v_k^{cl,R}(0) + v_k^{cl,R}(\frac{p}{2}+l) + v_k^{cl,R}(\frac{p}{2}-l) \Big) iF_k(l) B_k^R(-l) \Big(v_k^{cl,R}(\frac{-p}{2}-l) + v_k^{cl,R}(\frac{p}{2}-l) + v_k^{cl,R}(0) \Big) \notag \\
    +&\Big(v_k^{cl,R}(p) + v_k^{cl,R}(l) + v_k^{cl,R}(-l) \Big) G_k^R(\frac{p}{2}+l) B_k^R(\frac{p}{2}-l) \Big(v_k^{anom}(l) + v_k^{anom}(l)\Big) \notag \\
    +&\Big(v_k^{cl,R}(0) + v_k^{cl,R}(\frac{p}{2}+l) + v_k^{cl,R}(\frac{p}{2}-l) \Big) G_k^R(l) B_k^R(-l) \Big(v_k^{anom}(\frac{p}{2}+l) + v_k^{anom}(\frac{-p}{2}+l) \Big) \notag \\
    +& \left.\Big(v_k^{cl,R}(0) + v_k^{cl,R}(\frac{p}{2}+l) + v_k^{cl,R}(\frac{p}{2}-l) \Big) G_k^R(l) B_k^R(-l) \Big(v_k^{anom}(\frac{-p}{2}+l) + v_k^{anom}(\frac{p}{2}+l) \Big) \notag \right \rbrace \; . 
\end{align}
 where there is no distinction between diagonal and off-diagonal index structures for the single component theory.  When performing calculations with self-consistently determined vertex functions, we will employ the one-loop vertex functions in Eq.~(\ref{eq:one-loop-vertex}) evaluated at the UV scale as initial condition for the flow equation (\ref{eq:vert-flow-N1}).
 
 Besides $N=1$ another relatively simple case is the limit $N \to \infty$, where one can employ a $1/N$ expansion. Since the leading-order contributions to a vertex always come from the diagonal vertex functions $(v^{diag} \sim 1/N)$, one can simply drop all terms containing off-diagonal vertex functions $(v^{off} \sim 1/N^2)$ to leading order in the 1/N expansion of the flow-equation (see e.g. Eq.~(\ref{1l-4pt})). Evaluating the remaining terms one finds that, due to the contraction of O(N) indices in the one-loop diagrams in Eq.~(\ref{4pt-flow}), the subset of diagrams where the flavor index flow is identical to the momentum flow will be enhanced by a factor of $N$ relative to all other diagrams. By collecting the leading $\mathcal{O}(1/N)$ contributions, the flow-equation Eq.~(\ref{4pt-flow}) then takes the following form in the large $N$ limit
\begin{align}
    &\partial_k [v_{cl,R,k}^{diag}(p) \delta_{ab}\delta_{\bar{a} \bar{b}} + v_{cl,R,k}^{diag}(0) (\delta_{a \bar{a}}\delta_{b \bar{b}} +\delta_{a \bar b}\delta_{\bar{a} b}) ] = -i N \bigg \lbrace \\
    &\delta_{ab}\delta_{\bar{a} \bar{b}} v_{cl}^{diag}(p) \left[ \int_{l} G_{R}\left(\frac{p}{2}+l\right) B_{F}\left(\frac{p}{2}-l\right)  + B_{R}\left(\frac{p}{2}+l\right) iF\left(\frac{p}{2}-l\right)\right] v_{cl}^{diag}(p) \notag \\
    &+(\delta_{a \bar{a}}\delta_{b \bar{b}} +\delta_{a \bar b}\delta_{\bar{a} b}) v_{cl}^{diag}(0) \left[ \int_{l} G_{R}(l) B_{F}(-l)  + B_{R}(l) iF(-l)\right] v_{cl}^{diag}(0) \notag \bigg \rbrace \; .
\end{align}
By separating the different index structures, one obtains the final result
\begin{align}
\label{eq:vClLargeNFlow}
\partial_{k} v_{cl}^{diag}(p) = -i N v_{cl}^{diag}(p) \left[ \int_{l} G_{R}\left(\frac{p}{2}+l\right) B_{F}\left(\frac{p}{2}-l\right)  + B_{R}\left(\frac{p}{2}+l\right) iF\left(\frac{p}{2}-l\right)\right] v_{cl}^{diag}(p) \; . \notag
\end{align}
such that in contrast to Eq.~(\ref{eq:vert-flow-N1}), the flow of the vertex function in the large $N$ limit is local in momentum space, in the sense that all vertex functions in Eq.~(\ref{eq:vClLargeNFlow}) are evaluated at the same momentum.
We further note that for special choices of the regulator function one can show that the right-hand side of the flow equation simplifies to a total differential \cite{Berges:2008sr,Gasenzer:2007za,Gasenzer:2010rq,Corell:2019jxh}
\begin{align}
\partial_{k} v_{cl}^{diag}(p) &\simeq -i N v_{cl}^{diag}(p) \left[\partial_{k} \int_{l} G_{R}\left(\frac{p}{2}+l\right) iF\left(\frac{p}{2}-l\right) \right] v_{cl}^{diag}(p) \; .
\end{align}
which can be solved directly by separation of variables
\begin{align}
\int_{k_{\rm UV}}^{k_{\rm IR}} dk  \frac{\partial_{k} v_{cl}^{diag}(p)}{\Big(v_{cl}^{diag}(p)\Big)^2} &= -i N \int_{k_{\rm UV}}^{k_{\rm IR}} dk~\partial_{k} \int_{l} G_{R,k}\left(\frac{p}{2}+l\right) iF_{k}\left(\frac{p}{2}-l\right) \; ,\\
\frac{1}{\lambda_{cl}}- \frac{1}{v_{cl}^{diag}(p)} &= -iN \int_l G_{R}\left(\frac{p}{2}+l\right) iF\left(\frac{p}{2}-l\right)\; ,
\end{align}
eventually yielding the familiar result of the 2PI 1/N expansion to next-to-leading order \cite{Aarts:2002dj,Berges:2008sr}
\begin{align}
v_{cl}^{diag}(p)&= \frac{\lambda_{cl}}{1+iN\lambda_{cl} \int_l G_{R}\left(\frac{p}{2}+l\right) iF\left(\frac{p}{2}-l\right)} \; ,
\end{align}
which corresponds to an infinite resummation of one-loop bubble chains. Based on this analysis, we therefore conclude that the above truncation of the real-time FRG flow equations not only encompasses the correct two-loop perturbative behavior of the spectral function for generic $N$, but also includes all contributions up to next-to-leading order of the 2PI 1/N expansion in the large $N$ limit. We further note that the interplay of the 2PI approach and the FRG in Euclidean time has been explored in the literature, e.g. the use of 2PI truncations in FRG calculations \cite{Blaizot:2010zx} or the use of the FRG to perform the complicated renormalization of 2PI calculations \cite{Carrington:2017lry,Berges:2005hc}, and we expect the interplay of the approaches to be similarly useful for real-time calculations.

\section{Numerical implementation}
\label{sec:Numerics}
Due to to the nested one-loop structure of the real-time FRG flow equations, it is benefitial to employ (pseudo-)spectral methods to solve the functional differential equations numerically. We have explored two different schemes, with the first one based on a straightforward lattice discretization of frequencies, where for an arbitrary function $G(\omega)$, we store the information at a discrete set of frequencies $\omega_{i}$
\begin{eqnarray}
G(\omega_i)= G_{i}^{(\omega)}\;, \qquad \omega_{i}=\frac{2\pi i}{N_t a_t} \;, \qquad i=0,\cdots,N-1\;,
\end{eqnarray}
Similarly, the corresponding function $G(t)$ in coordinate space is obtained at a discrete set of points $t_{i}$ 
\begin{eqnarray}
G(t_{i})=G_{i}^{(t)}\;, \qquad t_{i}=\left\{\begin{matrix} i \Delta t & i\leq N/2 \\ (i-N) \Delta t & i>N/2 \end{matrix}\right.\;,
\end{eqnarray}
by a Fast-Fourier Transformation (FFT) $G_{i}^{(t)}=\frac{1}{N_t a_t} \sum_{j} G_{j}^{(\omega)} e^{i \omega_{j} t_{i}}$. Clearly, the advantage of this method is that the right-hand sides of the flow equations are simple products and sums in position space, whereas in momentum space we would need to compute convolution integrals. Similarly, also the tadpole term can be computed efficiently using the FFT method by employing
\begin{align}
  \int_{y} \left[ N v_{cl,A,k}^{diag}(y) + 2 v_{cl,A,k}^{off}(y)  \right] B_{F}(0) =   
  \left[ N \tilde v_{cl,A,k}^{diag}(p=0) + 2 \tilde v_{cl,A,k}^{off}(p=0)  \right] B_{F}(x=0) \; .
\end{align}
While the evaluation of the right hand sides of the flow equation for the two point functions then becomes straightforward, the situation is different for the flow equation for the four point function, where the integral on the right-hand side of the flow-equation for the vertex function (\ref{eq:vert-flow-N1}) can not be solved by using FFT, and we instead use the lattice sum to approximate the integral. Subsequently, the FRG flow equations themselves are solved numerically using a fourth order Runge-Kutta scheme.

Secondly, we have also explored the possibility of numerically solving the FRG flow equations based on an expansion in Hermite functions, whereby the propagators $G(t)$ and $G(\omega)$ in the time and frequency domain are expanded according to
\begin{eqnarray}
G(t)=\sum_{k=0}^{N-1} g^{(t)}_{k} \Psi_{k}(t/a_t)\;, \qquad  G(\omega)=\sum_{k=0}^{N-1} g^{(\omega)}_{k} \Psi_{k}(\omega/a_{\omega})\;,
\end{eqnarray}
with the spacing $a_{\omega}=1/a_t$ adjusted to properly resolve the propagators at all relevant scales. Numerically, we keep track of the expansion coefficients $g^{(t)}_{k}$ and $g^{(\omega)}_{k}$, as well as the values of the propagators at times $t=x_{k} a_t$ and frequencies $\omega=x_{k} a_\omega$, where $x_{i}$ are the Gauss Hermite points, for which $G(t)$ and $G(\omega)$ are simply given by
\begin{eqnarray}
G(t=x_{i} a_t)=\sum_{k=0}^{N} g^{(t)}_{k} \Psi_{k}(x_i)\;, \qquad G(\omega=x_{i} a_\omega)=\sum_{k=0}^{N-1} g^{(\omega)}_{k} \Psi_{k}(x_i)\;, 
\end{eqnarray}
which then allows for efficient calculations of basic products and sums of the various functions. By use of the orthonormality relation $\int_{-\infty}^{\infty} dx~\Psi_{i}(x) \Psi_{j}(x)=\delta_{ij}$ for Hermite functions, the expansion coefficients $g^{(t)}_{k}$ and $g^{(\omega)}_{k}$ can be obtained from the integrals $g^{(t)}_{k}=\int dt/a_t G(t) \Psi_{k}(t/a_t)$ and $g^{(\omega)}_{k}=\int d\omega/a_\omega G(\omega) \Psi_{k}(\omega/a_\omega)$, where in practice we employ Gauss-Hermite quadrature, such that
\begin{eqnarray}
g^{(t)}_{k}=\sum_{i=0}^{N-1} \Psi_{k}(x_{i}) G(x_{i} a_t) w_{i}\;, \qquad g^{(\omega)}_{k}=\sum_{i=0}^{N-1} \Psi_{k}(x_{i}) G(x_{i} a_\omega) w_{i}\;, 
\end{eqnarray}
where $w_{i}= \frac{1}{N \psi^2_{N-1}(x_{i})}$ are the corresponding quadrature weights. Based on the following relations between the expansion coefficients $g^{(t)}_{k}$ and $g^{(\omega)}_{k}$,
\begin{eqnarray}
g^{(t)}_{k}=\frac{a_{\omega}}{\sqrt{2\pi}} (-i)^{k} g^{(\omega)}_{k}\;, \qquad  g^{(\omega)}_{k}=a_{t} \sqrt{2\pi} (+i)^{k} g^{(t)}_{k}\;.
\end{eqnarray}
it is then straightforward to perform Fourier transformations, in order to efficiently calculate the right hand side of the flow equations for the two-point functions. Similarly to the FFT method, we employ a Gauss Hermite quadrature when evaluating the integrals on the right hand side of the flow equation for the four point function, and for simplicity we resort to a forward Euler scheme when solving the FRG flow equations.

\section{Benchmarks and case studies}
\label{sec:Bench}
We will benchmark the method at the example of the anharmonic oscillator, which corresponds to the scalar field theory in $d=0$ dimensions. Generally the results for such an anharmonic oscillator depend on the three dimensionless combinations of parameters
\begin{eqnarray}
\frac{\lambda}{m^3}\;, \qquad \frac{\lambda}{\beta m^4}\;, \qquad \frac{\gamma}{\beta m}\;.
\end{eqnarray}
as well as on the number of field components $N$, which we will set to $N=1$. However, it is well known that in the classical-statistical theory the coupling constant and temperature dependence are related, such that upon performing a re-scaling of the classical field equations of motion with
\begin{eqnarray}
x^{0}\to m x^{0}\;,\qquad \phi \to \sqrt{\frac{\lambda}{m^3}} \sqrt{m} \phi\;, \qquad \eta \to \eta/\sqrt{m}\;,
\end{eqnarray}
the dependence on $\frac{\lambda}{m^3}$ can be eliminated from the classical-statistical field theory. Of course, this is not the case in the corresponding quantum theory, such that for fixed values of the (dimensionless) thermal interaction strength $\frac{\lambda}{\beta m^4}$, the dimensionless parameter $\lambda/m^3$ effectively describes the quantum interaction strength, with $\lambda/m^3=0$ corresponding to the classical-statistical limit.

\begin{figure}
    \centering
    \begin{minipage}[t]{0.45\textwidth}
        \includegraphics[width=\textwidth]{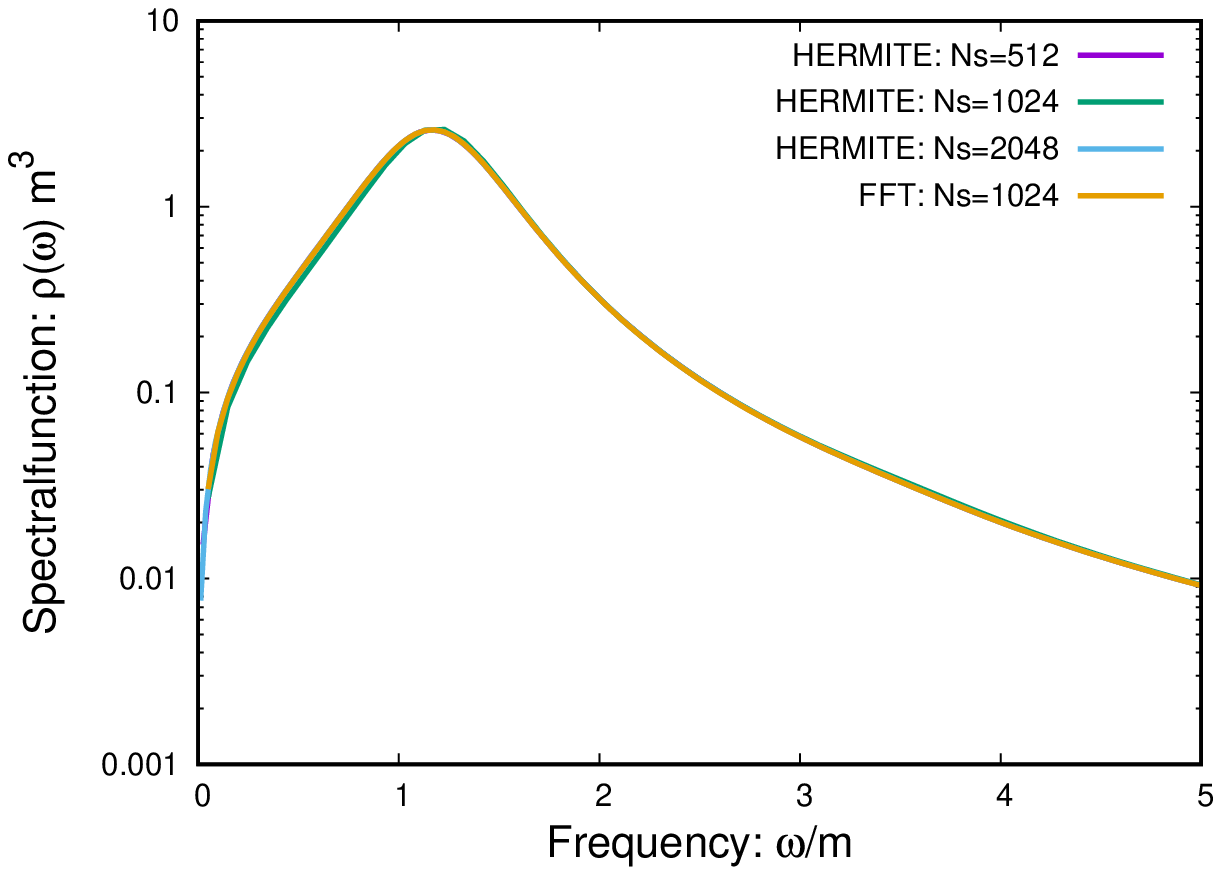}
        \caption{Comparison of spectral functions obtained by the two different numerical methods with different discretizations. Simulations were done with parameters $m=1.0$, $\frac{\lambda}{\beta m^4}=2.0$, $\frac{\gamma}{\beta m}=0.5$ in the classical limit. }
        \label{fig:comp_num}
    \end{minipage}
    \hspace*{0.5cm}
    \begin{minipage}[t]{0.45\textwidth}
        \includegraphics[width=\textwidth]{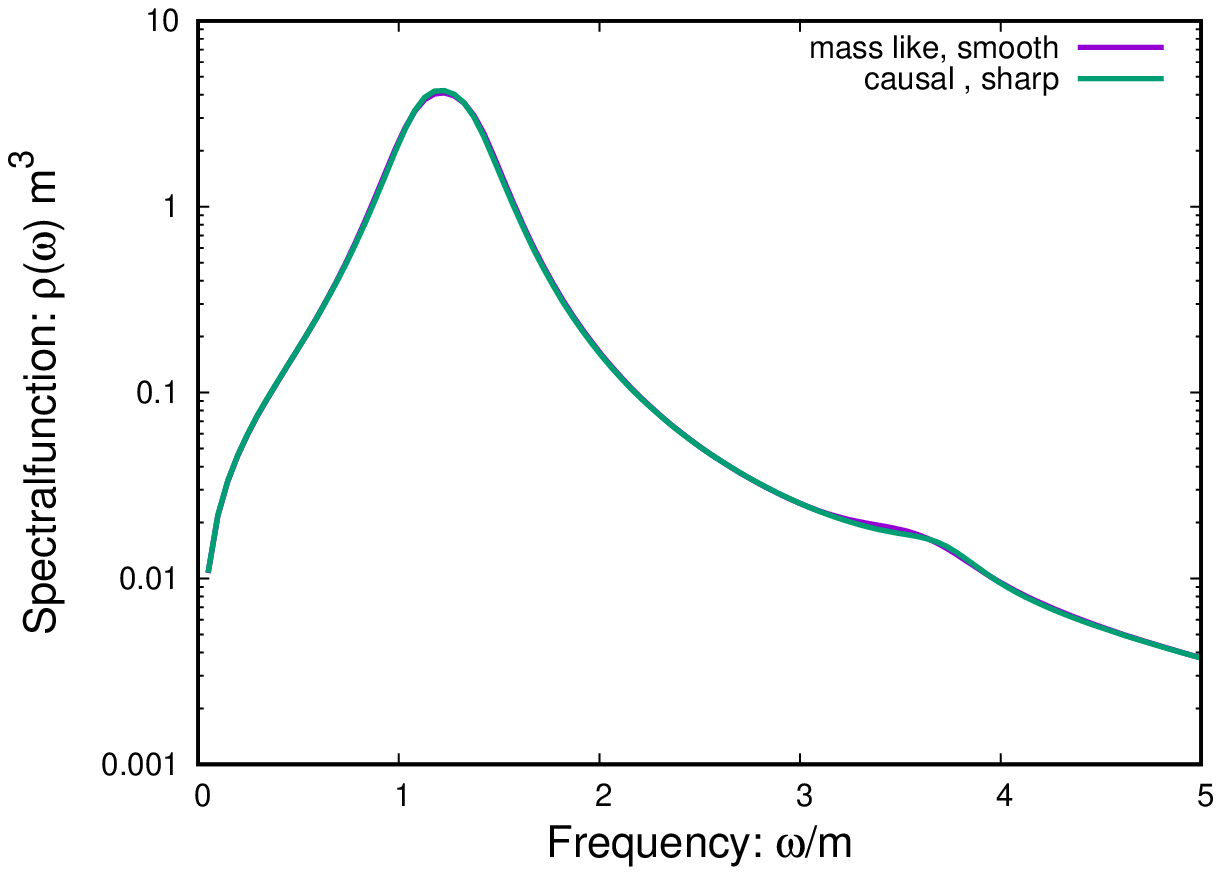}
        \caption{Comparison of spectral functions using different regulator schemes. Simulations were done with parameters $m=1.0$, $\frac{\lambda}{\beta m^4}=2.0$, $\frac{\gamma}{\beta m}=0.2$, $\frac{\lambda}{m^3}=2.0$.}
        \label{fig: comp_reg}
    \end{minipage}
\end{figure}

Before we present a series of results of real-time FRG calculations for classical and quantum systems, some sanity checks are in order, to instil confidence in our numerics. Evidently, a first important check is to compare the results obtained by different numerical methods, as shown in Fig.~\ref{fig:comp_num}, where we compare results for the spectral functions computed using the Discrete Fourier Grid and Gauss-Hermite Representation methods described in Sec.~\ref{sec:Numerics}. Excellent agreement between the two approaches is observed when sufficiently many discretization points are taken into account, indicating the common convergence to the correct result. Despite significant differences in the underlying implementation, we also find that within our implementation both approaches have comparable execution times on the order of 40s for calculations with one-loop vertices and 20m for calculations with self-consistent vertices, when employing the same number of $N=1024$ discretization points. \footnote{Calculations were performed on a single node equipped with an Intel Xeon Silver 4110, the UV cutoff was chosen to be $\Lambda=10$ and the stepsize $dk=0.05$.} However, we note that the convergence of the results generally appears to be somewhat better for the Discrete Fourier Grid.

Besides the convergence of the discrete representation of functions, another important sanity check for any FRG calculation is the comparison of results using different regulator schemes as the physical results in the infrared should be independent of the regulator choice. Fig.~\ref{fig: comp_reg} shows a comparison of spectral functions calculated with the one-loop form for the vertex functions. Our first regulator choice is mass-like with a sharp regulator function, i.e. choosing $\gamma_k = 0$ and $r_k$ being the optimized Litim regulator in Eq.~(\ref{eq:complicated_reg}). Our second regulator scheme is using the d+1 dimensional regulator obeying causality i.e. choosing $\mu_k$ and $\gamma_k$ according to Eq.~(\ref{eq:causal_reg}) and a smooth double exponential cutoff for $r_k$. The spectral function from both regulator schemes match almost perfecty, giving us good confidence in our methods.

\subsection{Benchmarks in the classical-statistical limit}
We begin with the study of classical-statistical dissipative systems, which in accordance with our discussion in Sec.~\ref{sec:FRG} can be compared against exact numerical results from classical-statistical simulations~\cite{Aarts:2001yx,Berges:2009jz,Schlichting:2019tbr,Schweitzer:2020noq}. With regards to the classical-statistical simulations, we follow the methodology of previous works~\cite{Schlichting:2019tbr,Schweitzer:2020noq} and simulate the time evolution of an ensemble of $N_{samples}=128$ independent realizations by solving the discretized classical evolution equations using an Euler-Maruyama scheme with step width $m \Delta t=0.01$. We then calculate the classical-statistical equilibrium spectral function from the un-equal time correlation function
\begin{eqnarray}
\rho_{\rm s}(t,t')= \frac{\beta}{2} \left\langle \phi(t)\pi(t')- \pi(t) \phi(t')\right\rangle_{\rm class.stat.}
\end{eqnarray} 
where $\left\langle .\right\rangle_{\rm class.stat.}$ denotes the average over the classical-statistical ensemble. Subsequently, we perform a discrete sine transformation to obtain the classical-statistical spectral function $\rho_{\rm cs}(\omega)$ in frequency space, which can then be compared directly with the classical-statistical real-time FRG calculations.

\begin{figure}
    \centering
    \includegraphics[width=0.8\textwidth]{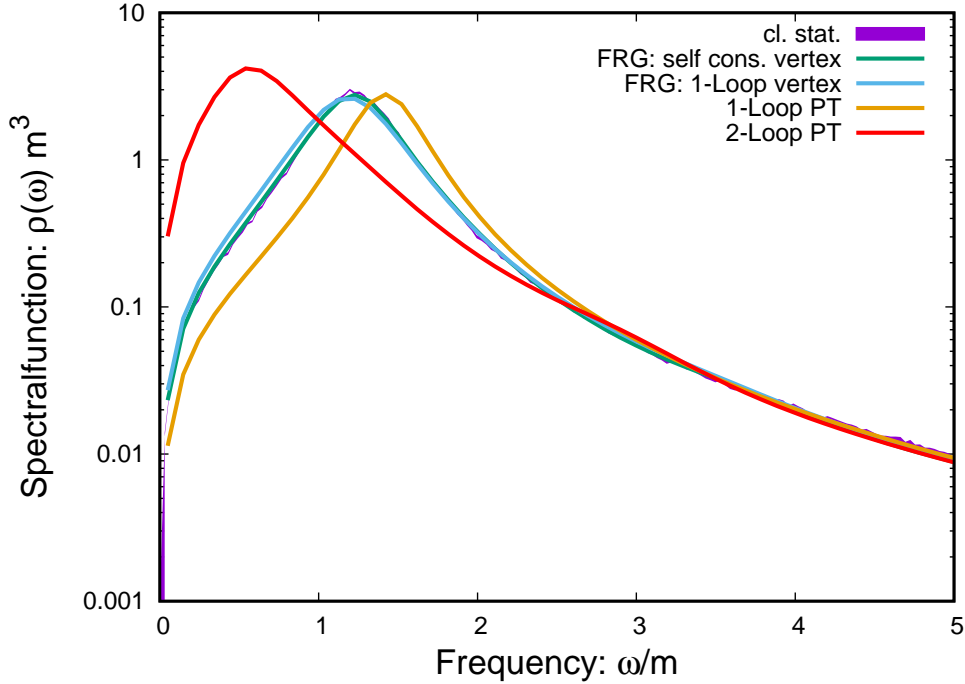}
    \caption{Comparison of spectral functions obtained by classical-statistical simulations, perturbation theory and FRG calculations with parameters $m=1.0$, $\frac{\lambda}{\beta m^4}=2.0$, $\frac{\gamma}{\beta m}=0.5$ }
    \label{fig:comp_cl}
\end{figure}

 We provide an example of such a comparison in Fig.~\ref{fig:comp_cl}, where results for the spectral functions from classical-statistical simulations, are compared to one and two-loop perturbation theory (c.f. Eqns.~(\ref{eq:DeltaGammaOneTwoPoint}) and (\ref{eq:Delta2Gamma2TwoLoopBubble},\ref{eq:Delta2Gamma2TwoLoopSunset})), as well as to FRG calculations with the one-loop vertex and fully self-consistent FRG calculations with a flowing vertex function. We see that for the particular choice of parameters $\lambda/\beta m^4=2$ and $\gamma/\beta m=1/2$ in Fig.~\ref{fig:comp_cl} the system can not be sufficently described by perturbation theory; while the one loop result over-estimates the thermal mass shift, the two loop results over-corrects this behavior, further indicating a poor convergence pattern. Conversely, the real-time FRG calculations are able to reproduce the classical-statistical results, such that even with the one-loop vertex ansatz the position and width of the peak are rather well described. By including the self consistent determination of the vertex functions, the spectral function only exhibits minor changes with a slight shift and narrowing of the broad resonance peak. Nevertheless, it is encouraging to observe that the inclusion of the self consistent vertex flow does improve the agreement with the exact result from classical-statistical simulations.  \newline

\begin{figure}
    \centering
    \includegraphics[width=0.8\textwidth]{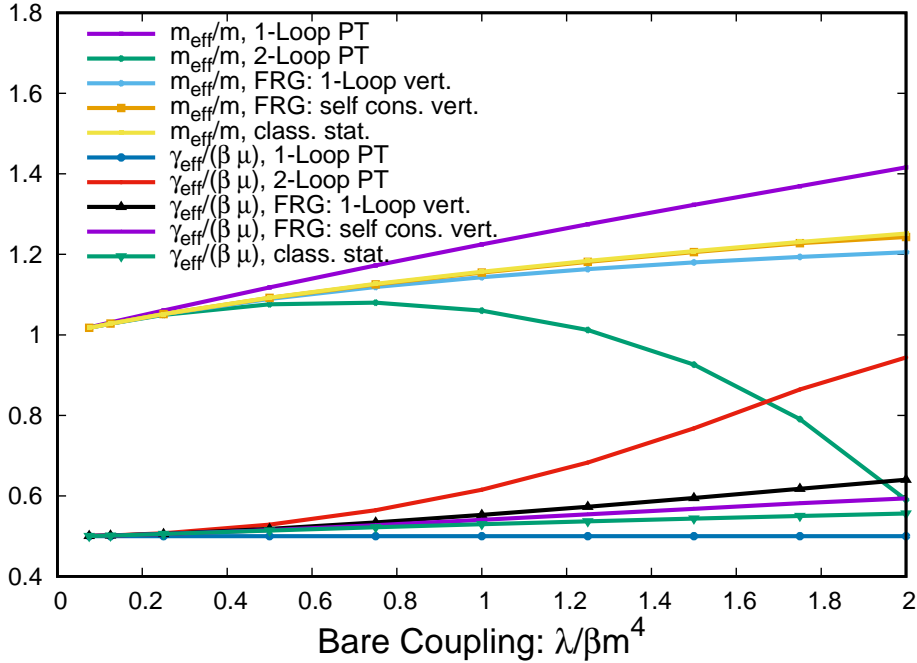}
    \caption{Effective masses and damping rates extracted from fits to a Breit-Wigner function. Simulations were performed with $m=1.0$ and $\gamma/\beta m=0.5$ in the classical limit. All values have a fit error of less than one percent and the fits gave a  $\chi^2_{red}<1$.}
    \label{fig:breit_wigner}
\end{figure}

Next, in order to further quantify the performance of different approaches, we have extracted the masses $m_{\rm eff}$ and widths $\gamma_{\rm eff}$ of the main peak of the spectral function by performing a fit to a Breit-Wigner Ansatz. Our results are compactly summarized in Fig.~\ref{fig:breit_wigner}, where we compare the results of the different approaches as a function of the (thermal) coupling strength $\lambda/\beta m^4$. Evidently, for small couplings we find a good agreement between all methods, while for larger couplings perturbation theory becomes unreliable as the LO result seems to overestimate the mass shift and does not capture the broadening of the peak, while the NLO result underestimates the mass shift and overestimates the broadening. Conversely, the one-loop FRG results and the data from fully self-consistent FRG simulations are comparable to each other and in general in good agreement with the classical-statistical results up to the largest investigated coupling in Fig.~\ref{fig:breit_wigner}.

\begin{figure}[]
    \centering
    \includegraphics[width=0.8\textwidth]{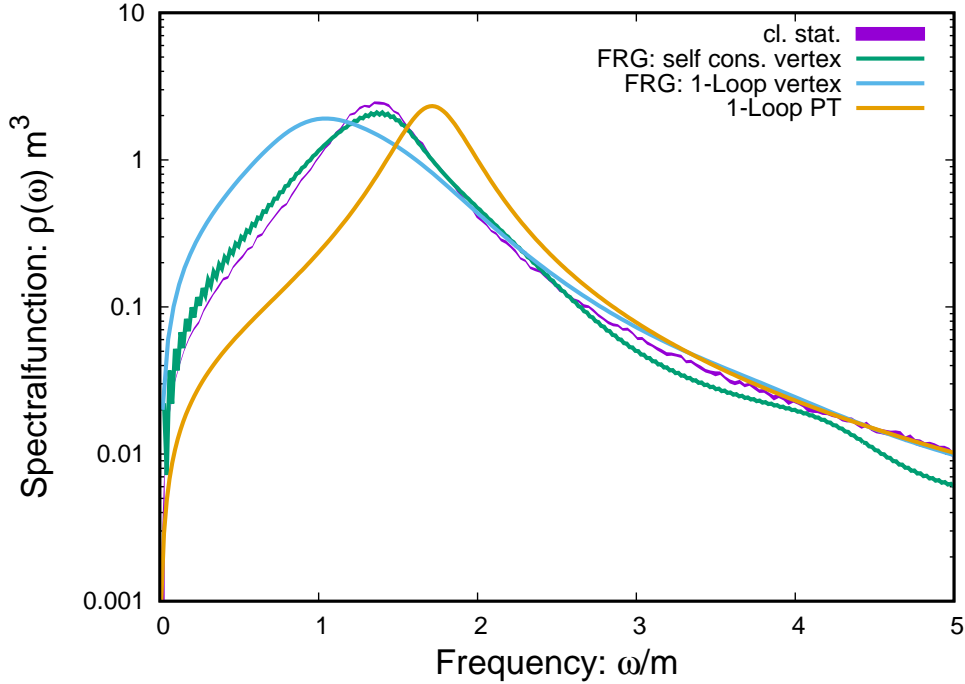}
    \caption{Comparison of spectral functions obtained by classical-statistical simulations, perturbation theory and FRG calculations with parameters $m=1.0$, $\frac{\lambda}{\beta m^4}=4.0$, $\frac{\gamma}{\beta m}=0.5$}
    \label{fig:large_coupl}
\end{figure}

Eventually, for even larger values of the coupling constant the spectral functions from the FRG calculations also deviate substantially from the classical-statistical results as can be seen from Fig.~\ref{fig:large_coupl}, where we present the results for $\lambda/\beta m^4=4$. Strongly coupled classical-statistical calculations in Fig.~\ref{fig:large_coupl} still produce a rather narrow quasi particle peak, whereas the FRG calcuation with the one-loop vertex overestimate the broadening resulting in large infrared contributions for the spectral function. The spectral function from the FRG calculation with the self-consistently determined vertex matches the spectral function best, however the data show some spurious oscillations in the spectral function. Eventually for $\lambda \gtrsim 4$, the FRG calculations fail to produce stable and sensible results for the spectral functions. We note that the point where the FRG with the one-loop vertex becomes unreliable can be readily estimated by looking at Eq.~(\ref{eq:one-loop-vertex}). Since this is a perturbative expression for some given coupling $\lambda$ the corrections of the bare vertex become of the same order of the bare vertex itself and -- similar to perturbation theory -- our results become unreliable. Even though one could expect that the inclusion of self-consistent vertices improves the behavior in the regime of large coupling strength, we find that for large couplings the calculations with self-consistent vertices become numerically unstable and we have not succeeded in obtaining physical results for the spectral function for significantly larger coupling strengths than in Fig.~\ref{fig:large_coupl}. 

\begin{figure}[t]
    \centering
    \includegraphics[width=0.8\textwidth]{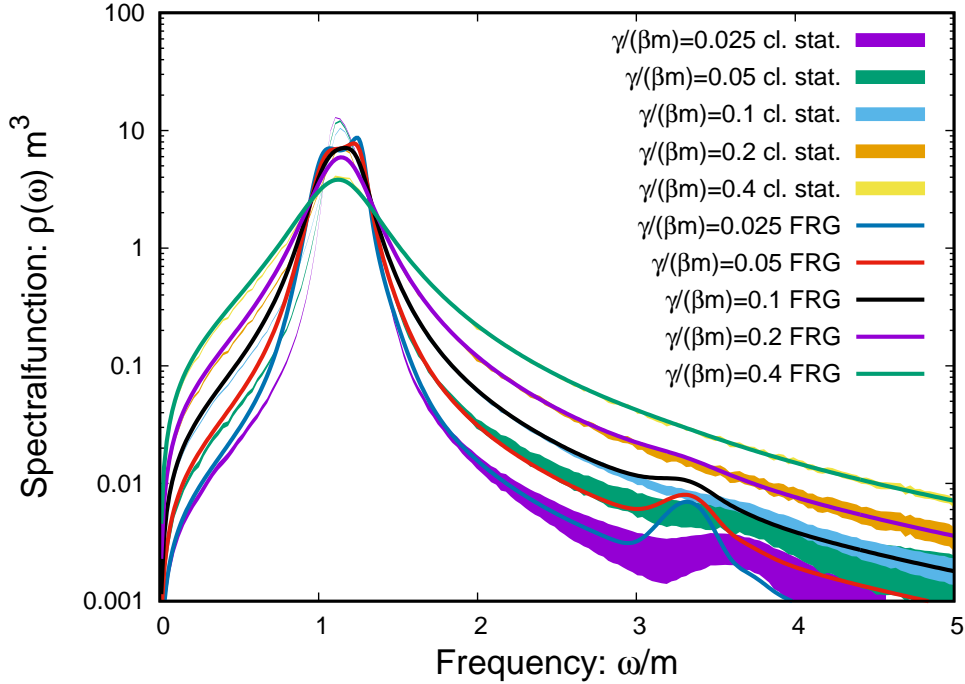}
    \caption{Comparison of spectral functions obtained by classical-statistical simulations and FRG calculations using the one-loop vertex with parameters $m=1.0$, $\frac{\lambda}{\beta m^4}=1.0$ for different damping rates in the classical limit. }
    \label{fig:broadening}
\end{figure}

So far we have investigated the spectral functions for a strongly dissipative anharmonic oscillator ($\gamma/\beta m=0.5$) and we will now study the effect of reducing the dissipative coupling to the heat bath. Before we proceed, we briefly note that the effect of the dissipative coupling $\gamma/\beta m$ is somewhat peculiar in $0+1$d as, in contrast to higher dimensional theories, we expect to recover a discrete spectrum in the limit of a closed system $\gamma/\beta m \to 0$, and the behavior could be qualitatively different in higher dimensions. Fig.~\ref{fig:broadening} shows a comparison of spectral functions obtained by classical-statistical simulations and FRG calculations with one-loop vertices in the classical limit. We observe that the deviations from the classical-statistical results are increasing when we decrease the dissipative coupling $\gamma/\beta m$, as may be expected due to the fact that the longer lived excitations can interact with each other over a larger time scale. While for $\gamma/\beta m=0.2$ the FRG calculation with one-loop vertex functions still provides a rather accurate description of the classical statistical result, the agreement becomes gradually worse with decreasing $\gamma/\beta m$. Especially for very small values of the dissipative coupling $\gamma/\beta m < 0.05$, the quasi-particle peak of the spectral function splits into a double peak, which is clearly not observed in the classical-statistical data. Similarly, also the strength of the $1\leftrightarrow 3$ resonance peak around located $\omega \sim 3m $ generally tends to be over-estimated by the FRG calculations.

\begin{figure}[t]
    \centering
    \begin{minipage}[t]{0.45\textwidth}
        \includegraphics[width=\textwidth]{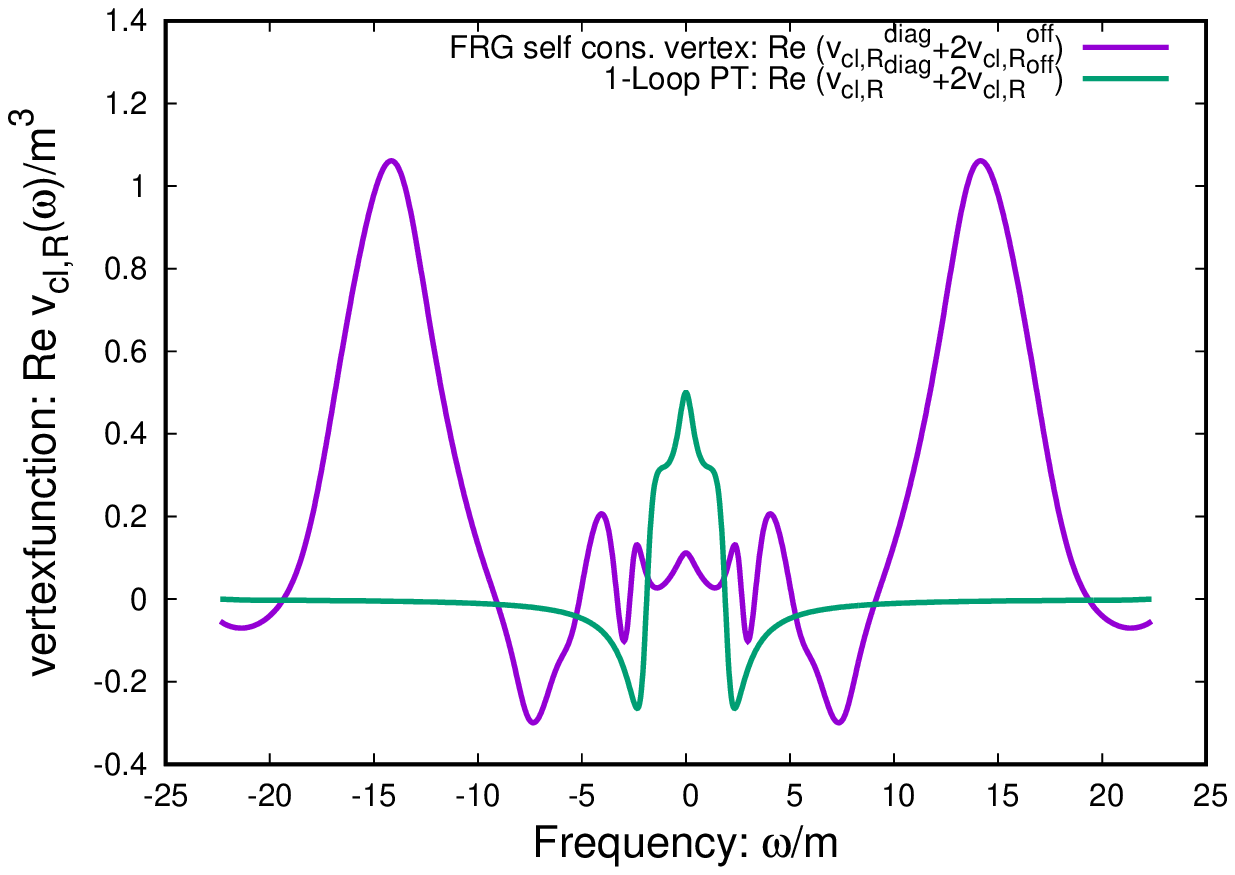}
    \end{minipage}
    \hspace*{0.5cm}
    \begin{minipage}[t]{0.45\textwidth}
        \includegraphics[width=\textwidth]{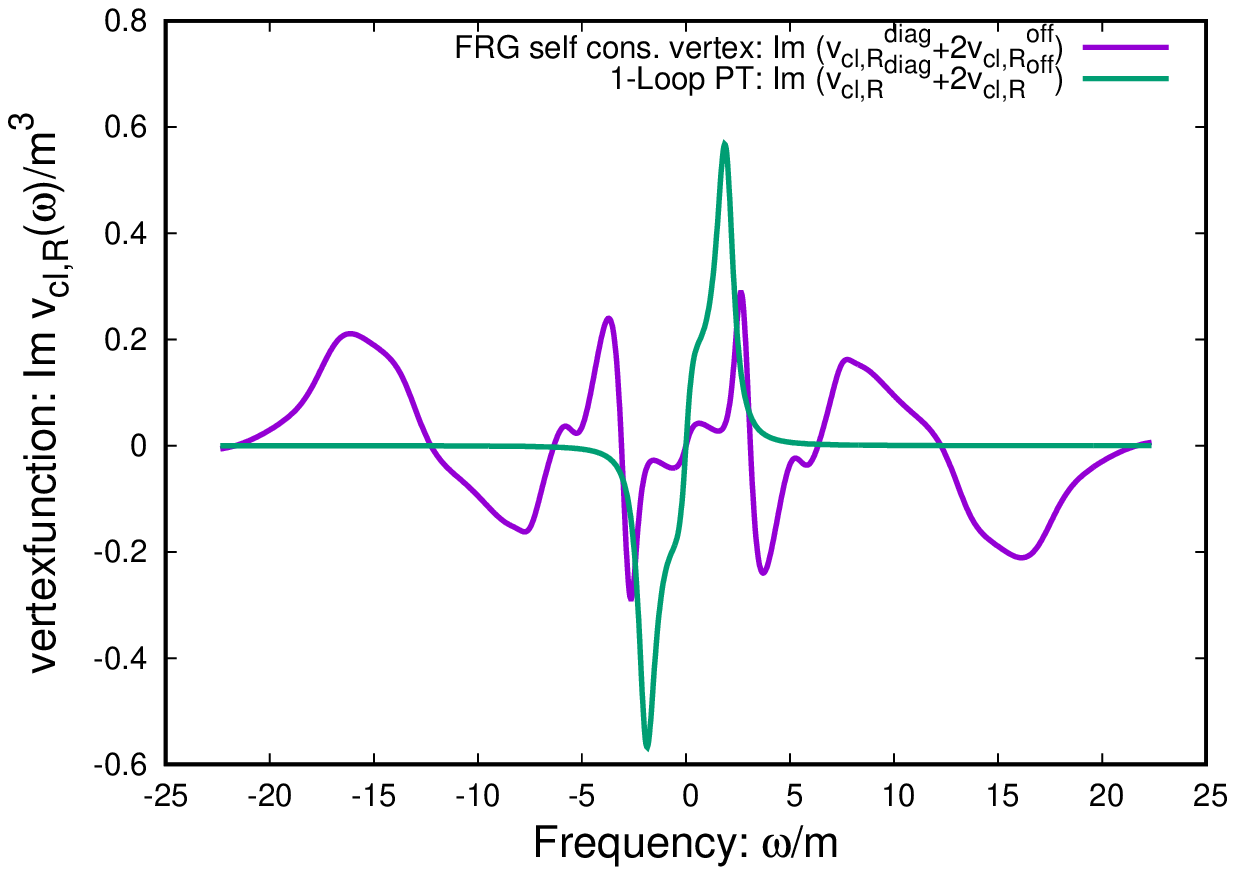}
    \end{minipage}
    \caption{Comparison of the perturbative vertex function at one-loop level and the self-consistently determined vertex function for parameters $m=1.0$, $\frac{\lambda}{\beta m^4}=2.0$, $\frac{\gamma}{\beta m}=0.5$, $\frac{\lambda}{m^3}=2.0$ in the classical limit. Left: comparison of the real parts. Right: comparison of the imaginary parts. \label{fig:comp_vertex}}
\end{figure}

We note that the FRG calculations with scale-dependent vertex functions also become unstable for small dissipative coupling $\gamma/\beta m$ than shown in Fig.\ref{fig:broadening}. In order to further investigate the instability of the self-constistent FRG method at large couplings (i.e. small dissipative couplings), we can now look at the momentum dependence of the classical, retarded vertex function $v_{cl,R}$. Fig.~\ref{fig:comp_vertex} shows a comparison of the self-consistently determined data for $v_{cl,R}$ with the perturbative one-loop result for the same parameters as in Fig.~\ref{fig:comp_cl}. We recognize that for small frequencies the self-consistent vertex function behaves as expected, as both real- and imaginary parts of the vertex functions are suppressed compared to the perturbative result. However, for larger frequencies we find large enhancements of the self-consistently determined vertex function over the perturbative result. Due to the rather complicated structure of the flow equation for the four point function, we are currently not sure about the exact origins of these spurious enhancements, which may be connected to the particular situation in 0+1 dimension and we hope that our procedure will work out better in higher dimensions. Besides additional studies of this behavior, it would also be useful to extract the corresponding vertex functions directly from classical-statistical simulations, which is clearly beyond the scope of this work but could potentially be achieved along the lines of \cite{Prufer:2019kak}.

\subsection{Spectral functions in the quantum theory}
\begin{figure}[htbp]
    \hspace*{-0.3cm}
    \begin{minipage}{0.08\textwidth}
    \vspace*{3cm}
    $\frac{\lambda}{m^3}=0$ \vspace*{2.5cm}\newline
    $\frac{\lambda}{m^3}=\frac{1}{8}$\vspace*{2.5cm}\newline
    $\frac{\lambda}{m^3}=\frac{1}{2}$\vspace*{2.55cm}\newline
    $\frac{\lambda}{m^3}=2$\vspace*{2.cm}\newline

    \end{minipage}
    \hspace*{-0.6cm}
    \begin{minipage}{0.08\textwidth}
    \rotatebox{90}{Spectralfunction: $\rho(\omega)~m^3$}
    \end{minipage}
    \hspace*{-0.7cm}
    \begin{minipage}{0.29\textwidth} 

    \centering $\frac{\lambda}{\beta m^4}=\frac{1}{8} $
    
    \includegraphics[width=\textwidth]{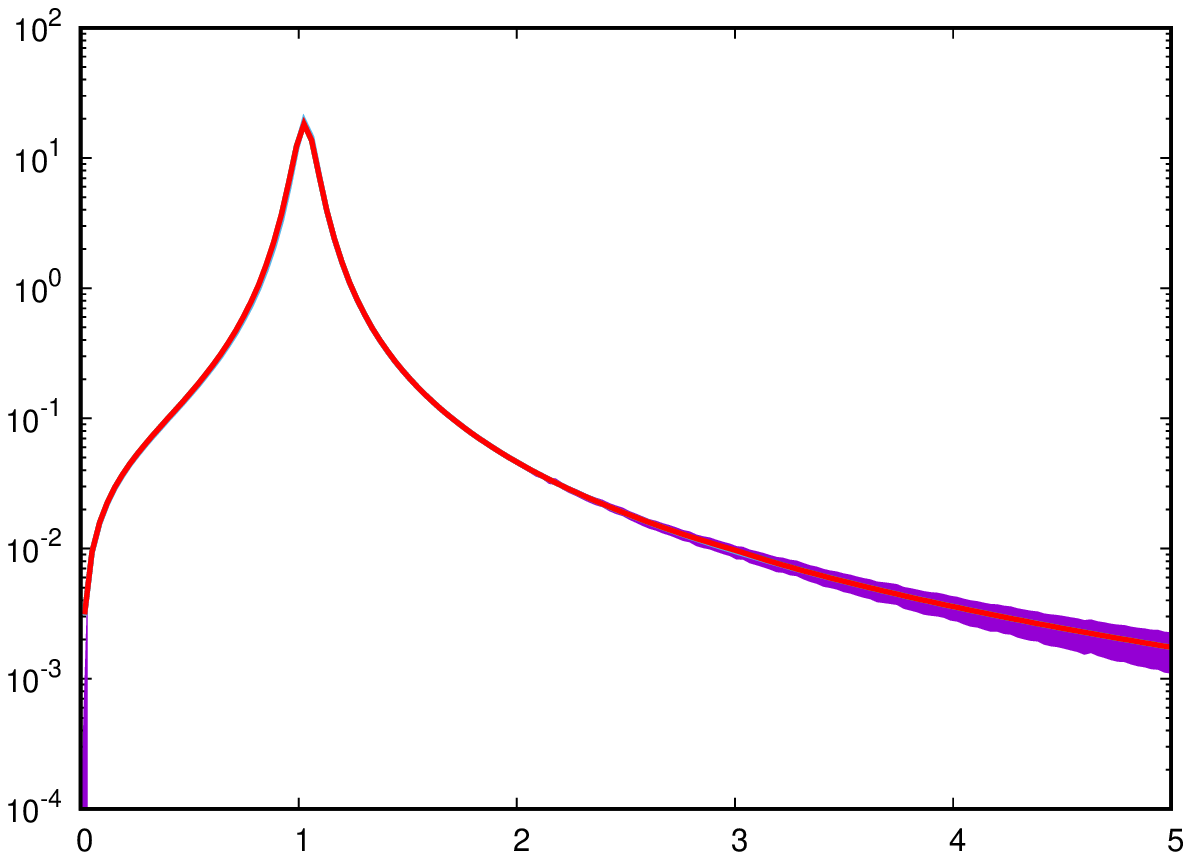}
    \includegraphics[width=\textwidth]{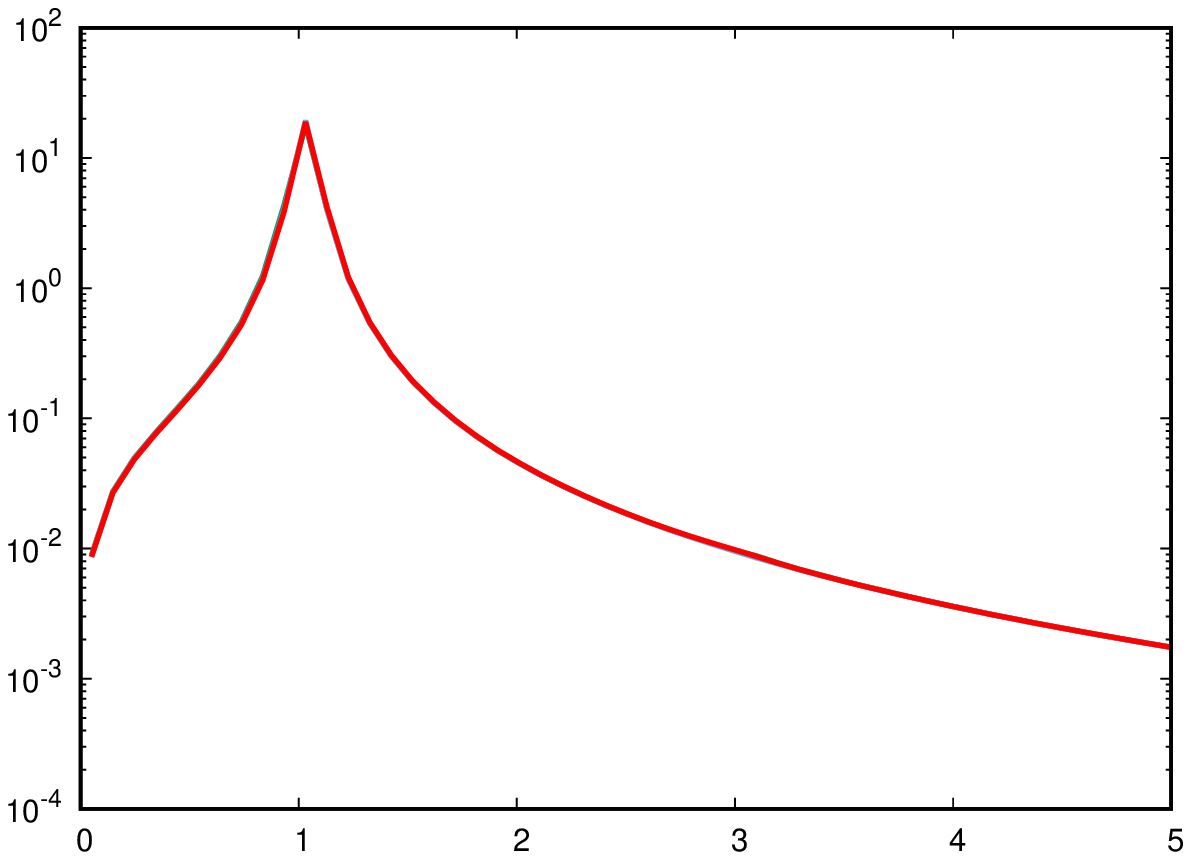}
    \includegraphics[width=\textwidth]{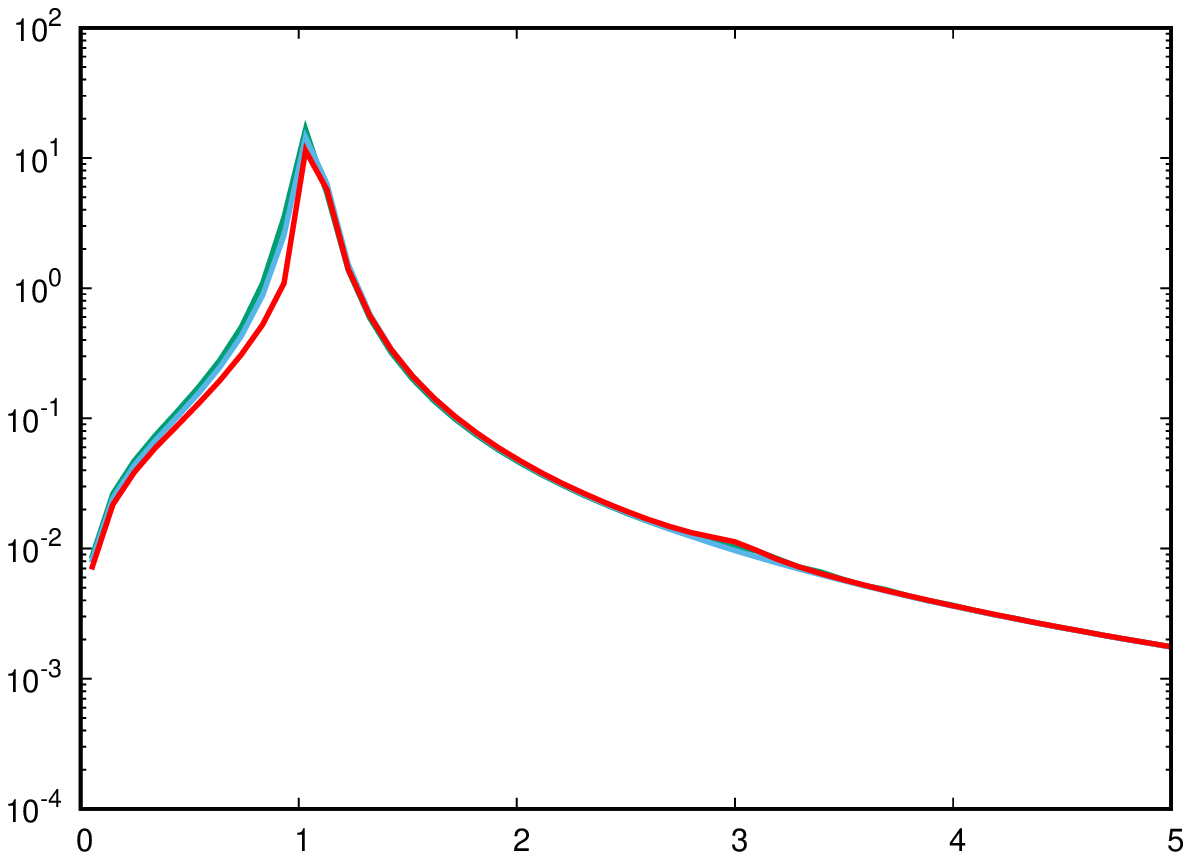}
    \includegraphics[width=\textwidth]{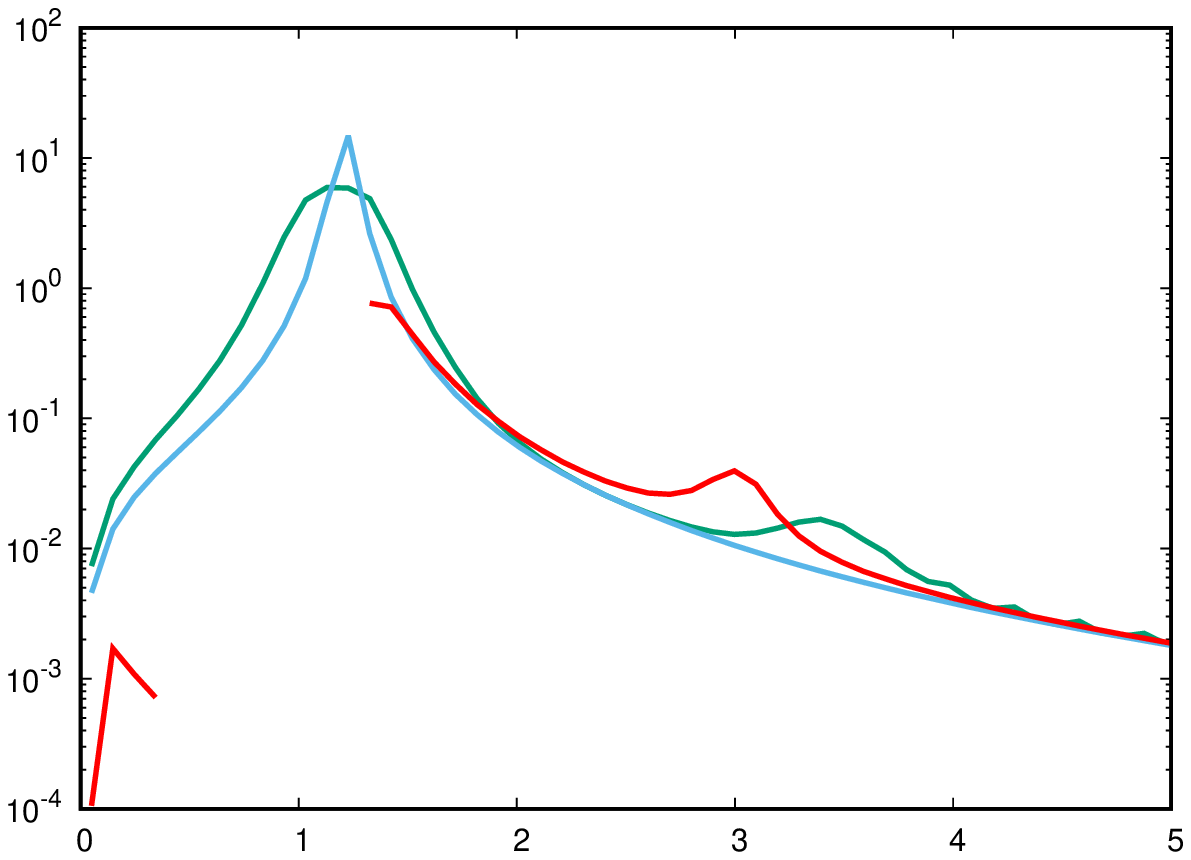}

    \end{minipage}
    \begin{minipage}{0.29\textwidth}
    \vspace*{0.9cm}
    \centering $\frac{\lambda}{\beta m^4}=\frac{1}{2} $

    \includegraphics[width=\textwidth]{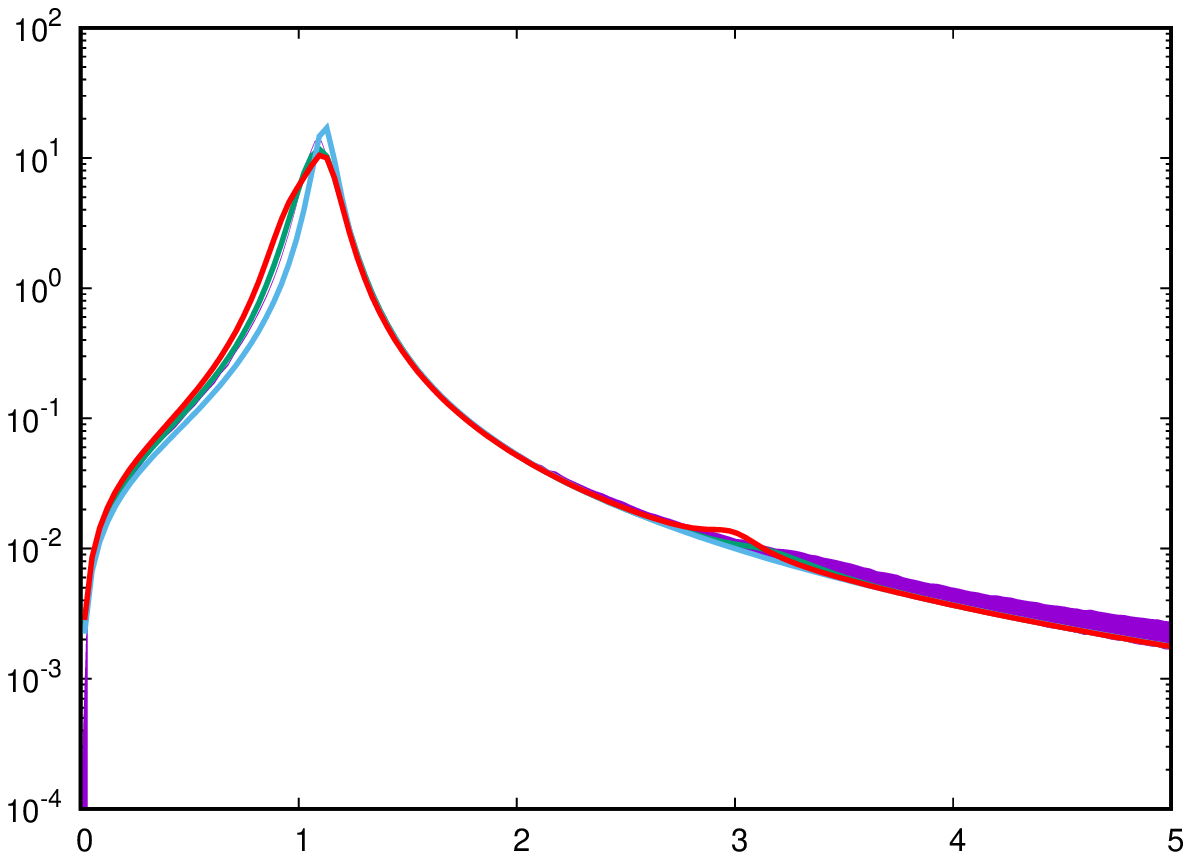}
    \includegraphics[width=\textwidth]{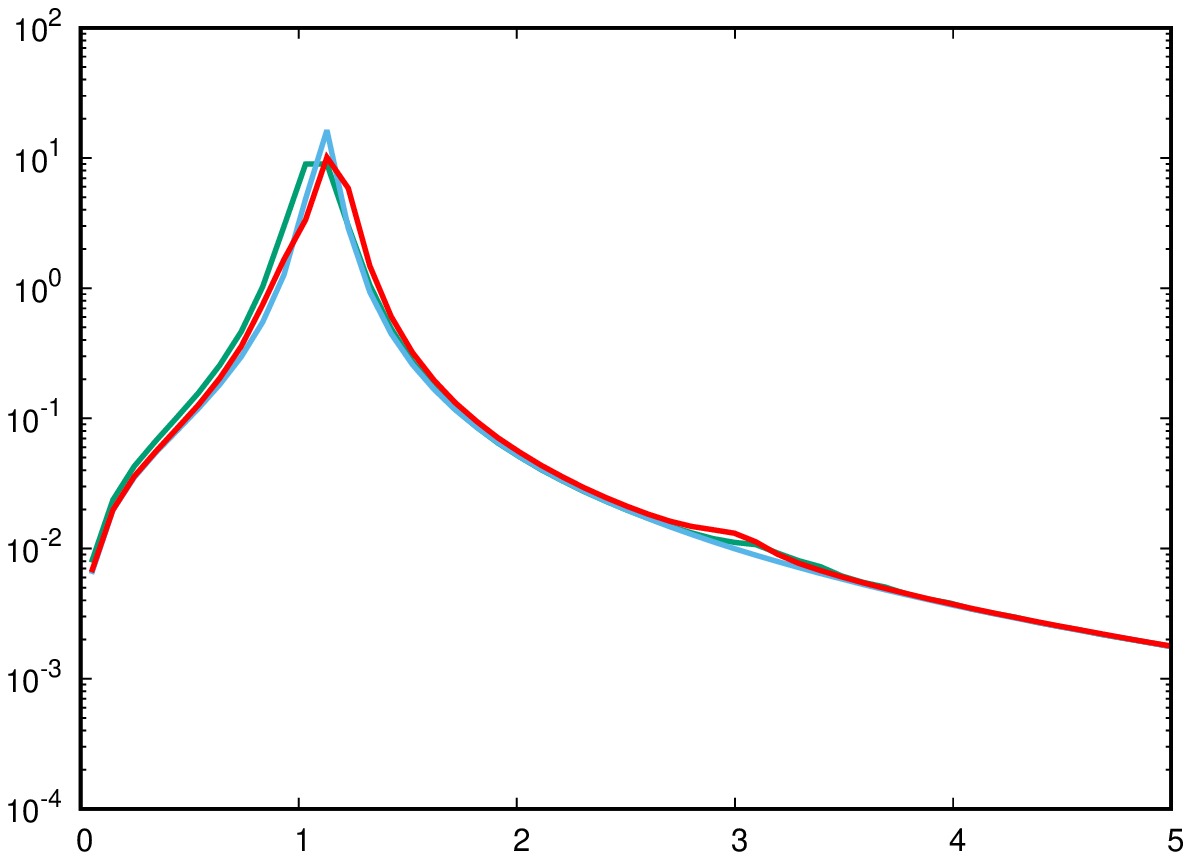}
    \includegraphics[width=\textwidth]{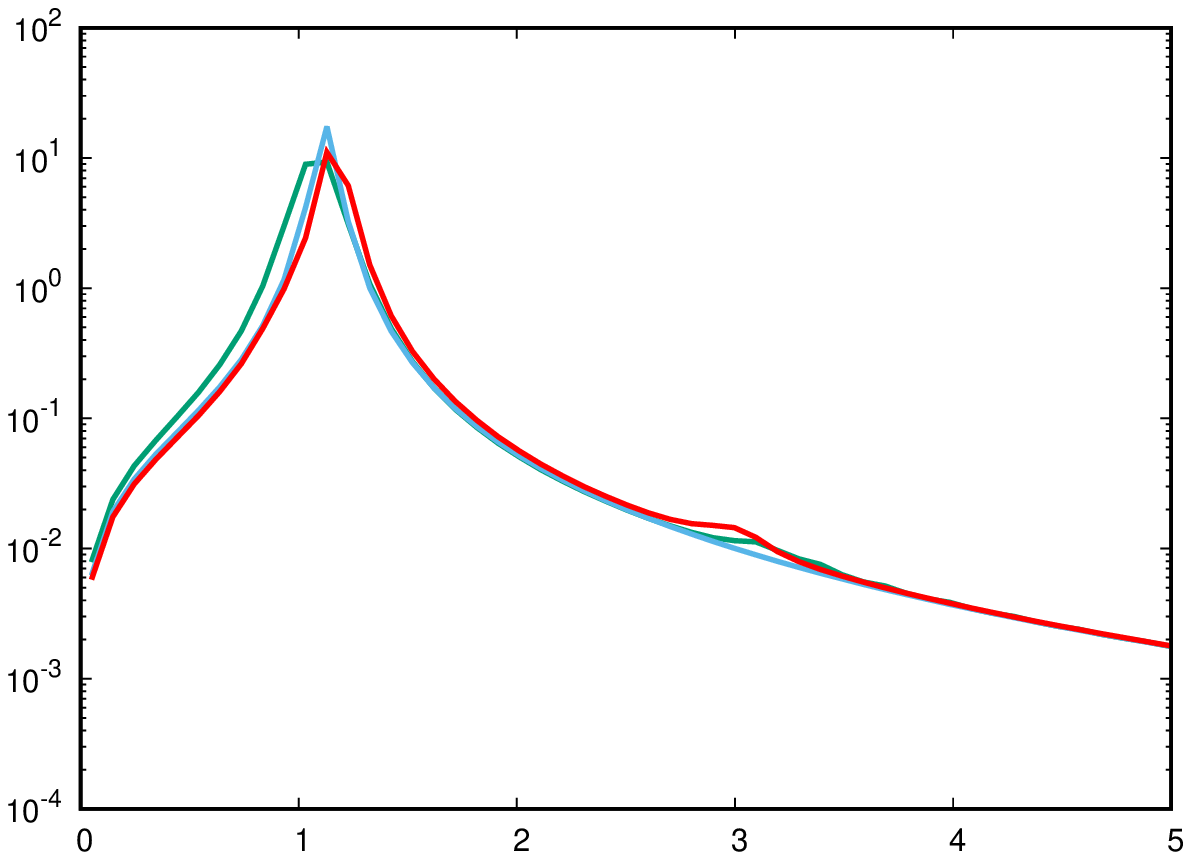}
    \includegraphics[width=\textwidth]{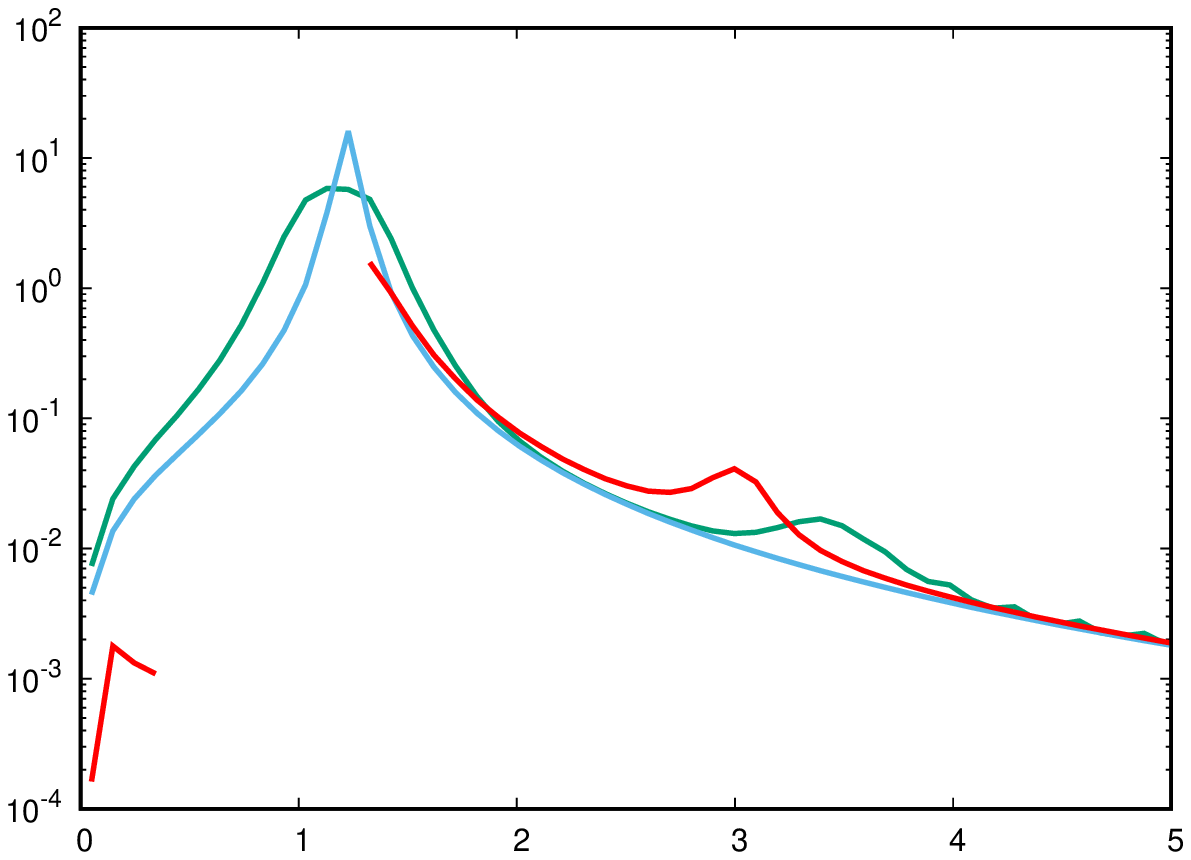}

    \centering
    Frequency: $\omega/m$
    \end{minipage}
    \begin{minipage}{0.29\textwidth}

    \centering $\frac{\lambda}{\beta m^4}=2 $
    
    \includegraphics[width=\textwidth]{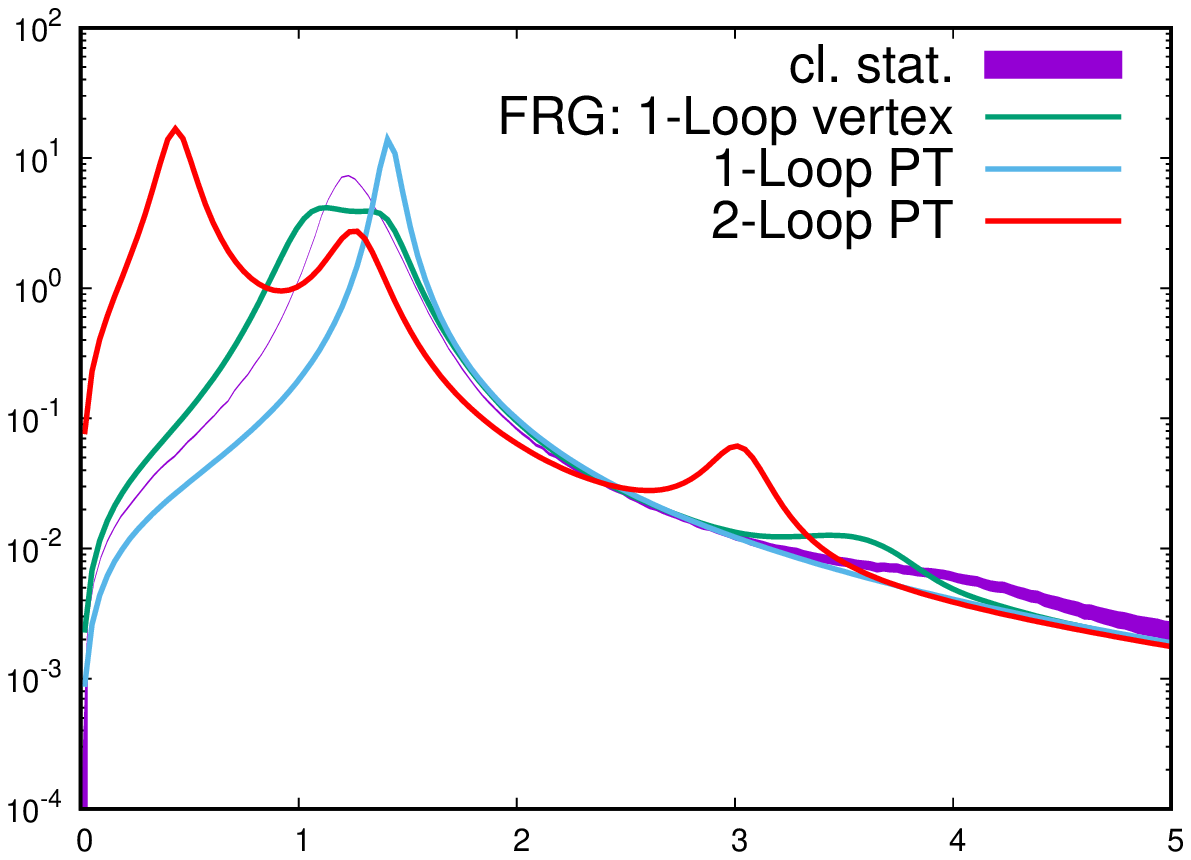}
    \includegraphics[width=\textwidth]{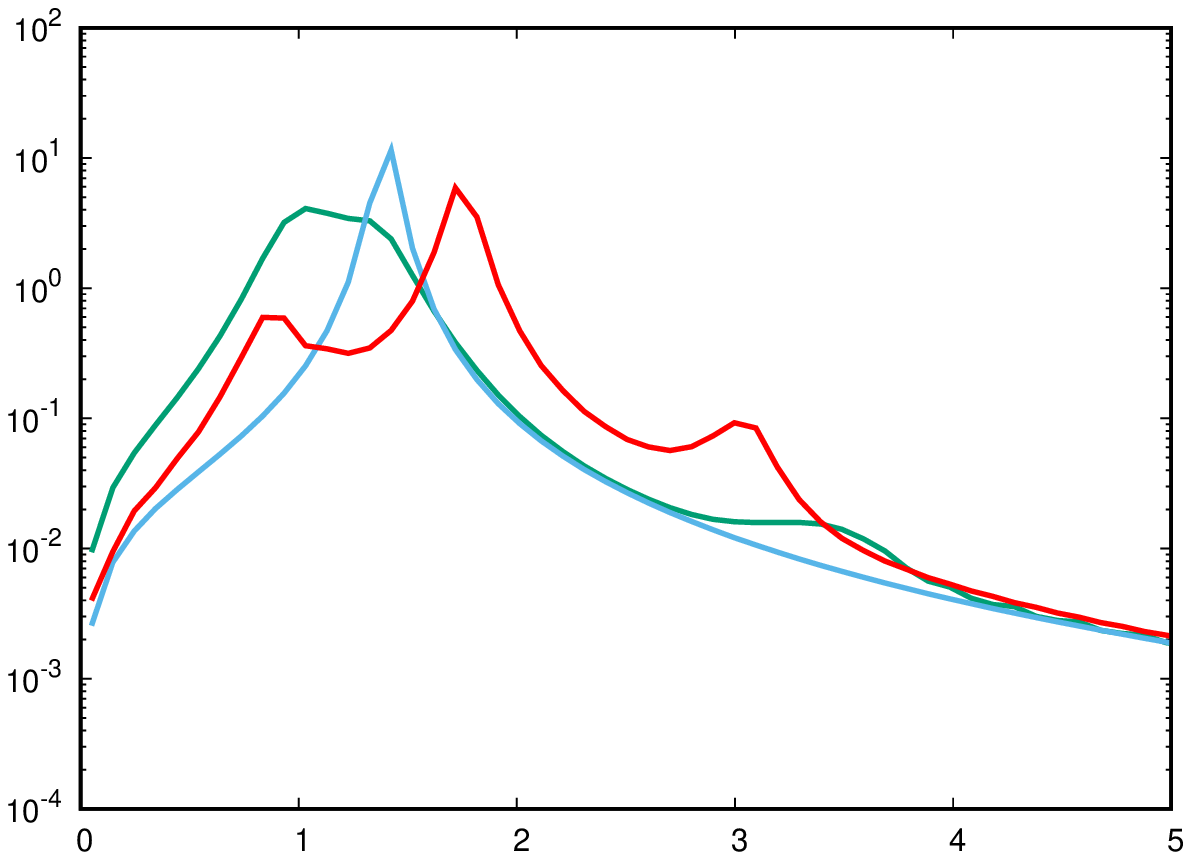}
    \includegraphics[width=\textwidth]{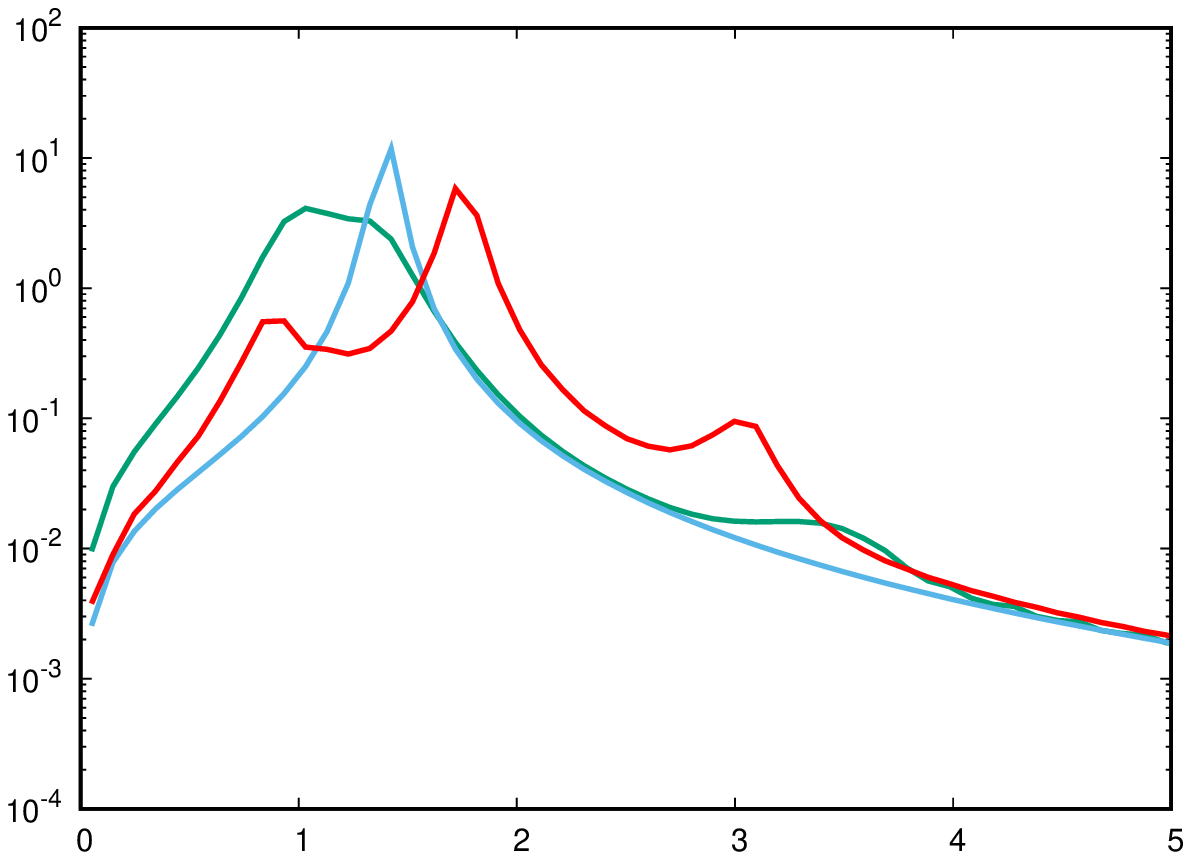}
    \includegraphics[width=\textwidth]{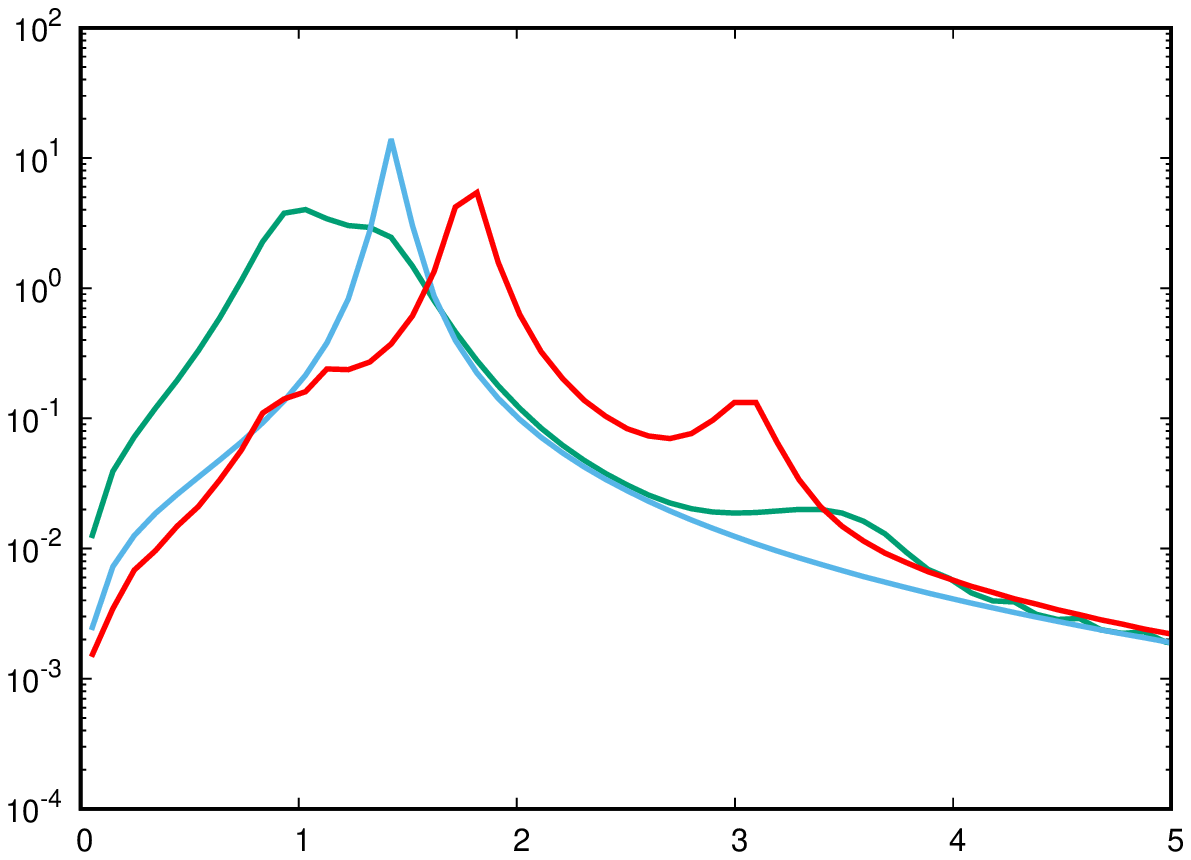}

    \end{minipage}
    \caption{Comparison of spectral functions obtained by FRG calculations compared to other methods at various values of the dimensionless couplings. The damping rate for all plot is $\gamma/\beta m = 0.1$ \label{fig_table_comp}}
\end{figure}

Now, that we have benchmarked and assessed the range of applicability of the method at the hand of the classical-statistical theory, we can continue to investigate spectral functions in the corresponding quantum theory. A compact summary of our results is provide in Fig.~\ref{fig_table_comp}, where we shows a comparison of spectral functions from the FRG with the one loop vertex with results from classical-statistical simulations and perturbative calculations for different values of the thermal and quantum coupling strength. We see that for small coupling all methods agree very well. When we increase the coupling we see a second peak emerging at roughly $3m$ due to the $1 \leftrightarrow 3$ processes in the one-loop correction to the four-point function. As there is no vertex correction at the perturbative one-loop level also the one-loop spectral function fails to capture this feature. When we increase the coupling either by increasing the dimensionless combination of coupling and temperature or by driving the system more towards a strongly coupled quantum system we see that perturbation theory becomes unreliable rather quickly as there are large differences between the LO and NLO results. Specifically for large couplings, the perturbative spectral functions at the two-loop level show additional spurious peaks or may even become negative. Conversely, the FRG results remain much more well behaved throughout the observed parameter range, except perhaps for the largest combination of couplings shown in the bottom right panel. 

\newpage
\section{Conclusions \& Outlook}
\label{sec:Conclusion}
We have presented an overview over on how to employ the functional renormalization group approach on the Schwinger-Keldysh contour to extract real time spectral functions for scalar theories. We introduced a $d+1$ dimensional regulator that is compatible with the time ordering properties of the propagators opening the possibility of having a fully Lorentz symmetric regulator scheme. By introducing a novel diagrammatic representation of the n-point functions we were able to reduce the number of involved diagrams and simplify the derivation of flow equations significantly. We performed a careful perturbative analysis of the FRG flow equations, which revealed that local potential approximations of the effective action, which are commonly used in Euclidean FRG calculations, are insufficient for describing real-time dynamics as e.g. such truncations will never lead to a broadening of the spectral function in the symmetric phase. Based on our perturbative analysis, we developed a different truncation scheme for real-time FRG calculations based on a vertex expansion. By taking into account non-local contributions to the four-point function, all propagators in this scheme are two-loop complete and the FRG flow induces a finite decay width of the spectral functions. 

Based on this expansion we derived the relevant flow equations for the two-point functions. By employing a generalization of the perturbative one-loop expression for the four-point functions, we also derived the flow equations for the vertex functions, taking into account generalized fluctuation-dissipation relations and neglecting contributions involving higher n-point functions, which enable us to solve the truncated system self-consistently. We developed to different numerical procedures to solve the RG flow equations employing (pseudo-)spectral methods, based on  on a straight forwards lattice discretization using Fast-Fourier Transform (FFT) and an expansion in terms of Hermite functions.

We benchmarked our methods at the $0+1d$ example of the anharmonic oscillator where we compared results from perturbation theory, our FRG calulations with both truncation schemes and results from classical-statistical simulations. Since the real-time FRG framework can be formulated in essentially the same way for classical and quantum theories, the comparison to exact results from classical-statistical simulations proved to be an important benchmark to asses the range of applicability and performance of the method. Overall we find that the real-time FRG is able to reproduce the classical-statistical results much better than the perturbative calculations. Still, we find that a larger couplings also the FRG fails to reproduce the correct results. In case of the one-loop vertex trunctation this is connected to the perturbative origin of the vertex. In case of the fully self-consistent FRG calculations we found a spurious enhancement of the vertex functions at large momenta leading to a break down of the method for large couplings. The origin of this enhancement is still unclear but could particularly be a problem of the 0+1 dimensional theory. Another possible cause is the omission of six-point functions in our truncation. Similar to the requirement of a scheme with two-loop complete propagators to reproduce the broadening of the spectral functions we might also need a truncation with two-loop complete four-point functions to be able to correctly renormalize the vertices.\newline

While the formalism described in this work has been derived for $N$ component scalar field theories in $d+1$ dimensions, so far our numerical investigations have been limited to the 0+1 dimensional theory. Clearly a next important step would be to generalize our numerical investigations to higher dimensional systems, especially in 3+1d. Evidently, the comparison of real-time FRG calculations in the classical limit to classical-statistical simulations proved extremely insightful, and should also be pursued for studies in higher dimensions. We also expect that in higher dimensions it should be possible to take the limit of vanishing dissipative coupling $\gamma/\beta m \to 0$, which would further allow to compare real-time FRG calculations in the quantum theory to results from lattice Monte-Carlo simulations and/or analytically continued FRG calculations in Euclidean space-time. Eventually, we want to generalize our framework to include fermions opening the possibility of applying our framework to low energy effective theories of QCD like e.g. the Quark-Meson model.

\subsection*{Acknowledgement}
We thank L.~von~Smekal, D.~Schweitzer, F.~Rennecke, M.~Spier and J.~Pawlowski for insightful discussions throughout this project. This work is supported by the Deutsche Forschungsgemeinschaft (DFG, German Research Foundation) through the CRC-TR 211 ’Strong-interaction matter under extreme conditions’– project number 315477589 – TRR 211.

\newpage
\appendix
\section{Conventions for real-time propagators}
Below we summarize our conventions and also note some useful relations among the various real-time propagators. Based on the operator definitions one has
\begin{eqnarray}
F(x\bar{x})=\frac{1}{2} \langle \{ \hat{\phi}(x),\hat{\phi}(\bar{x}) \} \rangle\;, \qquad \rho(x\bar{x}) = i \langle [\hat{\phi}(x),\hat{\phi}(\bar{x}) ] \rangle\;,
\end{eqnarray}
along with
\begin{eqnarray}
\label{eq:GRGARhoRel}
G^{R}(x\bar{x})= + \theta(x_{0} -\bar{x}_{0}) \rho(x\bar{x})\;, \qquad G^{A}(x\bar{x})= - \theta(\bar{x}_{0}-x_{0}) \rho(x\bar{x})\;.
\end{eqnarray}
We also note for convenience the following relations between real and imaginary parts of the various correlation functions\footnote{Note that unlike other authors we do not introduce an additional factor of $-i$ in the Fourier transform of the spectral function. Hence the corresponding factor of $-i$ appears explicitly in the relation between the statistical function and the spectral function.}
\begin{eqnarray}
\rho(p)=2i \text{Im} G^{R}(p)=G^{R}(p)-G^{A}(p)\;, \qquad F(p)= -i n_{eff}(p) \rho(p)\;,
\end{eqnarray}
where in the quantum case we have $n_{eff}^{qu}=n_{BE} + 1/2$ with $n_{BE}(p)=1/(e^{\beta p_0}+1)$ being the Bose-Einstein distribution, such that
\begin{eqnarray}
n_{eff}^{qu}(p) = \left(n_{BE}(p)+1/2\right) = -\left(n_{BE}(-p)+1/2\right) = \frac{1}{2} \coth(\beta p_0/2)\;.
\end{eqnarray}
In the classical case we find 
\begin{equation}
   n_{eff}^{cl}(p) = \frac{1}{\beta p_0} \; ,
\end{equation}
to be the Rayleigh-Jeans distribution.
Our convention for the Fourier transformation reads
\begin{equation}
  \varphi(x) = \int \frac{d\omega}{2\pi} \int \frac{d^d\mathbf{p}}{(2\pi)^d} e^{+ipx} \varphi(p) \; .
\end{equation}
We further note the following symmetry relations
\begin{eqnarray}
\rho(p)=-\rho(-p)\;, \qquad F(p)=F(-p)\;, \qquad G_{R}(p)=G_{A}(-p)\;,
\end{eqnarray}
as well as the various relations for the real and imaginary parts
\begin{eqnarray}
\text{Re}(\rho(p))&=&0\;, \qquad \text{Im}(\rho(p))= +2  \text{Im} G^{R}(p) = -2  \text{Im} G^{A}(p)\;, \\
\text{Re}(F(p))&=&+2 n_{eff}(p_0) \text{Im} G^{R}(p)= -2n_{eff}(p_0) \text{Im} G^{A}(p)  \qquad \text{Im}(F(p))=0\;, \nonumber
\end{eqnarray}
along with
\begin{eqnarray}
\label{eq:KramersKronig}
\text{Im}G^{R}(p) &=& \frac{+1}{2i} \rho(p)\;, \qquad \text{Re}G^{R}(p)= P.V. \frac{1}{2\pi i} \int_{-\infty}^{\infty} d\omega  \frac{\rho(\omega,\vec{p})}{\omega-p_0}\;, \\
\text{Im}G^{A}(p) &=& \frac{-1}{2i} \rho(p)\;, \qquad \text{Re}G^{A}(p)= P.V. \frac{1}{2\pi i} \int_{-\infty}^{\infty} d\omega  \frac{\rho(\omega,\vec{p})}{\omega+p_0}
\end{eqnarray}
where we used the relations $\theta(x)=\frac{1}{2\pi i} \int_{-\infty}^{\infty} d\omega \frac{1}{\omega - i \epsilon} e^{i \omega x}$ along with $\frac{1}{\omega - i \epsilon}= P.V. \frac{1}{\omega} + i \pi \delta(\omega)$, such that the above idenitities follow directly from the respective time orderings.\\

\section{Evaluation of the real-time effective action in the limit $k \to \Lambda$}
Starting from the definition of the effective action in Eq.~(\ref{eq:Gammak}), it is convenient to perform a field shift $\varphi \to \phi +\varphi$ in the functional integration to separate off the contribution from the classical action. By exploiting the equations of motion to re-express the appearance of the sources $J,\tilde{J}$ in terms of derivatives of the effective action, Eq.~(\ref{eq:Gammak}) can then be recast into a functional integro-differential equation for the effective action
\begin{eqnarray}
\Gamma_{k}[\phi,\tilde \phi] = S[\phi,\tilde \phi] - i \log \Delta Z_{k}[\phi,\tilde \phi] \; , \label{eq:effActionB}
\end{eqnarray}
with $\Delta Z_{k}[\phi,\tilde \phi]$ given by the functional 
\begin{eqnarray}
\Delta Z_{k}[\phi,\tilde \phi]&=&\left. e^{i \Big(S_{\mathcal C}[\phi+\frac{\delta}{\delta \tilde j},\tilde \phi + \frac{\delta}{\delta j}] - S_{\mathcal C}[\phi,\tilde \phi] - \Gamma_{k}^{\phi}\otimes \frac{\delta}{\delta \tilde j} - \Gamma_{k}^{\tilde \phi}\otimes  \frac{\delta}{\delta j}  \Big) } \int \left[D\varphi D\tilde \varphi\right]_{k} e^{\tilde{j} \otimes \varphi + j \otimes \tilde \varphi} \right|_{j=\tilde{j}=0}\;, \label{eq:pathIntB}
\end{eqnarray}
where $\int \left[D\varphi D\tilde \varphi\right]_{k}$ we denotes the regulated path integral
\begin{eqnarray}
\int \left[D\varphi D\tilde \varphi\right]_{k} &=& \int D\varphi D\tilde \varphi~e^{i \Delta S_{k}[\varphi,\tilde \varphi]}\;.
\end{eqnarray}
which for a Gaussian regulator can be evaluated explicitly. Expressing the functional integrations in Fourier space, one finds (up to irrelevant pre-factors)
\begin{eqnarray}
\int \left[D\varphi D\tilde \varphi\right]_{k} &=& \prod_{a} \int_{-\infty}^{\infty} d\varphi_{a}^{(0)} \int_{-\infty}^{\infty} d\tilde\varphi_{a}^{(0)}  \left. e^{- \gamma_{k}(\omega) \omega n_{\rm eff}(\omega)   \left(\tilde \varphi_{a}^{(0)} \right)^2 - i \mu_{k}(\omega) \varphi_{a}^{(0)}\tilde{\varphi}_{a}^{(0)}} \right|_{\omega=0} \\
&& \prod_{\omega >0} \int_{-\infty}^{\infty} d\text{Re}\varphi_{a}(\omega) \int_{-\infty}^{\infty} d\text{Im}\varphi_{a}(\omega) \int_{-\infty}^{\infty} d\text{Re}\tilde\varphi_{a}(\omega) \int_{-\infty}^{\infty} d\text{Im}\tilde\varphi_{a}(\omega) \nonumber \\
&&  \times e^{- \gamma_{k}(\omega) \omega n_{\rm eff}(\omega) |\tilde \varphi_{a}(\omega)|^2 - i \mu_{k}(\omega) \frac{\varphi_{a}(\omega) \tilde{\varphi}^{*}_{a}(\omega) + \varphi^{*}_{a}(\omega) \tilde{\varphi}_{a}(\omega)}{2} -i\gamma_{k}(\omega)\frac{\varphi_{a}(\omega) \tilde{\varphi}^{*}_{a}(\omega) - \varphi^{*}_{a}(\omega) \tilde{\varphi}_{a}(\omega)}{2i}} \nonumber \; .
\end{eqnarray}
Evaluating the functional integral explicitly according to
\begin{eqnarray}
\int \left[D\varphi D\tilde \varphi\right]_{k} e^{\tilde{j} \otimes \varphi + j  \otimes  \tilde \varphi} = 
\prod_{a}  \frac{2\pi}{|\mu_{k}(\omega)|} e^{\frac{-i \mu_{k}(\omega)}{\mu_{k}^2(\omega)} \tilde{j}_{a}^{0} j_{a}^{0}}e^{\frac{\omega \gamma_{k}(\omega) n_{\rm eff}(\omega)}{\mu_{k}^2(\omega)} \tilde{j}_{a}^{0} \tilde{j}_{a}^{0}} \prod_{\omega >0} \frac{(2\pi)^2}{\mu_{k}^2(\omega)+\omega^2 \gamma_{k}^2(\omega)} \\
e^{\frac{2 \omega \gamma_{k}(\omega)}{\mu_{k}^2(\omega)+\omega^2 \gamma_{k}^2(\omega)} \left(j(-\omega)\tilde{j}(\omega)-j(\omega)\tilde{j}(-\omega) + 2 \tilde{j}(-\omega)\tilde{j}(\omega) n_{\rm eff}(\omega)\right)}
e^{- \frac{2 i \mu_{k}(\omega)}{\mu_{k}^2(\omega)+\omega^2 \gamma_{k}^2(\omega)}\left(j(-\omega)\tilde{j}(\omega)-j(\omega)\tilde{j}(-\omega)\right)} \; ,
\end{eqnarray}
one finds that in the limit $k \to \Lambda$, the relevant factors characterizing the variations w.r.t. the sources $j,\tilde{j}$ are inversely proportional to regulators, such that
\begin{eqnarray}
\lim_{k \to \Lambda} \frac{\omega \gamma_{k}(\omega)}{\mu_{k}^2(\omega)+\omega^2 \gamma_{k}^2(\omega)} = 0\;, \qquad  \lim_{k \to \Lambda} \frac{\mu_{k}(\omega)}{\mu_{k}^2(\omega)+\omega^2 \gamma_{k}^2(\omega)} = 0\;,
\end{eqnarray}
and the functional becomes independent of the sources $j,\tilde{j}$ in the vicinity of $j=\tilde{j}=0$ where derivatives are to be evaluated. One concludes, that in the limit $k \to \Lambda$ the effective action in Eq.~(\ref{eq:effActionB}) does not receive any additional contributions from the path integral in Eq.~(\ref{eq:pathIntB}) and thus reduces to 
\begin{eqnarray}
\lim_{k \to \Lambda} \Gamma_{k}[\phi,\tilde \phi]= S_{\mathcal{C}}[\phi,\tilde \phi]\;.
\end{eqnarray}

\section{Perturbative contribution to the damping rate}
Here we will evaluate the perturbative contribution to the damping rate, which is also useful to understand the differences between classical and quantum statistical processes and in establishing the comparison to the literature that is largely based on analytic continuations of Euclidean calculations. Based on Eq.~(\ref{eq:Delta2Gamma2TwoLoopSunset}) the perturbative contribution is obtained by evaluating the RHS with free propagators, which take the following form
 in momentum space 
\begin{eqnarray}
G^{R}_{0}(p)= \frac{-1}{\omega^2 -E_{p}^2 + i \gamma/\beta \omega}\;, \qquad G^{A}_{0}(p)= \frac{-1}{\omega^2 -E_{p}^2 - i \gamma/\beta \omega}\;.
\end{eqnarray}
Spectral function and  statistical function are then given by
\begin{eqnarray}
&&\rho_{0}(p) = G^{R}_{0}(p)-G^{A}_{0}(p) = \frac{2i (\gamma/\beta) \omega }{(\omega^2 -E_{p}^2)^2 + ( \gamma/\beta \omega)^2}\;, \\ 
&&\qquad F_{0}(p)=\frac{(\gamma/\beta) \omega 2 n_{neff}(\omega)}{(\omega^2 -E_{p}^2)^2 + ( \gamma/\beta \omega)^2}  \;,
\end{eqnarray}
allowing us to proceed directly with the evaluation of diagrams. By expressing all retarded/advanced propagators in terms the spectral function using Eq.~(\ref{eq:GRGARhoRel}), the retarded self-energy, defined as
\begin{eqnarray}
&&G^R(x \bar x) = G^{R}_0(x \bar x) + G^{R}_0(x y) \Sigma^{R}(y \bar y) G^R(\bar y \bar x) \; , \\
&&\Sigma^{R}(x \bar x) = \Sigma^{R}_{cl}(x\bar{x}) + \Sigma^{R}_{qu}(x\bar{x}) \; ,
\end{eqnarray}
can be expressed in the form
\begin{eqnarray}
\Sigma^{R}_{cl}(x\bar{x})&=&-\frac{3}{2}(N+2)\lambda_{cl}^2~\theta(x\bar{x})~iF(x\bar{x})~iF(x\bar{x})~\rho(x\bar{x})\;, \\
\Sigma^{R}_{qu}(x\bar{x})&=&\frac{-\frac{3}{2}(N+2)}{3} \lambda_{cl} \lambda_{qu}~\theta(x\bar{x})~\rho(x\bar{x})~\rho(x\bar{x})~\rho(x\bar{x})\;,
\end{eqnarray}
Since spectral and statistical correlation functions are purely real in coordinate space, and have well defined real and imaginary parts in momentum space, the real and imaginary parts of the can be readily evaluated, by use of relations as in (\ref{eq:KramersKronig}), which follow directly from the properties of the Heavyside step function. Since likewise the real-part can be re-constructed from Kramers-Kronig type relations, we will focus on the the imaginary part, which can be directly evaluated as
\begin{eqnarray}
\text{Im}\Sigma^{R}_{cl}(p)&=&-\frac{3}{2}(N+2)\lambda_{cl}^2~\int \frac{d^{d+1}k}{(2\pi)^{d+1}}~\int \frac{d^{d+1}q}{(2\pi)^{d+1}}~\frac{iF(k)~iF(q)~\rho(p-k-q)}{2i}\;, \\
\text{Im}\Sigma^{R}_{qu}(p)&=&\frac{-\frac{3}{2}(N+2)}{3} \lambda_{cl} \lambda_{qu}~\int \frac{d^{d+1}k}{(2\pi)^{d+1}}~\int \frac{d^{d+1}q}{(2\pi)^{d+1}}~\frac{\rho(k)~\rho(q)~\rho(p-k-q)}{2i}\;,
\end{eqnarray}
Specifically in the limit $\gamma \to 0$ of non-dissipative systems, the energy integrations can be performed using
\begin{eqnarray}
\rho_{0}(p)  &\stackrel{\gamma \to 0}{=}& 2\pi i~\text{sign}(\omega) ~\delta(\omega^2-E_{p}^2) = 2\pi i  \sum_{s_{p}} s_{p} \frac{\delta(\omega-s_p E_{p})}{2 E_{p}}\;, \\ 
F_{0}(p) &\stackrel{\gamma \to 0}{=}& 2\pi~\delta(\omega^2-E_{p}^2) n_{eff}(|\omega|) = 2\pi  \sum_{s_{p}}  n_{eff}(|\omega|) \frac{\delta(\omega-s_p E_{p})}{2 E_{p}}\;,
\end{eqnarray}
and the result can be compactly expressed in the form 
\begin{align}
\text{Im}\Sigma^{R}_{cl}(p)=&-\frac{3}{2}(N+2)\lambda_{cl}^2~\int_{\vec{q},\vec{k}}~\sum_{s_{k} s_{q}  s_{r}}~\frac{s_{r}n_{eff}(E_{k})n_{eff}(E_{q})}{8 E_{k} E_{q} E_{r} } \\
&\times \pi\delta(p^{0}-s_{k}E_{k}-s_{q}E_{q}-s_{r}E_{r})\;, \nonumber \\ \notag\\
\text{Im}\Sigma^{R}_{qu}(p)=&\frac{-\frac{3}{2}(N+2)}{3} \lambda_{cl} \lambda_{qu}~\int_{\vec{q},\vec{k}}~\sum_{s_{k} s_{q} s_{r}}~\frac{s_{q}s_{k} s_{r}}{8 E_{k} E_{q} E_{r} }~\pi\delta(p^{0}-s_{k}E_{k}-s_{q}E_{q}-s_{r}E_{r})\;,
\end{align}
where the summation over $s_{i}=\pm$ collects the positive and negative frequency contributions and we denote $\int_{\vec{q},\vec{k}}=\int \frac{d^{d}k}{(2\pi)^{d}}~\int \frac{d^{d}q}{(2\pi)^{d}}$ as well as $r=p-k-q$ to lighten the notation. By further symmetrizing the integrand of $\text{Im}\Sigma^{R}_{cl}(p)$, the expressions can be re-cast
in the form
\begin{align}
&\text{Im}\Sigma^{R}_{cl}(p)=\frac{-\frac{3\pi}{2}(N+2)}{3}\lambda_{cl}^2~\int_{\vec{q},\vec{k}}~\sum_{s_{k} s_{q} s_{r}}~\frac{\delta(p^{0}-s_{k}E_{k}-s_{q}E_{q}-s_{r}E_{r})}{8 E_{k} E_{q} E_{r} } \\
 &\left[s_{r}n_{eff}(E_{k})n_{eff}(E_{q}) + s_{q}n_{eff}(E_{k})n_{eff}(E_{r}) + s_{k}(n_{eff}(E_{q}) n_{eff}(E_{r}) \right] \nonumber  \;, \\
&\text{Im}\Sigma^{R}_{qu}(p)=\frac{-\frac{3\pi}{2}(N+2)}{3} \lambda_{cl} \lambda_{qu}~\int_{\vec{q},\vec{k}}~\sum_{s_{k} s_{q} s_{r}}~\frac{\delta(p^{0}-s_{k}E_{k}-s_{q}E_{q}-s_{r}E_{r})}{8 E_{k} E_{q} E_{r} } s_{k} s_{q} s_{r}\;,
\end{align}
where the pre-factors of the two terms are equal except for the different appearances of the coupling constants $\lambda_{cl}$ and $\lambda_{qu}$. We will now concentrate on a quantum theory where we can set $n_{eff}(E)=n(E)+1/2$ -- with the Bose-Einstein distribution $n(E)$ -- and exploit the relation $\lambda_{qu}=\lambda_{cl}/4$ between the classical and quantum (tree level) vertices,  it is straightforward to show that the above terms can be combined in the following way
\begin{align}
\frac{s_q s_k s_r}{4} &+ s_{r}\left(n(E_{k})+\frac{1}{2}\right)\left(n(E_{q})+\frac{1}{2}\right) + s_{q}\left(n(E_{k})+\frac{1}{2}\right)\left(n(E_{r})+\frac{1}{2}\right) \notag\\
&+ s_{k}\left(n(E_{q})+\frac{1}{2}\right)\left(n(E_{r})+\frac{1}{2}\right) = \\
& \hspace{2cm} \Big[n(E_{k})+\frac{1+s_{k}}{2}\Big] \Big[n(E_{q})+\frac{1+s_{q}}{2}\Big] \Big[n(E_{r})+\frac{1+s_{r}}{2}\Big]\notag \\
& \hspace{2cm}- \Big [n(E_{k})+\frac{1-s_{k}}{2}\Big] \Big[n(E_{q})+\frac{1-s_{q}}{2}\Big] \Big[n(E_{r})+\frac{1-s_{r}}{2}\Big]\;,\notag
\end{align}
which contain the usual quantum statistical factors for in/out-going particles in a scattering process. Collecting everything, the imaginary part of the retarded self-energy $\text{Im}\Sigma^{R}(p)=\text{Im}\Sigma^{R}_{cl}(p)+\text{Im}\Sigma^{R}_{qu}(p)$ can then be compactly expressed as
\begin{align}
\text{Im}\Sigma^{R}(p)=&-\frac{\pi(N+2)}{2}\lambda_{cl}^2~\int \frac{d^{d}k}{(2\pi)^{d}}~\int \frac{d^{d}q}{(2\pi)^{d}}~\sum_{s_{k} s_{q} s_{r}}~\bigg \lbrace\frac{\delta(p^{0}-s_{k}E_{k}-s_{q}E_{q}-s_{r}E_{r})}{8 E_{k} E_{q} E_{r} } \notag\\
&[n(E_{k})+\frac{1+s_k}{2}] [n(E_{q})+\frac{1+s_q}{2}] [n(E_{r})+\frac{1+s_r}{2}] \\ -  &[n(E_{k})+\frac{1-s_k}{2}] [n(E_{q})+\frac{1-s_q}{2}] [n(E_{r})+\frac{1-s_r}{2}] \bigg \rbrace \;.\notag
\end{align}
in agreement with the standard result in ref.~\cite{Wang:1995qf}. Vice versa, in the classical-statistical theory the contribution proportional to $\lambda_{qu}$ vanishes identically, and the occupancy factors $n(E_{p})+1/2$ are to be replaced by the Rayleigh-Jeans distribution $n_{cl}(E_p)=1/\beta E_{p}$, such that 
\begin{align}
\text{Im}\Sigma^{R}_{cl}(p)=&\frac{-\frac{3\pi}{2}(N+2)}{3}\lambda_{cl}^2~\int_{\vec{q},\vec{k}}~\sum_{s_{k} s_{q} s_{r}}~\frac{\delta(p^{0}-s_{k}E_{k}-s_{q}E_{q}-s_{r}E_{r})}{8 E_{k} E_{q} E_{r} }  \\
& ~\left[s_{r}n_{cl}(E_{k})n_{cl}(E_{q}) + s_{q}n_{cl}(E_{k})n_{cl}(E_{r}) + s_{k}n_{cl}(E_{q})n_{cl}(E_{r}) \right]\;,\notag
\end{align}
now yielding the classical statistical factors for in/out-going particles in a scattering process.  We have thus verified explicitly that with a suitable truncation which properly accounts for the non-local vertex structure generated at the one loop level, the real-time FRG approach correctly captures the collisional broadening of the spectral function at the two loop level.

Since for $\beta E_p \ll 1$ the statistical factors $n(E_{p})+\frac{1+s_p}{2} \approx 1/\beta E_{p}$ agree approximately the classical statistical theory is expected to accurately capture the relevant contributions of excitations with energies much smaller then the temperature. However, one crucial difference is that the classical statistical theory only allows for interactions between physically occupied excitations of the system. Due to the statistical factors, the classical-statistical result behaves as $\text{Im}\Sigma^{R}_{cl}(p) \sim T^3$, such that in the limit $T\to0$ where classically no states are physically occupied, all contributions to the self-energy vanish identically, which is of course not the case in the corresponding quantum theory.

\section{Fluctuation-dissipation relation}
Below we demonstrate explicitly, that the flow equations of the two-point functions satisfy the relation
\begin{eqnarray}
\partial_{k} \Gamma_{k}^{\tilde{\phi}\tilde{\phi}}(p) = n_\text{eff}(p_0) \left( \partial_{k} \Gamma^{\tilde{\phi}\phi}_{k}(p) - \partial_{k} \Gamma^{\phi\tilde{\phi}}_{k}(p) \right)\;. \label{fluc-dis1}
\end{eqnarray}
A quick calculation with the one-loop forms for the vertices show that at one-loop level we have
\begin{equation}
  v_{an}(p) = n_\text{eff}(p_0) (v_{cl,R,k}(p)-v_{cl,A,k}(p)) \;.
\end{equation}
And thus also 
\begin{equation}
  \Gamma^{\tilde{\phi}\tilde{\phi}}(p) = n_\text{eff}(p_0) (\Gamma^{\tilde{\phi}\phi}(p)- \Gamma^{\phi\tilde{\phi}}(p)) \; . \label{fluc-dis}
\end{equation}
Now, in the case of the O(N) model beyond one-loop we have to introduce diagonal and off-diagonal vertex functions.
Here we will make an assumption for the structure of the diagonal as well as the off-diagonal parts of the vertex function, namely
\begin{equation}
  v_{qu,R}^X(x \bar x) B_A(x \bar x) = v_{qu,A}^X(x \bar x) B_R(x \bar x) = 0 \; . \label{gen-fluc-dis}
\end{equation}
This equation just tells us that $v_{qu,R/A}$ is an retarded/advanced function\footnote{This is not correct in a strict sense as $v_{qu,R/A}$ is not necessarily zero for $x - \bar x=0 $ but its good enough for our purposes.}, thus this assumption should better be true. Let us go to momentum space and write down
the flow equations for the quantities in Eq.~(\ref{fluc-dis}).
\begin{align*}
  \partial_k(\Gamma^{\tilde{\phi}\phi}(p)-\Gamma^{\phi\tilde{\phi}}(p)) =& - \frac{i}{2} \int_q \bigg \lbrace 
  2 [(v_{cl,R,k}^{diag}(p-q)-v_{cl,A,k}^{diag}(p-q)) \notag\\  &+ (N+1)(v_{cl,R,k}^{off}(p-q)-v_{cl,A,k}^{off}(p-q))] n_\text{eff}(q_0) B_\rho(q) \\
  &+ 2 [v_{an}^{diag}(p-q)+ (N+1)v_{an}^{off}(p-q)] B_\rho(q) \bigg\rbrace\; , \\
  \partial_k \Gamma^{\tilde{\phi}\tilde{\phi}}(p) =& - \frac{i}{2} \int_q \Big \lbrace 2 [v_{an}^{diag}(p-q)+ (N+1)v_{an}^{off}(p-q)] n_\text{eff}(q_0) B_\rho(q) \\
  &\hspace*{-2cm}+ 2 [(v_{qu,R}^{diag}(p-q)-v_{qu,A}^{diag}(p-q))+ (N+1)(v_{qu,R}^{off}(p-q)-v_{qu,A}^{off}(p-q))] B_\rho(q) \Big \rbrace \; ,
\end{align*}
where we have used the assumption from Eq.~(\ref{gen-fluc-dis}). We have further used that there is a fluctuation-dissipation relation for the $B$'s
\begin{equation}
    B_F(p) = n_{eff}(p) B_\rho(p)\; , \qquad \text{with} \qquad B_\rho(p)=B_R(p)-B_A(p)\; . \label{f-d-B}
\end{equation}
By compactifying
\begin{align*}
  v_{cl,R/A}(p) &=  v_{cl,R/A}^{diag}(p) + (N+1)v_{cl,R/A}^{off}(p) \; , \\
  v_{an}(p) &=  v_{an}^{diag}(p) + (N+1)v_{an}^{off}(p) \; ,
\end{align*}
the flow equations now read
\begin{align*}
  \partial_k(\Gamma^{\tilde{\phi}\phi}(p)-\Gamma^{\phi\tilde{\phi}}(p)) =& - \frac{i}{2} \int_q 
  2 [(v_{cl,R,k}(p-q)-v_{cl,A,k}(p-q)) n_\text{eff}(q_0) + 2 v_{an}(p-q)] B_\rho(q) \; , \\
  \partial_k \Gamma^{\tilde{\phi}\tilde{\phi}}(p) =& - \frac{i}{2} \int_q  [2 v_{an}(p-q) n_\text{eff}(q_0)
  + 2 (v_{qu,R}(p-q)-v_{qu,A}(p-q))] B_\rho(q) \; .
\end{align*}
Now, let us check if the fluctuation-dissipation relation does hold
\begin{align*}
  \partial_k \Gamma^{\tilde{\phi}\tilde{\phi}}(p) =& n_\text{eff}(p_0) \partial_k (\Gamma^{\tilde{\phi}\phi}(p)- \Gamma^{\phi\tilde{\phi}}(p)) \; ,  \\
  \int_q  [ v_{an}(p-q) n_\text{eff}(q_0) + (v_{qu,R}(p-q)&-v_{qu,A}(p-q))] B_\rho(q) = \\
 \int_q n_\text{eff}(p_0) [(v_{cl,R,k}(p-q)&-v_{cl,A,k}(p-q)) n_\text{eff}(q_0) +  v_{an}(p-q)] B_\rho(q)\;, \\
 \Rightarrow v_{an}(p-q) n_\text{eff}(q_0) + (v_{qu,R}(p-q)&-v_{qu,A}(p-q)) \\
 - n_\text{eff}(p_0) [(v_{cl,R,k}(p-q)&-v_{cl,A,k}(p-q)) n_\text{eff}(q_0) +  v_{an}(p-q)] = 0
\end{align*}
In case of a quantum theory, the relation between classical and quantum vertices is given by 
\begin{equation}
  v_{cl,R/A}(p) = 4 v_{qu,R/A}(p) \; ,
\end{equation}
for all $k$. So we find
\begin{align}
  (v_{cl,R,k}(p-q)-v_{cl,A,k}(p-q))& \left( \frac{1}{4} - n_\text{eff}(q_0) n_\text{eff}(p_0) \right) + v_{an}(p-q) (n_\text{eff}(q_0) - n_\text{eff}(p_0)) = 0 \; , \notag \\
  v_{an}(p-q) &= (v_{cl,R,k}(p-q)-v_{cl,A,k}(p-q)) \frac{n_\text{eff}(q_0) n_\text{eff}(p_0) - \frac{1}{4}}{n_\text{eff}(q_0) - n_\text{eff}(p_0)} \; .
  \label{eq:proof-fluc-dis}
\end{align}
By using the addition theorem for the coth or respectively the effective occupation numbers we arrive at 
\begin{equation}
  v_{an}(p-q) = n_\text{eff}(p_0-q_0) (v_{cl,R,k}(p-q)-v_{cl,A,k}(p-q)) \; . \label{eq:fluc-dis-vertexf}
\end{equation}
In the case of a classical theory, we have $v_{qu,R/A}(p)=0$, i.e. we can just drop the $-1/4$ in Eq.~(\ref{eq:proof-fluc-dis}). By plugging in the Rayleigh-Jeans distribution we again find Eq.~(\ref{eq:fluc-dis-vertexf}).
Going through the calculation in reverse order proofs that the existence of a generalized fluctuation-dissipation relation for the vertex functions beyond one-loop leads to a fluctuation-dissipation relation for the two-point functions. The only assumption that goes into the proof in Eq.~(\ref{gen-fluc-dis}) is a generalization of the one-loop result and if violated will lead to a violation of causality in the two-point functions.\\ 
Another subtlety is the use of Eq.~(\ref{f-d-B}). A quick calculation shows that Eq.~(\ref{f-d-B}) holds if the propagators fulfill the fluctuation-dissipation relation, but obviously this is only true at all $k$ if the differential equation Eq.~(\ref{fluc-dis1}) holds at all $k$. Which is what we wanted to show. However, at the UV-cutoff $k=\Lambda$ the propagators and therefore the $B$'s fulfill the fluctuation-dissipation relation and so does the differential equation Eq.~(\ref{fluc-dis1}). But that means, that the $B$'s at $k-dk$ obey Eq.~(\ref{f-d-B}) and eventually the $B$'s fulfill the fluctuation-dissipation relation for all $k$.

\section{Vertex flow equation for arbitrary N}
In the flow equation for the vertex function for arbitrary $N$ in Eq.~(\ref{4pt-flow}) there are in total nine contributing diagrams on the right-hand side  
\begin{align*}
&\partial_k \Big[ \left(v_{cl,R,k}^{diag}(p) + 2 v_{cl,R,k}^{off}(0)\right)\delta_{ab} \delta_{\bar{a}\bar{b}} +  (v_{cl,R,k}^{off}(p) + v_{cl,R,k}^{diag}(0) + v_{cl,R,k}^{off}(0)) \left(\delta_{a\bar{a}} \delta_{b\bar{b}} + \delta_{a\bar{b}} \delta_{\bar{a}b}\right) \Big] = \notag \\ &-i\, \Bigg \lbrace \quad
\begin{picture}(110,40) (170,-37)
      \SetWidth{1.0}
      \SetColor{Black}
      \Arc[color=Blue, arrow](210,-33)(15,0,90)     \Arc[color=Blue, arrow](210,-33)(15,90,180)        
      \Arc[color=Blue](210,-33)(15,180,270)
      \Arc[color=Red, arrow, clockwise, arrowpos=1](210,-33)(15,0,270)
      \FBox(207,-15)(213,-21)
      \Line[color=Blue](185,-25)(194,-33)
      \Line[color=Red](185,-41)(194,-33)
      \Line[color=Blue](235,-25)(226,-33)
      \Line[color=Blue](235,-41)(226,-33)
      \Vertex(194,-33){3.4}
      \Vertex(226,-33){3.4}
      \Text(165,-50)[lb]{\scriptsize $+p/2,a$}
      \Text(165,-23)[lb]{\scriptsize $+p/2,b$}
      \Text(227,-23)[lb]{\scriptsize $-p/2,\bar b$}
      \Text(227,-50)[lb]{\scriptsize $-p/2,\bar a$}
\end{picture}\hspace*{-0.7cm}+
\begin{picture}(110,40) (170,-37)
      \SetWidth{1.0}
      \SetColor{Black}
      \Arc[color=Red, arrow](210,-33)(15,0,90)       \Arc[color=Blue, arrow](210,-33)(15,90,180)       \Arc[color=Blue, arrow, clockwise](210,-33)(15,0,180)
      \FBox(207,-15)(213,-21)
      \Line[color=Blue](185,-25)(194,-33)
      \Line[color=Red](185,-41)(194,-33)
      \Line[color=Blue](235,-25)(226,-33)
      \Line[color=Blue](235,-41)(226,-33)
      \Vertex(194,-33){3.4}
      \Vertex(226,-33){3.4}
      \Text(165,-50)[lb]{\scriptsize $+p/2,a$}
      \Text(165,-23)[lb]{\scriptsize $-p/2,\bar a$}
      \Text(227,-23)[lb]{\scriptsize $+p/2,b$}
      \Text(227,-50)[lb]{\scriptsize $-p/2,\bar b$}
\end{picture}\hspace*{-0.7cm}+
\begin{picture}(110,40) (170,-37)
      \SetWidth{1.0}
      \SetColor{Black}
      \Arc[color=Red, arrow](210,-33)(15,0,90)     \Arc[color=Blue, arrow](210,-33)(15,90,180)        
      \Arc[color=Blue](210,-33)(15,180,270)
      \Arc[color=Red, arrow, clockwise, arrowpos=1](210,-33)(15,0,270)
      \FBox(207,-15)(213,-21)
      \Line[color=Blue](185,-25)(194,-33)
      \Line[color=Red](185,-41)(194,-33)
      \Line[color=Blue](235,-25)(226,-33)
      \Line[color=Blue](235,-41)(226,-33)
      \Vertex(194,-33){3.4}
      \Vertex(226,-33){3.4}
      \Text(165,-50)[lb]{\scriptsize $+p/2,a$}
      \Text(165,-23)[lb]{\scriptsize $-p/2,\bar b$}
      \Text(227,-23)[lb]{\scriptsize $-p/2,\bar a$}
      \Text(227,-50)[lb]{\scriptsize $+p/2, b$}
\end{picture} \\
&\hspace*{1cm}+\begin{picture}(110,40) (170,-37)
      \SetWidth{1.0}
      \SetColor{Black}
      \Arc[color=Blue, arrow](210,-33)(15,0,90)     \Arc[color=Blue, arrow](210,-33)(15,90,180)        
      \Arc[color=Blue](210,-33)(15,180,270)
      \Arc[color=Red, arrow, clockwise, arrowpos=1](210,-33)(15,0,270)
      \FBox(207,-15)(213,-21)
      \Line[color=Blue](185,-25)(194,-33)
      \Line[color=Red](185,-41)(194,-33)
      \Line[color=Blue](235,-25)(226,-33)
      \Line[color=Blue](235,-41)(226,-33)
      \Vertex(194,-33){3.4}
      \Vertex(226,-33){3.4}
      \Text(165,-50)[lb]{\scriptsize $+p/2,a$}
      \Text(165,-23)[lb]{\scriptsize $-p/2,\bar a$}
      \Text(227,-23)[lb]{\scriptsize $+p/2, b$}
      \Text(227,-50)[lb]{\scriptsize $-p/2,\bar b$}
\end{picture}\hspace*{-0.7cm}+
\begin{picture}(110,40) (170,-37)
      \SetWidth{1.0}
      \SetColor{Black}
      \Arc[color=Red, arrow](210,-33)(15,0,90)       \Arc[color=Blue, arrow](210,-33)(15,90,180)       \Arc[color=Blue, arrow, clockwise](210,-33)(15,0,180)
      \FBox(207,-15)(213,-21)
      \Line[color=Blue](185,-25)(194,-33)
      \Line[color=Red](185,-41)(194,-33)
      \Line[color=Blue](235,-25)(226,-33)
      \Line[color=Blue](235,-41)(226,-33)
      \Vertex(194,-33){3.4}
      \Vertex(226,-33){3.4}
      \Text(165,-50)[lb]{\scriptsize $+p/2,a$}
      \Text(165,-23)[lb]{\scriptsize $-p/2,\bar a$}
      \Text(227,-23)[lb]{\scriptsize $+p/2, b$}
      \Text(227,-50)[lb]{\scriptsize $-p/2,\bar b$}
\end{picture}\hspace*{-0.7cm}+
\begin{picture}(110,40) (170,-37)
      \SetWidth{1.0}
      \SetColor{Black}
      \Arc[color=Red, arrow](210,-33)(15,0,90)     \Arc[color=Blue, arrow](210,-33)(15,90,180)        
      \Arc[color=Blue](210,-33)(15,180,270)
      \Arc[color=Red, arrow, clockwise, arrowpos=1](210,-33)(15,0,270)
      \FBox(207,-15)(213,-21)
      \Line[color=Blue](185,-25)(194,-33)
      \Line[color=Red](185,-41)(194,-33)
      \Line[color=Blue](235,-25)(226,-33)
      \Line[color=Blue](235,-41)(226,-33)
      \Vertex(194,-33){3.4}
      \Vertex(226,-33){3.4}
      \Text(165,-50)[lb]{\scriptsize $+p/2,a$}
      \Text(165,-23)[lb]{\scriptsize $-p/2,\bar a$}
      \Text(227,-23)[lb]{\scriptsize $+p/2, b$}
      \Text(227,-50)[lb]{\scriptsize $-p/2,\bar b$}
\end{picture}\\
&\hspace*{1cm}+\begin{picture}(110,40) (170,-37)
      \SetWidth{1.0}
      \SetColor{Black}
      \Arc[color=Blue, arrow](210,-33)(15,0,90)     \Arc[color=Blue, arrow](210,-33)(15,90,180)        
      \Arc[color=Blue](210,-33)(15,180,270)
      \Arc[color=Red, arrow, clockwise, arrowpos=1](210,-33)(15,0,270)
      \FBox(207,-15)(213,-21)
      \Line[color=Blue](185,-25)(194,-33)
      \Line[color=Red](185,-41)(194,-33)
      \Line[color=Blue](235,-25)(226,-33)
      \Line[color=Blue](235,-41)(226,-33)
      \Vertex(194,-33){3.4}
      \Vertex(226,-33){3.4}
      \Text(165,-50)[lb]{\scriptsize $+p/2,a$}
      \Text(165,-23)[lb]{\scriptsize $-p/2,\bar b$}
      \Text(227,-23)[lb]{\scriptsize $-p/2,\bar a$}
      \Text(227,-50)[lb]{\scriptsize $+p/2, b$}
\end{picture}\hspace*{-0.7cm}+
\begin{picture}(110,40) (170,-37)
      \SetWidth{1.0}
      \SetColor{Black}
      \Arc[color=Red, arrow](210,-33)(15,0,90)       \Arc[color=Blue, arrow](210,-33)(15,90,180)       \Arc[color=Blue, arrow, clockwise](210,-33)(15,0,180)
      \FBox(207,-15)(213,-21)
      \Line[color=Blue](185,-25)(194,-33)
      \Line[color=Red](185,-41)(194,-33)
      \Line[color=Blue](235,-25)(226,-33)
      \Line[color=Blue](235,-41)(226,-33)
      \Vertex(194,-33){3.4}
      \Vertex(226,-33){3.4}
      \Text(165,-50)[lb]{\scriptsize $+p/2,a$}
      \Text(165,-23)[lb]{\scriptsize $-p/2,\bar b$}
      \Text(227,-23)[lb]{\scriptsize $-p/2,\bar a$}
      \Text(227,-50)[lb]{\scriptsize $+p/2, b$}
\end{picture}\hspace*{-0.7cm}+
\begin{picture}(110,40) (170,-37)
      \SetWidth{1.0}
      \SetColor{Black}
      \Arc[color=Red, arrow](210,-33)(15,0,90)     \Arc[color=Blue, arrow](210,-33)(15,90,180)        
      \Arc[color=Blue](210,-33)(15,180,270)
      \Arc[color=Red, arrow, clockwise, arrowpos=1](210,-33)(15,0,270)
      \FBox(207,-15)(213,-21)
      \Line[color=Blue](185,-25)(194,-33)
      \Line[color=Red](185,-41)(194,-33)
      \Line[color=Blue](235,-25)(226,-33)
      \Line[color=Blue](235,-41)(226,-33)
      \Vertex(194,-33){3.4}
      \Vertex(226,-33){3.4}
      \Text(165,-50)[lb]{\scriptsize $+p/2,a$}
      \Text(165,-23)[lb]{\scriptsize $-p/2,\bar b$}
      \Text(227,-23)[lb]{\scriptsize $-p/2,\bar a$}
      \Text(227,-50)[lb]{\scriptsize $+p/2, b$}
\end{picture} \Bigg \rbrace \; ,
\end{align*}
where a line with a box stands for the according $B$'s introduced in Eq.~(\ref{eq:def-B}) and the arrows indicate the direction of momentum flow with all external momenta taken as incoming. Denoting the different $O(N)$ index structures as $\hat s = \delta_{ab} \delta_{\bar{a} \bar{b}}$ , $\hat t = \delta_{a \bar{a}} \delta_{b \bar{b}}$ and $\hat u = \delta_{a \bar{b}} \delta_{\bar{a} b}$, the contributions of the individual diagrams are then given by
\begin{align*}
&\begin{picture}(90,40) (170,-37)
      \SetWidth{1.0}
      \SetColor{Black}
      \Arc[color=Blue, arrow](210,-33)(15,0,90)     \Arc[color=Blue, arrow](210,-33)(15,90,180)        
      \Arc[color=Blue](210,-33)(15,180,270)
      \Arc[color=Red, arrow, clockwise, arrowpos=1](210,-33)(15,0,270)
      \FBox(207,-15)(213,-21)
      \Line[color=Blue](185,-25)(194,-33)
      \Line[color=Red](185,-41)(194,-33)
      \Line[color=Blue](235,-25)(226,-33)
      \Line[color=Blue](235,-41)(226,-33)
      \Vertex(194,-33){3.4}
      \Vertex(226,-33){3.4}
      \Text(165,-50)[lb]{\scriptsize $+p/2,a$}
      \Text(165,-23)[lb]{\scriptsize $+p/2,b$}
      \Text(227,-23)[lb]{\scriptsize $-p/2,\bar b$}
      \Text(227,-50)[lb]{\scriptsize $-p/2,\bar a$}
\end{picture}
  = \int_l B_F(p/2-l) G_R(p/2+l) \Big \lbrace \\ \\
  &\hat s \textcolor{red}{\Big \lbrace} [v_{cl,R}^\text{diag}(p) + v_{cl,R}^\text{off}(l)  + v_{cl,R}^\text{off}(-l)  ]
  \textcolor{green}{\Big [} N [v_{cl,R}^\text{diag}(p) + v_{cl,R}^\text{off}(l) + v_{cl,R}^\text{off}(l) ] \\
  &+  [v_{cl,R}^\text{off}(p)  + v_{cl,R}^\text{diag}(l) + v_{cl,R}^\text{off}(l) ]
  + [v_{cl,R}^\text{off}(p) + v_{cl,R}^\text{off}(l) + v_{cl,R}^\text{diag}(l) ] \textcolor{green}{\Big ]} \\
  &+ \textcolor{green}{\Big [} [v_{cl,R}^\text{off}(p) + v_{cl,R}^\text{diag}(l) + v_{cl,R}^\text{off}(-l)] + [v_{cl,R}^\text{off}(p) + v_{cl,R}^\text{off}(l) + v_{cl,R}^\text{diag}(-l)] \textcolor{green}{\Big ]}\\
  &\times [v_{cl,R}^\text{diag}(p) + v_{cl,R}^\text{off}(l)+ v_{cl,R}^\text{off}(l)] \textcolor{red}{\Big \rbrace} \\
  + &\hat t \textcolor{red}{\Big \lbrace} [v_{cl,R}^\text{off}(p)  + v_{cl,R}^\text{diag}(l) + v_{cl,R}^\text{off}(-l) ] [v_{cl,R}^\text{off}(p)  + v_{cl,R}^\text{diag}(l) + v_{cl,R}^\text{off}(l) ]\\
  &+ [v_{cl,R}^\text{off}(p)  + v_{cl,R}^\text{off}(l) + v_{cl,R}^\text{diag}(-l) ] [v_{cl,R}^\text{off}(p)  + v_{cl,R}^\text{off}(l) + v_{cl,R}^\text{diag}(l) ] \textcolor{red}{\Big \rbrace}\\
  +& \hat u \textcolor{red}{\Big \lbrace} [v_{cl,R}^\text{off}(p)  + v_{cl,R}^\text{diag}(l) + v_{cl,R}^\text{off}(-l) ] [v_{cl,R}^\text{off}(p)  + v_{cl,R}^\text{off}(l) + v_{cl,R}^\text{diag}(l) ]\\
  &+ [v_{cl,R}^\text{off}(p)  + v_{cl,R}^\text{off}(l) + v_{cl,R}^\text{diag}(-l)] [v_{cl,R}^\text{off}(p)  + v_{cl,R}^\text{diag}(l) + v_{cl,R}^\text{off}(l) ] \textcolor{red}{\Big \rbrace} \Big \rbrace \; ,
\end{align*}

\begin{align*}
&\begin{picture}(90,10) (170,-37)
      \SetWidth{1.0}
      \SetColor{Black}
      \Arc[color=Blue, arrow](210,-33)(15,0,90)     \Arc[color=Blue, arrow](210,-33)(15,90,180)        
      \Arc[color=Blue](210,-33)(15,180,270)
      \Arc[color=Red, arrow, clockwise, arrowpos=1](210,-33)(15,0,270)
      \FBox(207,-15)(213,-21)
      \Line[color=Blue](185,-25)(194,-33)
      \Line[color=Red](185,-41)(194,-33)
      \Line[color=Blue](235,-25)(226,-33)
      \Line[color=Blue](235,-41)(226,-33)
      \Vertex(194,-33){3.4}
      \Vertex(226,-33){3.4}
      \Text(165,-50)[lb]{\scriptsize $+p/2,a$}
      \Text(165,-23)[lb]{\scriptsize $-p/2,\bar a$}
      \Text(227,-23)[lb]{\scriptsize $+p/2, b$}
      \Text(227,-50)[lb]{\scriptsize $-p/2,\bar b$}
\end{picture}
  =\int_l B_F(-l) G_R(l) \Big \lbrace  \\ \quad \\
  &\hat s \textcolor{red}{\Big \lbrace} [v_{cl,R}^\text{off}(0) + v_{cl,R}^\text{off}(p/2-l)  + v_{cl,R}^\text{diag}(p/2+l)] [v_{cl,R}^\text{off}(0) + v_{cl,R}^\text{diag}(l-p/2) + v_{cl,R}^\text{off}(l+p/2) ]\\
  &+ [v_{cl,R}^\text{off}(0) + v_{cl,R}^\text{diag}(p/2-l) + v_{cl,R}^\text{off}(p/2+l)] [v_{cl,R}^\text{off}(0) + v_{cl,R}^\text{off}(l-p/2) + v_{cl,R}^\text{diag}(l+p/2) ] \textcolor{red}{\Big \rbrace} \\
  +&\hat t \textcolor{red}{\Big \lbrace} [v_{cl,R}^\text{diag}(0) + v_{cl,R}^\text{off}(p/2-l) + v_{cl,R}^\text{off}(p/2+l)] \textcolor{green}{\Big [} N [v_{cl,R}^\text{diag}(0) + v_{cl,R}^\text{off}(l-p/2) + v_{cl,R}^\text{off}(l+p/2) ]\\
  &+ [v_{cl,R}^\text{off}(0) + v_{cl,R}^\text{diag}(l-p/2) + v_{cl,R}^\text{off}(l+p/2) ] + [v_{cl,R}^\text{off}(0) + v_{cl,R}^\text{off}(l-p/2) + v_{cl,R}^\text{diag}(l+p/2) ] \textcolor{green}{\Big ]}\\
  &+\textcolor{green}{\Big [} [v_{cl,R}^\text{off}(0) + v_{cl,R}^\text{diag}(p/2-l) + v_{cl,R}^\text{off}(p/2+l)] + [v_{cl,R}^\text{off}(0) + v_{cl,R}^\text{off}(p/2-l) + v_{cl,R}^\text{diag}(p/2+l)] \textcolor{green}{\Big ]}\\
  &\times [v_{cl,R}^\text{diag}(0) + v_{cl,R}^\text{off}(l-p/2) + v_{cl,R}^\text{off}(l+p/2) ] \textcolor{red}{\Big \rbrace} \\
  +&\hat u \textcolor{red}{\Big \lbrace} [v_{cl,R}^\text{off}(0) + v_{cl,R}^\text{off}(p/2-l)  + v_{cl,R}^\text{diag}(p/2+l)] [v_{cl,R}^\text{off}(0) + v_{cl,R}^\text{off}(l-p/2) + v_{cl,R}^\text{diag}(l+p/2) ]\\
  &+ [v_{cl,R}^\text{off}(0) + v_{cl,R}^\text{diag}(p/2-l) + v_{cl,R}^\text{off}(p/2+l)] [v_{cl,R}^\text{off}(0) + v_{cl,R}^\text{diag}(l-p/2) + v_{cl,R}^\text{off}(l+p/2) ] \textcolor{red}{\Big \rbrace} \Big \rbrace \; ,
\end{align*}

\begin{align*}
&\begin{picture}(110,40) (170,-37)
      \SetWidth{1.0}
      \SetColor{Black}
      \Arc[color=Blue, arrow](210,-33)(15,0,90)     \Arc[color=Blue, arrow](210,-33)(15,90,180)        
      \Arc[color=Blue](210,-33)(15,180,270)
      \Arc[color=Red, arrow, clockwise, arrowpos=1](210,-33)(15,0,270)
      \FBox(207,-15)(213,-21)
      \Line[color=Blue](185,-25)(194,-33)
      \Line[color=Red](185,-41)(194,-33)
      \Line[color=Blue](235,-25)(226,-33)
      \Line[color=Blue](235,-41)(226,-33)
      \Vertex(194,-33){3.4}
      \Vertex(226,-33){3.4}
      \Text(165,-50)[lb]{\scriptsize $+p/2,a$}
      \Text(165,-23)[lb]{\scriptsize $-p/2,\bar b$}
      \Text(227,-23)[lb]{\scriptsize $-p/2,\bar a$}
      \Text(227,-50)[lb]{\scriptsize $+p/2, b$}
\end{picture}
  =\int_l B_F(-l) G_R(l) \Big \lbrace \\  \\
  &\hat s \textcolor{red}{\Big \lbrace} [v_{cl,R}^\text{off}(0) + v_{cl,R}^\text{off}(p/2-l)  + v_{cl,R}^\text{diag}(p/2+l)] [v_{cl,R}^\text{off}(0) + v_{cl,R}^\text{off}(p/2+l)  + v_{cl,R}^\text{diag}(p/2-l)] \\
  &+ [v_{cl,R}^\text{off}(0) + v_{cl,R}^\text{diag}(p/2-l) + v_{cl,R}^\text{off}(p/2+l)] [v_{cl,R}^\text{off}(0) + v_{cl,R}^\text{diag}(p/2+l) + v_{cl,R}^\text{off}(p/2-l)]  \textcolor{red}{\Big \rbrace} \\
  +&\hat t \textcolor{red}{\Big \lbrace} [v_{cl,R}^\text{off}(0) + v_{cl,R}^\text{diag}(p/2-l) + v_{cl,R}^\text{off}(p/2+l)] [v_{cl,R}^\text{off}(0) + v_{cl,R}^\text{off}(p/2+l)  + v_{cl,R}^\text{diag}(p/2-l)] \\
  &+ [v_{cl,R}^\text{off}(0) + v_{cl,R}^\text{off}(p/2-l)  + v_{cl,R}^\text{diag}(p/2+l)] [v_{cl,R}^\text{off}(0) + v_{cl,R}^\text{diag}(p/2+l) + v_{cl,R}^\text{off}(p/2-l)] \textcolor{red}{\Big \rbrace} \\
  +&\hat u \textcolor{red}{\Big \lbrace} [v_{cl,R}^\text{diag}(0) + v_{cl,R}^\text{off}(p/2-l)  + v_{cl,R}^\text{off}(p/2+l) ] \textcolor{green}{\Big [} N [v_{cl,R}^\text{diag}(0) + v_{cl,R}^\text{off}(p/2+l)  + v_{cl,R}^\text{off}(p/2-l)] \\
  &+ [v_{cl,R}^\text{off}(0) + v_{cl,R}^\text{diag}(p/2+l) + v_{cl,R}^\text{off}(p/2-l)] + [v_{cl,R}^\text{off}(0) + v_{cl,R}^\text{off}(p/2+l)  + v_{cl,R}^\text{diag}(p/2-l)] \textcolor{green}{\Big ]}\\
  &+ \textcolor{green}{\Big [} [v_{cl,R}^\text{off}(0) + v_{cl,R}^\text{diag}(p/2-l) + v_{cl,R}^\text{off}(p/2+l)] + [v_{cl,R}^\text{off}(0) + v_{cl,R}^\text{off}(p/2-l)  + v_{cl,R}^\text{diag}(p/2+l)] \textcolor{green}{\Big ]}\\
  &\times  [v_{cl,R}^\text{off}(0) + v_{cl,R}^\text{off}(p/2+l)  + v_{cl,R}^\text{diag}(p/2-l)] \textcolor{red}{\Big \rbrace} \Big \rbrace \; ,
\end{align*}

\begin{align*}
&\begin{picture}(110,40) (170,-37)
      \SetWidth{1.0}
      \SetColor{Black}
      \Arc[color=Red, arrow](210,-33)(15,0,90)       \Arc[color=Blue, arrow](210,-33)(15,90,180)       \Arc[color=Blue, arrow, clockwise](210,-33)(15,0,180)
      \FBox(207,-15)(213,-21)
      \Line[color=Blue](185,-25)(194,-33)
      \Line[color=Red](185,-41)(194,-33)
      \Line[color=Blue](235,-25)(226,-33)
      \Line[color=Blue](235,-41)(226,-33)
      \Vertex(194,-33){3.4}
      \Vertex(226,-33){3.4}
      \Text(165,-50)[lb]{\scriptsize $+p/2,a$}
      \Text(165,-23)[lb]{\scriptsize $-p/2,\bar a$}
      \Text(227,-23)[lb]{\scriptsize $+p/2,b$}
      \Text(227,-50)[lb]{\scriptsize $-p/2,\bar b$}
\end{picture}
  =\int_l B_R(p/2-l) iF(p/2+l) \Big \lbrace \\ \\
  &\hat s \textcolor{red}{\Big \lbrace} [v_{cl,R}^\text{diag}(p) + v_{cl,R}^\text{off}(l)  + v_{cl,R}^\text{off}(-l)  ]
  \textcolor{green}{\Big [} N [v_{cl,R}^\text{diag}(p) + v_{cl,R}^\text{off}(-l)  + v_{cl,R}^\text{off}(-l)  ] \\
  &+  [v_{cl,R}^\text{off}(p)  + v_{cl,R}^\text{diag}(-l) + v_{cl,R}^\text{off}(-l)  ]
  + [v_{cl,R}^\text{off}(p) + v_{cl,R}^\text{off}(-l)  + v_{cl,R}^\text{diag}(-l) ] \textcolor{green}{\Big ]} \\
  &+ \textcolor{green}{\Big [} [v_{cl,R}^\text{off}(p)  + v_{cl,R}^\text{diag}(l) + v_{cl,R}^\text{off}(-l) ] + [v_{cl,R}^\text{off}(p)  + v_{cl,R}^\text{off}(l)  + v_{cl,R}^\text{diag}(-l)] \textcolor{green}{\Big ]}\\
  &\times [ [v_{cl,R}^\text{diag}(p) + v_{cl,R}^\text{off}(-l)+ v_{cl,R}^\text{off}(-l) ] \textcolor{red}{\Big \rbrace} \\
  + &\hat t \textcolor{red}{\Big \lbrace} [v_{cl,R}^\text{off}(p)  + v_{cl,R}^\text{diag}(l) + v_{cl,R}^\text{off}(-l)  ] [v_{cl,R}^\text{off}(p)  + v_{cl,R}^\text{diag}(-l) + v_{cl,R}^\text{off}(-l) ]\\
  &+ [v_{cl,R}^\text{off}(p)  + v_{cl,R}^\text{off}(l)  + v_{cl,R}^\text{diag}(-l) ] [v_{cl,R}^\text{off}(p)  + v_{cl,R}^\text{off}(-l)  + v_{cl,R}^\text{diag}(-l) ] \textcolor{red}{\Big \rbrace}\\
  +& \hat u \textcolor{red}{\Big \lbrace} [v_{cl,R}^\text{off}(p)  + v_{cl,R}^\text{diag}(l) + v_{cl,R}^\text{off}(-l)  ] [v_{cl,R}^\text{off}(p)  + v_{cl,R}^\text{off}(-l)  + v_{cl,R}^\text{diag}(-l) ]\\
  &+ [v_{cl,R}^\text{off}(p)  + v_{cl,R}^\text{off}(l)  + v_{cl,R}^\text{diag}(-l) ] [v_{cl,R}^\text{off}(p)  + v_{cl,R}^\text{diag}(-l) + v_{cl,R}^\text{off}(-l)  ] \textcolor{red}{\Big \rbrace} \Big \rbrace \; ,
\end{align*}

\begin{align*}
&\begin{picture}(110,40) (170,-37)
      \SetWidth{1.0}
      \SetColor{Black}
      \Arc[color=Red, arrow](210,-33)(15,0,90)       \Arc[color=Blue, arrow](210,-33)(15,90,180)       \Arc[color=Blue, arrow, clockwise](210,-33)(15,0,180)
      \FBox(207,-15)(213,-21)
      \Line[color=Blue](185,-25)(194,-33)
      \Line[color=Red](185,-41)(194,-33)
      \Line[color=Blue](235,-25)(226,-33)
      \Line[color=Blue](235,-41)(226,-33)
      \Vertex(194,-33){3.4}
      \Vertex(226,-33){3.4}
      \Text(165,-50)[lb]{\scriptsize $+p/2,a$}
      \Text(165,-23)[lb]{\scriptsize $-p/2,\bar a$}
      \Text(227,-23)[lb]{\scriptsize $+p/2, b$}
      \Text(227,-50)[lb]{\scriptsize $-p/2,\bar b$}
\end{picture}
  =\int_l B_R(-l) iF(l) \Big \lbrace \\ \\
  &\hat s \textcolor{red}{\Big \lbrace} [v_{cl,R}^\text{off}(0) + v_{cl,R}^\text{off}(p/2-l)  + v_{cl,R}^\text{diag}(p/2+l)] [v_{cl,R}^\text{off}(0) + v_{cl,R}^\text{diag}(p/2-l) + v_{cl,R}^\text{off}(-p/2-l) ]\\
  &+ [v_{cl,R}^\text{off}(0) + v_{cl,R}^\text{diag}(p/2-l) + v_{cl,R}^\text{off}(p/2+l)] [v_{cl,R}^\text{off}(0) + v_{cl,R}^\text{off}(p/2-l) + v_{cl,R}^\text{diag}(-p/2-l) ] \textcolor{red}{\Big \rbrace} \\
  +&\hat t \textcolor{red}{\Big \lbrace} [v_{cl,R}^\text{diag}(0) + v_{cl,R}^\text{off}(p/2-l)  + v_{cl,R}^\text{off}(p/2+l) ] \textcolor{green}{\Big [} N [v_{cl,R}^\text{diag}(0) + v_{cl,R}^\text{off}(p/2-l) + v_{cl,R}^\text{off}(-p/2-l) ]\\
  &+ [v_{cl,R}^\text{off}(0) + v_{cl,R}^\text{diag}(p/2-l) + v_{cl,R}^\text{off}(-p/2-l) ] + [v_{cl,R}^\text{off}(0) + v_{cl,R}^\text{off}(p/2-l) + v_{cl,R}^\text{diag}(-p/2-l) ] \textcolor{green}{\Big ]}\\
  &+\textcolor{green}{\Big [} [v_{cl,R}^\text{off}(0) + v_{cl,R}^\text{diag}(p/2-l) + v_{cl,R}^\text{off}(p/2+l)] + [v_{cl,R}^\text{off}(0) + v_{cl,R}^\text{off}(p/2-l)  + v_{cl,R}^\text{diag}(p/2+l)] \textcolor{green}{\Big ]}\\
  &\times [v_{cl,R}^\text{diag}(0) + v_{cl,R}^\text{off}(p/2-l) + v_{cl,R}^\text{off}(-p/2-l) ] \textcolor{red}{\Big \rbrace} \\
  +&\hat u \textcolor{red}{\Big \lbrace} [v_{cl,R}^\text{off}(0) + v_{cl,R}^\text{off}(p/2-l)  + v_{cl,R}^\text{diag}(p/2+l)] [v_{cl,R}^\text{off}(0) + v_{cl,R}^\text{off}(p/2-l) + v_{cl,R}^\text{diag}(-p/2-l) ]\\
  &+ [v_{cl,R}^\text{off}(0) + v_{cl,R}^\text{diag}(p/2-l) + v_{cl,R}^\text{off}(p/2+l)] [v_{cl,R}^\text{off}(0) + v_{cl,R}^\text{diag}(p/2-l) + v_{cl,R}^\text{off}(-p/2-l) ] \textcolor{red}{\Big \rbrace} \Big \rbrace \; ,
\end{align*}

\begin{align*}
&\begin{picture}(110,40) (170,-37)
      \SetWidth{1.0}
      \SetColor{Black}
      \Arc[color=Red, arrow](210,-33)(15,0,90)       \Arc[color=Blue, arrow](210,-33)(15,90,180)       \Arc[color=Blue, arrow, clockwise](210,-33)(15,0,180)
      \FBox(207,-15)(213,-21)
      \Line[color=Blue](185,-25)(194,-33)
      \Line[color=Red](185,-41)(194,-33)
      \Line[color=Blue](235,-25)(226,-33)
      \Line[color=Blue](235,-41)(226,-33)
      \Vertex(194,-33){3.4}
      \Vertex(226,-33){3.4}
      \Text(165,-50)[lb]{\scriptsize $+p/2,a$}
      \Text(165,-23)[lb]{\scriptsize $-p/2,\bar b$}
      \Text(227,-23)[lb]{\scriptsize $-p/2,\bar a$}
      \Text(227,-50)[lb]{\scriptsize $+p/2, b$}
\end{picture}
  =\int_l B_R(-l) iF(l) \Big \lbrace \\ \\
  &\hat s \textcolor{red}{\Big \lbrace} [v_{cl,R}^\text{off}(0) + v_{cl,R}^\text{off}(p/2-l)  + v_{cl,R}^\text{diag}(p/2+l)] [v_{cl,R}^\text{off}(0) + v_{cl,R}^\text{off}(-p/2-l)  + v_{cl,R}^\text{diag}(p/2-l)] \\
  &+ [v_{cl,R}^\text{off}(0) + v_{cl,R}^\text{diag}(p/2-l) + v_{cl,R}^\text{off}(p/2+l)] [v_{cl,R}^\text{off}(0) + v_{cl,R}^\text{diag}(-p/2-l) + v_{cl,R}^\text{off}(p/2-l)]  \textcolor{red}{\Big \rbrace} \\
  +&\hat t \textcolor{red}{\Big \lbrace} [v_{cl,R}^\text{off}(0) + v_{cl,R}^\text{diag}(p/2-l) + v_{cl,R}^\text{off}(p/2+l)] [v_{cl,R}^\text{off}(0) + v_{cl,R}^\text{off}(-p/2-l)  + v_{cl,R}^\text{diag}(p/2-l)] \\
  &+ [v_{cl,R}^\text{off}(0) + v_{cl,R}^\text{off}(p/2-l)  + v_{cl,R}^\text{diag}(p/2+l)] [v_{cl,R}^\text{off}(0) + v_{cl,R}^\text{diag}(-p/2-l) + v_{cl,R}^\text{off}(p/2-l)] \textcolor{red}{\Big \rbrace} \\
  +&\hat u \textcolor{red}{\Big \lbrace} [v_{cl,R}^\text{diag}(0) + v_{cl,R}^\text{off}(p/2-l)  + v_{cl,R}^\text{off}(p/2+l) ] \textcolor{green}{\Big [} N [v_{cl,R}^\text{diag}(0) + v_{cl,R}^\text{off}(-p/2-l)  + v_{cl,R}^\text{off}(p/2-l)] \\
  &+ [v_{cl,R}^\text{off}(0) + v_{cl,R}^\text{diag}(-p/2-l) + v_{cl,R}^\text{off}(p/2-l)] + [v_{cl,R}^\text{off}(0) + v_{cl,R}^\text{off}(-p/2-l)  + v_{cl,R}^\text{diag}(p/2-l)] \textcolor{green}{\Big ]}\\
  &+ \textcolor{green}{\Big [} [v_{cl,R}^\text{off}(0) + v_{cl,R}^\text{diag}(p/2-l) + v_{cl,R}^\text{off}(p/2+l)] + [v_{cl,R}^\text{off}(0) + v_{cl,R}^\text{off}(p/2-l)  + v_{cl,R}^\text{diag}(p/2+l)] \textcolor{green}{\Big ]}\\
  &\times  [v_{cl,R}^\text{off}(0) + v_{cl,R}^\text{off}(-p/2-l)  + v_{cl,R}^\text{diag}(p/2-l)] \textcolor{red}{\Big \rbrace} \Big \rbrace \; ,
\end{align*}

\begin{align*}
&\begin{picture}(110,40) (170,-37)
      \SetWidth{1.0}
      \SetColor{Black}
      \Arc[color=Red, arrow](210,-33)(15,0,90)     \Arc[color=Blue, arrow](210,-33)(15,90,180)        
      \Arc[color=Blue](210,-33)(15,180,270)
      \Arc[color=Red, arrow, clockwise, arrowpos=1](210,-33)(15,0,270)
      \FBox(207,-15)(213,-21)
      \Line[color=Blue](185,-25)(194,-33)
      \Line[color=Red](185,-41)(194,-33)
      \Line[color=Blue](235,-25)(226,-33)
      \Line[color=Blue](235,-41)(226,-33)
      \Vertex(194,-33){3.4}
      \Vertex(226,-33){3.4}
      \Text(165,-50)[lb]{\scriptsize $+p/2,a$}
      \Text(165,-23)[lb]{\scriptsize $-p/2,\bar b$}
      \Text(227,-23)[lb]{\scriptsize $-p/2,\bar a$}
      \Text(227,-50)[lb]{\scriptsize $+p/2, b$}
\end{picture}
  =\int_l B_R(p/2-l) G_R(p/2+l) \Big \lbrace \\ \\
  &\hat s \textcolor{red}{\Big \lbrace} [v_{cl,R}^\text{diag}(p) + v_{cl,R}^\text{off}(l)  + v_{cl,R}^\text{off}(-l)  ]
  \textcolor{green}{\Big [} N 2 v_{an}^\text{off}(l) +2 (v_{an}^\text{off}(l)+ v_{an}^\text{diag}(l) )\textcolor{green}{\Big ]} \\
  &+ \textcolor{green}{\Big [} [v_{cl,R}^\text{off}(p) + v_{cl,R}^\text{diag}(l) + v_{cl,R}^\text{off}(-l) ] + [v_{cl,R}^\text{off}(p)  + v_{cl,R}^\text{off}(l)  + v_{cl,R}^\text{diag}(-l)] \textcolor{green}{\Big ]}\\
  &\times 2 v_{an}^\text{off}(l) \textcolor{red}{\Big \rbrace} \\
  + &\hat t \textcolor{red}{\Big \lbrace} [v_{cl,R}^\text{off}(p)  + v_{cl,R}^\text{diag}(l) + v_{cl,R}^\text{off}(-l) ] (v_{an}^\text{diag}(l)+ v_{an}^\text{off}(l) )\\
  &+ [v_{cl,R}^\text{off}(p)  + v_{cl,R}^\text{off}(l) + v_{cl,R}^\text{diag}(-l) ] (v_{an}^\text{off}(l)+ v_{an}^\text{diag}(l) ) \textcolor{red}{\Big \rbrace}\\
  +& \hat u \textcolor{red}{\Big \lbrace} [v_{cl,R}^\text{off}(p)  + v_{cl,R}^\text{diag}(l) + v_{cl,R}^\text{off}(-l) ] (v_{an}^\text{off}(l)+ v_{an}^\text{diag}(l) )\\
  &+ [v_{cl,R}^\text{off}(p)  + v_{cl,R}^\text{off}(l)  + v_{cl,R}^\text{diag}(-l) ] (v_{an}^\text{off}(l)+ v_{an}^\text{diag}(l) ) \textcolor{red}{\Big \rbrace} \Big \rbrace \; ,
\end{align*}

\begin{align*}
&\begin{picture}(110,40) (170,-37)
      \SetWidth{1.0}
      \SetColor{Black}
      \Arc[color=Red, arrow](210,-33)(15,0,90)     \Arc[color=Blue, arrow](210,-33)(15,90,180)        
      \Arc[color=Blue](210,-33)(15,180,270)
      \Arc[color=Red, arrow, clockwise, arrowpos=1](210,-33)(15,0,270)
      \FBox(207,-15)(213,-21)
      \Line[color=Blue](185,-25)(194,-33)
      \Line[color=Red](185,-41)(194,-33)
      \Line[color=Blue](235,-25)(226,-33)
      \Line[color=Blue](235,-41)(226,-33)
      \Vertex(194,-33){3.4}
      \Vertex(226,-33){3.4}
      \Text(165,-50)[lb]{\scriptsize $+p/2,a$}
      \Text(165,-23)[lb]{\scriptsize $-p/2,\bar a$}
      \Text(227,-23)[lb]{\scriptsize $+p/2, b$}
      \Text(227,-50)[lb]{\scriptsize $-p/2,\bar b$}
\end{picture}
  =\int_l B_R(-l) G_R(l) \Big \lbrace \\ \\
  &\hat s \textcolor{red}{\Big \lbrace} [v_{cl,R}^\text{off}(0) + v_{cl,R}^\text{off}(p/2-l)  + v_{cl,R}^\text{diag}(p/2+l)] [v_{an}^\text{diag}(l-p/2) + v_{an}^\text{off}(l+p/2)  ]\\
  &+ [v_{cl,R}^\text{off}(0) + v_{cl,R}^\text{diag}(p/2-l) + v_{cl,R}^\text{off}(p/2+l)2] [v_{an}^\text{off}(l-p/2) + v_{an}^\text{diag}(l+p/2) ] \textcolor{red}{\Big \rbrace} \\
  +&\hat t \textcolor{red}{\Big \lbrace} [v_{cl,R}^\text{diag}(0) + v_{cl,R}^\text{off}(p/2-l)  + v_{cl,R}^\text{off}(p/2+l) ] \textcolor{green}{\Big [} N [ v_{an}^\text{off}(l-p/2) + v_{an}^\text{off}(l+p/2) ]\\
  &+ v_{an}^\text{diag}(l-p/2) + v_{an}^\text{off}(l+p/2) ] + [v_{an}^\text{off}(l-p/2) + v_{an}^\text{diag}(l+p/2) ] \textcolor{green}{\Big ]}\\
  &+\textcolor{green}{\Big [} [v_{cl,R}^\text{off}(0) + v_{cl,R}^\text{diag}(p/2-l) + v_{cl,R}^\text{off}(p/2+l)2] + [v_{cl,R}^\text{off}(0) + v_{cl,R}^\text{off}(p/2-l)  + v_{cl,R}^\text{diag}(p/2+l)] \textcolor{green}{\Big ]}\\
  &\times [v_{an}^\text{off}(l-p/2) + v_{an}^\text{off}(l+p/2) ] \textcolor{red}{\Big \rbrace} \\
  +&\hat u \textcolor{red}{\Big \lbrace} [v_{cl,R}^\text{off}(0) + v_{cl,R}^\text{off}(p/2-l)  + v_{cl,R}^\text{diag}(p/2+l)] [v_{an}^\text{off}(l-p/2) + v_{an}^\text{diag}(l+p/2) ]\\
  &+ [v_{cl,R}^\text{off}(0) + v_{cl,R}^\text{diag}(p/2-l) + v_{cl,R}^\text{off}(p/2+l)] [v_{an}^\text{diag}(l-p/2) + v_{an}^\text{off}(l+p/2) ] \textcolor{red}{\Big \rbrace} \Big \rbrace \; ,
\end{align*}

\begin{align*}
&\begin{picture}(110,40) (170,-37)
      \SetWidth{1.0}
      \SetColor{Black}
      \Arc[color=Red, arrow](210,-33)(15,0,90)     \Arc[color=Blue, arrow](210,-33)(15,90,180)        
      \Arc[color=Blue](210,-33)(15,180,270)
      \Arc[color=Red, arrow, clockwise, arrowpos=1](210,-33)(15,0,270)
      \FBox(207,-15)(213,-21)
      \Line[color=Blue](185,-25)(194,-33)
      \Line[color=Red](185,-41)(194,-33)
      \Line[color=Blue](235,-25)(226,-33)
      \Line[color=Blue](235,-41)(226,-33)
      \Vertex(194,-33){3.4}
      \Vertex(226,-33){3.4}
      \Text(165,-50)[lb]{\scriptsize $+p/2,a$}
      \Text(165,-23)[lb]{\scriptsize $-p/2,\bar b$}
      \Text(227,-23)[lb]{\scriptsize $-p/2,\bar a$}
      \Text(227,-50)[lb]{\scriptsize $+p/2, b$}
\end{picture}
  =\int_l B_R(-l) G_R(l) \Big \lbrace \\ \\
  &\hat s \textcolor{red}{\Big \lbrace} [v_{cl,R}^\text{off}(0) + v_{cl,R}^\text{off}(p/2-l)  + v_{cl,R}^\text{diag}(p/2+l)] [v_{an}^\text{off}(p/2+l)  + v_{an}^\text{diag}(p/2-l)] \\
  &+ [v_{cl,R}^\text{off}(0) + v_{cl,R}^\text{diag}(p/2-l) + v_{cl,R}^\text{off}(p/2+l)] [v_{an}^\text{diag}(p/2+l) + v_{an}^\text{off}(p/2-l)]  \textcolor{red}{\Big \rbrace} \\
  +&\hat t \textcolor{red}{\Big \lbrace} [v_{cl,R}^\text{off}(0) + v_{cl,R}^\text{diag}(p/2-l) + v_{cl,R}^\text{off}(p/2+l)] [v_{an}^\text{off}(p/2+l)  + v_{an}^\text{diag}(p/2-l)] \\
  &+ [v_{cl,R}^\text{off}(0) + v_{cl,R}^\text{off}(p/2-l)  + v_{cl,R}^\text{diag}(p/2+l)] [v_{an}^\text{diag}(p/2+l) + v_{an}^\text{off}(p/2-l)] \textcolor{red}{\Big \rbrace} \\
  +&\hat u \textcolor{red}{\Big \lbrace} [v_{cl,R}^\text{diag}(0) + v_{cl,R}^\text{off}(p/2-l)  + v_{cl,R}^\text{off}(p/2+l) ] \textcolor{green}{\Big [} N [v_{an}^\text{off}(p/2+l)  + v_{an}^\text{off}(p/2-l)] \\
  &+ [v_{an}^\text{diag}(p/2+l) + v_{an}^\text{off}(p/2-l)] + [v_{an}^\text{off}(p/2+l)  + v_{an}^\text{diag}(p/2-l)] \textcolor{green}{\Big ]}\\
  &+ \textcolor{green}{\Big [} [v_{cl,R}^\text{off}(0) + v_{cl,R}^\text{diag}(p/2-l) + v_{cl,R}^\text{off}(p/2+l)] + [v_{cl,R}^\text{off}(0) + v_{cl,R}^\text{off}(p/2-l)  + v_{cl,R}^\text{diag}(p/2+l)] \textcolor{green}{\Big ]}\\
  &\times  [v_{an}^\text{off}(p/2+l) + v_{an}^\text{diag}(p/2-l)] \textcolor{red}{\Big \rbrace} \Big \rbrace \; .
\end{align*}

\bibliography{mybib}

\begin{thebibliography}{61}
\expandafter\ifx\csname natexlab\endcsname\relax\def\natexlab#1{#1}\fi
\expandafter\ifx\csname bibnamefont\endcsname\relax
  \def\bibnamefont#1{#1}\fi
\expandafter\ifx\csname bibfnamefont\endcsname\relax
  \def\bibfnamefont#1{#1}\fi
\expandafter\ifx\csname citenamefont\endcsname\relax
  \def\citenamefont#1{#1}\fi
\expandafter\ifx\csname url\endcsname\relax
  \def\url#1{\texttt{#1}}\fi
\expandafter\ifx\csname urlprefix\endcsname\relax\def\urlprefix{URL }\fi
\providecommand{\bibinfo}[2]{#2}
\providecommand{\eprint}[2][]{\url{#2}}

\bibitem[{\citenamefont{Meyer}(2007)}]{Meyer:2007ic}
\bibinfo{author}{\bibfnamefont{H.~B.} \bibnamefont{Meyer}},
  \bibinfo{journal}{Phys. Rev. D} \textbf{\bibinfo{volume}{76}},
  \bibinfo{pages}{101701} (\bibinfo{year}{2007}), \eprint{0704.1801}.

\bibitem[{\citenamefont{Brandt et~al.}(2016)\citenamefont{Brandt, Francis,
  Jäger, and Meyer}}]{Brandt:2015aqk}
\bibinfo{author}{\bibfnamefont{B.~B.} \bibnamefont{Brandt}},
  \bibinfo{author}{\bibfnamefont{A.}~\bibnamefont{Francis}},
  \bibinfo{author}{\bibfnamefont{B.}~\bibnamefont{Jäger}}, \bibnamefont{and}
  \bibinfo{author}{\bibfnamefont{H.~B.} \bibnamefont{Meyer}},
  \bibinfo{journal}{Phys. Rev. D} \textbf{\bibinfo{volume}{93}},
  \bibinfo{pages}{054510} (\bibinfo{year}{2016}), \eprint{1512.07249}.

\bibitem[{\citenamefont{Ding et~al.}(2019)\citenamefont{Ding, Kaczmarek, Kruse,
  Larsen, Mazur, Mukherjee, Ohno, Sandmeyer, and Shu}}]{Ding:2018uhl}
\bibinfo{author}{\bibfnamefont{H.-T.} \bibnamefont{Ding}},
  \bibinfo{author}{\bibfnamefont{O.}~\bibnamefont{Kaczmarek}},
  \bibinfo{author}{\bibfnamefont{A.-L.} \bibnamefont{Kruse}},
  \bibinfo{author}{\bibfnamefont{R.}~\bibnamefont{Larsen}},
  \bibinfo{author}{\bibfnamefont{L.}~\bibnamefont{Mazur}},
  \bibinfo{author}{\bibfnamefont{S.}~\bibnamefont{Mukherjee}},
  \bibinfo{author}{\bibfnamefont{H.}~\bibnamefont{Ohno}},
  \bibinfo{author}{\bibfnamefont{H.}~\bibnamefont{Sandmeyer}},
  \bibnamefont{and} \bibinfo{author}{\bibfnamefont{H.-T.} \bibnamefont{Shu}},
  \bibinfo{journal}{Nucl. Phys. A} \textbf{\bibinfo{volume}{982}},
  \bibinfo{pages}{715} (\bibinfo{year}{2019}), \eprint{1807.06315}.

\bibitem[{\citenamefont{Aarts et~al.}(2011)\citenamefont{Aarts, Allton, Kim,
  Lombardo, Oktay, Ryan, Sinclair, and Skullerud}}]{Aarts:2011sm}
\bibinfo{author}{\bibfnamefont{G.}~\bibnamefont{Aarts}},
  \bibinfo{author}{\bibfnamefont{C.}~\bibnamefont{Allton}},
  \bibinfo{author}{\bibfnamefont{S.}~\bibnamefont{Kim}},
  \bibinfo{author}{\bibfnamefont{M.}~\bibnamefont{Lombardo}},
  \bibinfo{author}{\bibfnamefont{M.}~\bibnamefont{Oktay}},
  \bibinfo{author}{\bibfnamefont{S.}~\bibnamefont{Ryan}},
  \bibinfo{author}{\bibfnamefont{D.}~\bibnamefont{Sinclair}}, \bibnamefont{and}
  \bibinfo{author}{\bibfnamefont{J.}~\bibnamefont{Skullerud}},
  \bibinfo{journal}{JHEP} \textbf{\bibinfo{volume}{11}}, \bibinfo{pages}{103}
  (\bibinfo{year}{2011}), \eprint{1109.4496}.

\bibitem[{\citenamefont{Aarts et~al.}(2015)\citenamefont{Aarts, Allton, Amato,
  Giudice, Hands, and Skullerud}}]{Aarts:2014nba}
\bibinfo{author}{\bibfnamefont{G.}~\bibnamefont{Aarts}},
  \bibinfo{author}{\bibfnamefont{C.}~\bibnamefont{Allton}},
  \bibinfo{author}{\bibfnamefont{A.}~\bibnamefont{Amato}},
  \bibinfo{author}{\bibfnamefont{P.}~\bibnamefont{Giudice}},
  \bibinfo{author}{\bibfnamefont{S.}~\bibnamefont{Hands}}, \bibnamefont{and}
  \bibinfo{author}{\bibfnamefont{J.-I.} \bibnamefont{Skullerud}},
  \bibinfo{journal}{JHEP} \textbf{\bibinfo{volume}{02}}, \bibinfo{pages}{186}
  (\bibinfo{year}{2015}), \eprint{1412.6411}.

\bibitem[{\citenamefont{Rapp et~al.}(1997)\citenamefont{Rapp, Chanfray, and
  Wambach}}]{Rapp:1997fs}
\bibinfo{author}{\bibfnamefont{R.}~\bibnamefont{Rapp}},
  \bibinfo{author}{\bibfnamefont{G.}~\bibnamefont{Chanfray}}, \bibnamefont{and}
  \bibinfo{author}{\bibfnamefont{J.}~\bibnamefont{Wambach}},
  \bibinfo{journal}{Nucl. Phys. A} \textbf{\bibinfo{volume}{617}},
  \bibinfo{pages}{472} (\bibinfo{year}{1997}), \eprint{hep-ph/9702210}.

\bibitem[{\citenamefont{Urban et~al.}(1998)\citenamefont{Urban, Buballa, Rapp,
  and Wambach}}]{Urban:1998eg}
\bibinfo{author}{\bibfnamefont{M.}~\bibnamefont{Urban}},
  \bibinfo{author}{\bibfnamefont{M.}~\bibnamefont{Buballa}},
  \bibinfo{author}{\bibfnamefont{R.}~\bibnamefont{Rapp}}, \bibnamefont{and}
  \bibinfo{author}{\bibfnamefont{J.}~\bibnamefont{Wambach}},
  \bibinfo{journal}{Nucl. Phys. A} \textbf{\bibinfo{volume}{641}},
  \bibinfo{pages}{433} (\bibinfo{year}{1998}), \eprint{nucl-th/9806030}.

\bibitem[{\citenamefont{Roder et~al.}(2006)\citenamefont{Roder, Ruppert, and
  Rischke}}]{Roder:2005vt}
\bibinfo{author}{\bibfnamefont{D.}~\bibnamefont{Roder}},
  \bibinfo{author}{\bibfnamefont{J.}~\bibnamefont{Ruppert}}, \bibnamefont{and}
  \bibinfo{author}{\bibfnamefont{D.~H.} \bibnamefont{Rischke}},
  \bibinfo{journal}{Nucl. Phys. A} \textbf{\bibinfo{volume}{775}},
  \bibinfo{pages}{127} (\bibinfo{year}{2006}), \eprint{hep-ph/0503042}.

\bibitem[{\citenamefont{Liu and Rapp}(2018)}]{Liu:2017qah}
\bibinfo{author}{\bibfnamefont{S.~Y.~F.} \bibnamefont{Liu}} \bibnamefont{and}
  \bibinfo{author}{\bibfnamefont{R.}~\bibnamefont{Rapp}},
  \bibinfo{journal}{Phys. Rev. C} \textbf{\bibinfo{volume}{97}},
  \bibinfo{pages}{034918} (\bibinfo{year}{2018}), \eprint{1711.03282}.

\bibitem[{\citenamefont{Mueller et~al.}(2010)\citenamefont{Mueller, Fischer,
  and Nickel}}]{Mueller:2010ah}
\bibinfo{author}{\bibfnamefont{J.~A.} \bibnamefont{Mueller}},
  \bibinfo{author}{\bibfnamefont{C.~S.} \bibnamefont{Fischer}},
  \bibnamefont{and} \bibinfo{author}{\bibfnamefont{D.}~\bibnamefont{Nickel}},
  \bibinfo{journal}{Eur. Phys. J. C} \textbf{\bibinfo{volume}{70}},
  \bibinfo{pages}{1037} (\bibinfo{year}{2010}), \eprint{1009.3762}.

\bibitem[{\citenamefont{Fischer et~al.}(2018)\citenamefont{Fischer, Pawlowski,
  Rothkopf, and Welzbacher}}]{Fischer:2017kbq}
\bibinfo{author}{\bibfnamefont{C.~S.} \bibnamefont{Fischer}},
  \bibinfo{author}{\bibfnamefont{J.~M.} \bibnamefont{Pawlowski}},
  \bibinfo{author}{\bibfnamefont{A.}~\bibnamefont{Rothkopf}}, \bibnamefont{and}
  \bibinfo{author}{\bibfnamefont{C.~A.} \bibnamefont{Welzbacher}},
  \bibinfo{journal}{Phys. Rev. D} \textbf{\bibinfo{volume}{98}},
  \bibinfo{pages}{014009} (\bibinfo{year}{2018}), \eprint{1705.03207}.

\bibitem[{\citenamefont{Shen and Berges}(2020)}]{Shen:2019jhl}
\bibinfo{author}{\bibfnamefont{L.}~\bibnamefont{Shen}} \bibnamefont{and}
  \bibinfo{author}{\bibfnamefont{J.}~\bibnamefont{Berges}},
  \bibinfo{journal}{Phys. Rev. D} \textbf{\bibinfo{volume}{101}},
  \bibinfo{pages}{056009} (\bibinfo{year}{2020}), \eprint{1912.07565}.

\bibitem[{\citenamefont{Shen et~al.}(2020)\citenamefont{Shen, Berges,
  Pawlowski, and Rothkopf}}]{Shen:2020jya}
\bibinfo{author}{\bibfnamefont{L.}~\bibnamefont{Shen}},
  \bibinfo{author}{\bibfnamefont{J.}~\bibnamefont{Berges}},
  \bibinfo{author}{\bibfnamefont{J.~M.} \bibnamefont{Pawlowski}},
  \bibnamefont{and} \bibinfo{author}{\bibfnamefont{A.}~\bibnamefont{Rothkopf}},
  \bibinfo{journal}{Phys. Rev. D} \textbf{\bibinfo{volume}{102}},
  \bibinfo{pages}{016012} (\bibinfo{year}{2020}), \eprint{2003.03270}.

\bibitem[{\citenamefont{Patkos et~al.}(2002)\citenamefont{Patkos, Szep, and
  Szepfalusy}}]{Patkos:2002xb}
\bibinfo{author}{\bibfnamefont{A.}~\bibnamefont{Patkos}},
  \bibinfo{author}{\bibfnamefont{Z.}~\bibnamefont{Szep}}, \bibnamefont{and}
  \bibinfo{author}{\bibfnamefont{P.}~\bibnamefont{Szepfalusy}},
  \bibinfo{journal}{Phys. Lett. B} \textbf{\bibinfo{volume}{537}},
  \bibinfo{pages}{77} (\bibinfo{year}{2002}), \eprint{hep-ph/0202261}.

\bibitem[{\citenamefont{Floerchinger}(2012)}]{Floerchinger:2011sc}
\bibinfo{author}{\bibfnamefont{S.}~\bibnamefont{Floerchinger}},
  \bibinfo{journal}{JHEP} \textbf{\bibinfo{volume}{05}}, \bibinfo{pages}{021}
  (\bibinfo{year}{2012}), \eprint{1112.4374}.

\bibitem[{\citenamefont{Kamikado et~al.}(2014)\citenamefont{Kamikado,
  Strodthoff, von Smekal, and Wambach}}]{Kamikado:2013sia}
\bibinfo{author}{\bibfnamefont{K.}~\bibnamefont{Kamikado}},
  \bibinfo{author}{\bibfnamefont{N.}~\bibnamefont{Strodthoff}},
  \bibinfo{author}{\bibfnamefont{L.}~\bibnamefont{von Smekal}},
  \bibnamefont{and} \bibinfo{author}{\bibfnamefont{J.}~\bibnamefont{Wambach}},
  \bibinfo{journal}{Eur. Phys. J. C} \textbf{\bibinfo{volume}{74}},
  \bibinfo{pages}{2806} (\bibinfo{year}{2014}), \eprint{1302.6199}.

\bibitem[{\citenamefont{Tripolt
  et~al.}(2014{\natexlab{a}})\citenamefont{Tripolt, Strodthoff, von Smekal, and
  Wambach}}]{Tripolt:2013jra}
\bibinfo{author}{\bibfnamefont{R.-A.} \bibnamefont{Tripolt}},
  \bibinfo{author}{\bibfnamefont{N.}~\bibnamefont{Strodthoff}},
  \bibinfo{author}{\bibfnamefont{L.}~\bibnamefont{von Smekal}},
  \bibnamefont{and} \bibinfo{author}{\bibfnamefont{J.}~\bibnamefont{Wambach}},
  \bibinfo{journal}{Phys. Rev. D} \textbf{\bibinfo{volume}{89}},
  \bibinfo{pages}{034010} (\bibinfo{year}{2014}{\natexlab{a}}),
  \eprint{1311.0630}.

\bibitem[{\citenamefont{Pawlowski and Strodthoff}(2015)}]{Pawlowski:2015mia}
\bibinfo{author}{\bibfnamefont{J.~M.} \bibnamefont{Pawlowski}}
  \bibnamefont{and}
  \bibinfo{author}{\bibfnamefont{N.}~\bibnamefont{Strodthoff}},
  \bibinfo{journal}{Phys. Rev. D} \textbf{\bibinfo{volume}{92}},
  \bibinfo{pages}{094009} (\bibinfo{year}{2015}), \eprint{1508.01160}.

\bibitem[{\citenamefont{Tripolt
  et~al.}(2014{\natexlab{b}})\citenamefont{Tripolt, von Smekal, and
  Wambach}}]{Tripolt:2014wra}
\bibinfo{author}{\bibfnamefont{R.-A.} \bibnamefont{Tripolt}},
  \bibinfo{author}{\bibfnamefont{L.}~\bibnamefont{von Smekal}},
  \bibnamefont{and} \bibinfo{author}{\bibfnamefont{J.}~\bibnamefont{Wambach}},
  \bibinfo{journal}{Phys. Rev. D} \textbf{\bibinfo{volume}{90}},
  \bibinfo{pages}{074031} (\bibinfo{year}{2014}{\natexlab{b}}),
  \eprint{1408.3512}.

\bibitem[{\citenamefont{Jung et~al.}(2017)\citenamefont{Jung, Rennecke,
  Tripolt, von Smekal, and Wambach}}]{Jung:2016yxl}
\bibinfo{author}{\bibfnamefont{C.}~\bibnamefont{Jung}},
  \bibinfo{author}{\bibfnamefont{F.}~\bibnamefont{Rennecke}},
  \bibinfo{author}{\bibfnamefont{R.-A.} \bibnamefont{Tripolt}},
  \bibinfo{author}{\bibfnamefont{L.}~\bibnamefont{von Smekal}},
  \bibnamefont{and} \bibinfo{author}{\bibfnamefont{J.}~\bibnamefont{Wambach}},
  \bibinfo{journal}{Phys. Rev. D} \textbf{\bibinfo{volume}{95}},
  \bibinfo{pages}{036020} (\bibinfo{year}{2017}), \eprint{1610.08754}.

\bibitem[{\citenamefont{Jung and von Smekal}(2019)}]{Jung:2019nnr}
\bibinfo{author}{\bibfnamefont{C.}~\bibnamefont{Jung}} \bibnamefont{and}
  \bibinfo{author}{\bibfnamefont{L.}~\bibnamefont{von Smekal}},
  \bibinfo{journal}{Phys. Rev. D} \textbf{\bibinfo{volume}{100}},
  \bibinfo{pages}{116009} (\bibinfo{year}{2019}), \eprint{1909.13712}.

\bibitem[{\citenamefont{Pawlowski et~al.}(2018)\citenamefont{Pawlowski,
  Strodthoff, and Wink}}]{Pawlowski:2017gxj}
\bibinfo{author}{\bibfnamefont{J.~M.} \bibnamefont{Pawlowski}},
  \bibinfo{author}{\bibfnamefont{N.}~\bibnamefont{Strodthoff}},
  \bibnamefont{and} \bibinfo{author}{\bibfnamefont{N.}~\bibnamefont{Wink}},
  \bibinfo{journal}{Phys. Rev. D} \textbf{\bibinfo{volume}{98}},
  \bibinfo{pages}{074008} (\bibinfo{year}{2018}), \eprint{1711.07444}.

\bibitem[{\citenamefont{Strodthoff}(2017)}]{Strodthoff:2016pxx}
\bibinfo{author}{\bibfnamefont{N.}~\bibnamefont{Strodthoff}},
  \bibinfo{journal}{Phys. Rev. D} \textbf{\bibinfo{volume}{95}},
  \bibinfo{pages}{076002} (\bibinfo{year}{2017}), \eprint{1611.05036}.

\bibitem[{\citenamefont{Wang and Zhuang}(2017)}]{Wang:2017vis}
\bibinfo{author}{\bibfnamefont{Z.}~\bibnamefont{Wang}} \bibnamefont{and}
  \bibinfo{author}{\bibfnamefont{P.}~\bibnamefont{Zhuang}},
  \bibinfo{journal}{Phys. Rev. D} \textbf{\bibinfo{volume}{96}},
  \bibinfo{pages}{014006} (\bibinfo{year}{2017}), \eprint{1703.01035}.

\bibitem[{\citenamefont{Yokota et~al.}(2016)\citenamefont{Yokota, Kunihiro, and
  Morita}}]{Yokota:2016tip}
\bibinfo{author}{\bibfnamefont{T.}~\bibnamefont{Yokota}},
  \bibinfo{author}{\bibfnamefont{T.}~\bibnamefont{Kunihiro}}, \bibnamefont{and}
  \bibinfo{author}{\bibfnamefont{K.}~\bibnamefont{Morita}},
  \bibinfo{journal}{PTEP} \textbf{\bibinfo{volume}{2016}},
  \bibinfo{pages}{073D01} (\bibinfo{year}{2016}), \eprint{1603.02147}.

\bibitem[{\citenamefont{Gasenzer and Pawlowski}(2008)}]{Gasenzer:2007za}
\bibinfo{author}{\bibfnamefont{T.}~\bibnamefont{Gasenzer}} \bibnamefont{and}
  \bibinfo{author}{\bibfnamefont{J.~M.} \bibnamefont{Pawlowski}},
  \bibinfo{journal}{Phys. Lett. B} \textbf{\bibinfo{volume}{670}},
  \bibinfo{pages}{135} (\bibinfo{year}{2008}), \eprint{0710.4627}.

\bibitem[{\citenamefont{Berges and Hoffmeister}(2009)}]{Berges:2008sr}
\bibinfo{author}{\bibfnamefont{J.}~\bibnamefont{Berges}} \bibnamefont{and}
  \bibinfo{author}{\bibfnamefont{G.}~\bibnamefont{Hoffmeister}},
  \bibinfo{journal}{Nucl. Phys. B} \textbf{\bibinfo{volume}{813}},
  \bibinfo{pages}{383} (\bibinfo{year}{2009}), \eprint{0809.5208}.

\bibitem[{\citenamefont{Berges and Mesterhazy}(2012)}]{Berges:2012ty}
\bibinfo{author}{\bibfnamefont{J.}~\bibnamefont{Berges}} \bibnamefont{and}
  \bibinfo{author}{\bibfnamefont{D.}~\bibnamefont{Mesterhazy}},
  \bibinfo{journal}{Nucl. Phys. B Proc. Suppl.} \textbf{\bibinfo{volume}{228}},
  \bibinfo{pages}{37} (\bibinfo{year}{2012}), \eprint{1204.1489}.

\bibitem[{\citenamefont{Pietroni}(2008)}]{Pietroni:2008jx}
\bibinfo{author}{\bibfnamefont{M.}~\bibnamefont{Pietroni}},
  \bibinfo{journal}{JCAP} \textbf{\bibinfo{volume}{10}}, \bibinfo{pages}{036}
  (\bibinfo{year}{2008}), \eprint{0806.0971}.

\bibitem[{\citenamefont{Schoeller and K\"onig}(2000)}]{PhysRevLett.84.3686}
\bibinfo{author}{\bibfnamefont{H.}~\bibnamefont{Schoeller}} \bibnamefont{and}
  \bibinfo{author}{\bibfnamefont{J.}~\bibnamefont{K\"onig}},
  \bibinfo{journal}{Phys. Rev. Lett.} \textbf{\bibinfo{volume}{84}},
  \bibinfo{pages}{3686} (\bibinfo{year}{2000}),
  \urlprefix\url{https://link.aps.org/doi/10.1103/PhysRevLett.84.3686}.

\bibitem[{\citenamefont{Canet et~al.}(2010)\citenamefont{Canet, Chate,
  Delamotte, and Wschebor}}]{Canet:2009vz}
\bibinfo{author}{\bibfnamefont{L.}~\bibnamefont{Canet}},
  \bibinfo{author}{\bibfnamefont{H.}~\bibnamefont{Chate}},
  \bibinfo{author}{\bibfnamefont{B.}~\bibnamefont{Delamotte}},
  \bibnamefont{and} \bibinfo{author}{\bibfnamefont{N.}~\bibnamefont{Wschebor}},
  \bibinfo{journal}{Phys. Rev. Lett.} \textbf{\bibinfo{volume}{104}},
  \bibinfo{pages}{150601} (\bibinfo{year}{2010}), \eprint{0905.1025}.

\bibitem[{\citenamefont{Delamotte and Canet}(2005)}]{Delamotte:2004zg}
\bibinfo{author}{\bibfnamefont{B.}~\bibnamefont{Delamotte}} \bibnamefont{and}
  \bibinfo{author}{\bibfnamefont{L.}~\bibnamefont{Canet}},
  \bibinfo{journal}{Condensed Matter Phys.} \textbf{\bibinfo{volume}{8}},
  \bibinfo{pages}{163} (\bibinfo{year}{2005}), \eprint{cond-mat/0412205}.

\bibitem[{\citenamefont{Canet and Chaté}(2007)}]{Canet_2007}
\bibinfo{author}{\bibfnamefont{L.}~\bibnamefont{Canet}} \bibnamefont{and}
  \bibinfo{author}{\bibfnamefont{H.}~\bibnamefont{Chaté}},
  \bibinfo{journal}{Journal of Physics A: Mathematical and Theoretical}
  \textbf{\bibinfo{volume}{40}}, \bibinfo{pages}{1937–1949}
  (\bibinfo{year}{2007}), ISSN \bibinfo{issn}{1751-8121},
  \urlprefix\url{http://dx.doi.org/10.1088/1751-8113/40/9/002}.

\bibitem[{\citenamefont{Karrasch et~al.}(2010)\citenamefont{Karrasch,
  Pletyukhov, Borda, and Meden}}]{Karrasch_2010}
\bibinfo{author}{\bibfnamefont{C.}~\bibnamefont{Karrasch}},
  \bibinfo{author}{\bibfnamefont{M.}~\bibnamefont{Pletyukhov}},
  \bibinfo{author}{\bibfnamefont{L.}~\bibnamefont{Borda}}, \bibnamefont{and}
  \bibinfo{author}{\bibfnamefont{V.}~\bibnamefont{Meden}},
  \bibinfo{journal}{Physical Review B} \textbf{\bibinfo{volume}{81}}
  (\bibinfo{year}{2010}), ISSN \bibinfo{issn}{1550-235X},
  \urlprefix\url{http://dx.doi.org/10.1103/PhysRevB.81.125122}.

\bibitem[{\citenamefont{Andergassen et~al.}(2011)\citenamefont{Andergassen,
  Pletyukhov, Schuricht, Schoeller, and Borda}}]{Andergassen_2011}
\bibinfo{author}{\bibfnamefont{S.}~\bibnamefont{Andergassen}},
  \bibinfo{author}{\bibfnamefont{M.}~\bibnamefont{Pletyukhov}},
  \bibinfo{author}{\bibfnamefont{D.}~\bibnamefont{Schuricht}},
  \bibinfo{author}{\bibfnamefont{H.}~\bibnamefont{Schoeller}},
  \bibnamefont{and} \bibinfo{author}{\bibfnamefont{L.}~\bibnamefont{Borda}},
  \bibinfo{journal}{Physical Review B} \textbf{\bibinfo{volume}{83}}
  (\bibinfo{year}{2011}), ISSN \bibinfo{issn}{1550-235X},
  \urlprefix\url{http://dx.doi.org/10.1103/PhysRevB.83.205103}.

\bibitem[{\citenamefont{Sieberer et~al.}(2014)\citenamefont{Sieberer, Huber,
  Altman, and Diehl}}]{PhysRevB.89.134310}
\bibinfo{author}{\bibfnamefont{L.~M.} \bibnamefont{Sieberer}},
  \bibinfo{author}{\bibfnamefont{S.~D.} \bibnamefont{Huber}},
  \bibinfo{author}{\bibfnamefont{E.}~\bibnamefont{Altman}}, \bibnamefont{and}
  \bibinfo{author}{\bibfnamefont{S.}~\bibnamefont{Diehl}},
  \bibinfo{journal}{Phys. Rev. B} \textbf{\bibinfo{volume}{89}},
  \bibinfo{pages}{134310} (\bibinfo{year}{2014}),
  \urlprefix\url{https://link.aps.org/doi/10.1103/PhysRevB.89.134310}.

\bibitem[{\citenamefont{Aarts}(2001)}]{Aarts:2001yx}
\bibinfo{author}{\bibfnamefont{G.}~\bibnamefont{Aarts}},
  \bibinfo{journal}{Phys. Lett. B} \textbf{\bibinfo{volume}{518}},
  \bibinfo{pages}{315} (\bibinfo{year}{2001}), \eprint{hep-ph/0108125}.

\bibitem[{\citenamefont{Berges et~al.}(2010)\citenamefont{Berges, Schlichting,
  and Sexty}}]{Berges:2009jz}
\bibinfo{author}{\bibfnamefont{J.}~\bibnamefont{Berges}},
  \bibinfo{author}{\bibfnamefont{S.}~\bibnamefont{Schlichting}},
  \bibnamefont{and} \bibinfo{author}{\bibfnamefont{D.}~\bibnamefont{Sexty}},
  \bibinfo{journal}{Nucl. Phys. B} \textbf{\bibinfo{volume}{832}},
  \bibinfo{pages}{228} (\bibinfo{year}{2010}), \eprint{0912.3135}.

\bibitem[{\citenamefont{Schlichting et~al.}(2020)\citenamefont{Schlichting,
  Smith, and von Smekal}}]{Schlichting:2019tbr}
\bibinfo{author}{\bibfnamefont{S.}~\bibnamefont{Schlichting}},
  \bibinfo{author}{\bibfnamefont{D.}~\bibnamefont{Smith}}, \bibnamefont{and}
  \bibinfo{author}{\bibfnamefont{L.}~\bibnamefont{von Smekal}},
  \bibinfo{journal}{Nucl. Phys. B} \textbf{\bibinfo{volume}{950}},
  \bibinfo{pages}{114868} (\bibinfo{year}{2020}), \eprint{1908.00912}.

\bibitem[{\citenamefont{Schweitzer et~al.}(2020)\citenamefont{Schweitzer,
  Schlichting, and von Smekal}}]{Schweitzer:2020noq}
\bibinfo{author}{\bibfnamefont{D.}~\bibnamefont{Schweitzer}},
  \bibinfo{author}{\bibfnamefont{S.}~\bibnamefont{Schlichting}},
  \bibnamefont{and} \bibinfo{author}{\bibfnamefont{L.}~\bibnamefont{von
  Smekal}} (\bibinfo{year}{2020}), \eprint{2007.03374}.

\bibitem[{\citenamefont{Berges}(2004)}]{Berges:2004yj}
\bibinfo{author}{\bibfnamefont{J.}~\bibnamefont{Berges}}, \bibinfo{journal}{AIP
  Conf. Proc.} \textbf{\bibinfo{volume}{739}}, \bibinfo{pages}{3}
  (\bibinfo{year}{2004}), \eprint{hep-ph/0409233}.

\bibitem[{\citenamefont{Sieberer et~al.}(2015)\citenamefont{Sieberer,
  Chiocchetta, Gambassi, Täuber, and Diehl}}]{Sieberer:2015hba}
\bibinfo{author}{\bibfnamefont{L.}~\bibnamefont{Sieberer}},
  \bibinfo{author}{\bibfnamefont{A.}~\bibnamefont{Chiocchetta}},
  \bibinfo{author}{\bibfnamefont{A.}~\bibnamefont{Gambassi}},
  \bibinfo{author}{\bibfnamefont{U.}~\bibnamefont{Täuber}}, \bibnamefont{and}
  \bibinfo{author}{\bibfnamefont{S.}~\bibnamefont{Diehl}},
  \bibinfo{journal}{Phys. Rev. B} \textbf{\bibinfo{volume}{92}},
  \bibinfo{pages}{134307} (\bibinfo{year}{2015}), \eprint{1505.00912}.

\bibitem[{\citenamefont{Hohenberg and Halperin}(1977)}]{Hohenberg:1977ym}
\bibinfo{author}{\bibfnamefont{P.}~\bibnamefont{Hohenberg}} \bibnamefont{and}
  \bibinfo{author}{\bibfnamefont{B.}~\bibnamefont{Halperin}},
  \bibinfo{journal}{Rev. Mod. Phys.} \textbf{\bibinfo{volume}{49}},
  \bibinfo{pages}{435} (\bibinfo{year}{1977}).

\bibitem[{\citenamefont{Rajagopal and Wilczek}(1993)}]{Rajagopal:1992qz}
\bibinfo{author}{\bibfnamefont{K.}~\bibnamefont{Rajagopal}} \bibnamefont{and}
  \bibinfo{author}{\bibfnamefont{F.}~\bibnamefont{Wilczek}},
  \bibinfo{journal}{Nucl. Phys. B} \textbf{\bibinfo{volume}{399}},
  \bibinfo{pages}{395} (\bibinfo{year}{1993}), \eprint{hep-ph/9210253}.

\bibitem[{\citenamefont{Martin et~al.}(1973)\citenamefont{Martin, Siggia, and
  Rose}}]{Martin:1973zz}
\bibinfo{author}{\bibfnamefont{P.}~\bibnamefont{Martin}},
  \bibinfo{author}{\bibfnamefont{E.}~\bibnamefont{Siggia}}, \bibnamefont{and}
  \bibinfo{author}{\bibfnamefont{H.}~\bibnamefont{Rose}},
  \bibinfo{journal}{Phys. Rev. A} \textbf{\bibinfo{volume}{8}},
  \bibinfo{pages}{423} (\bibinfo{year}{1973}).

\bibitem[{\citenamefont{Hertz et~al.}(2017)\citenamefont{Hertz, Roudi, and
  Sollich}}]{Hertz:2016vpy}
\bibinfo{author}{\bibfnamefont{J.~A.} \bibnamefont{Hertz}},
  \bibinfo{author}{\bibfnamefont{Y.}~\bibnamefont{Roudi}}, \bibnamefont{and}
  \bibinfo{author}{\bibfnamefont{P.}~\bibnamefont{Sollich}},
  \bibinfo{journal}{J. Phys. A} \textbf{\bibinfo{volume}{50}},
  \bibinfo{pages}{033001} (\bibinfo{year}{2017}), \eprint{1604.05775}.

\bibitem[{\citenamefont{Aarts and Smit}(1997)}]{Aarts:1996qi}
\bibinfo{author}{\bibfnamefont{G.}~\bibnamefont{Aarts}} \bibnamefont{and}
  \bibinfo{author}{\bibfnamefont{J.}~\bibnamefont{Smit}},
  \bibinfo{journal}{Phys. Lett. B} \textbf{\bibinfo{volume}{393}},
  \bibinfo{pages}{395} (\bibinfo{year}{1997}), \eprint{hep-ph/9610415}.

\bibitem[{\citenamefont{Aarts and Smit}(1998)}]{Aarts:1997kp}
\bibinfo{author}{\bibfnamefont{G.}~\bibnamefont{Aarts}} \bibnamefont{and}
  \bibinfo{author}{\bibfnamefont{J.}~\bibnamefont{Smit}},
  \bibinfo{journal}{Nucl. Phys. B} \textbf{\bibinfo{volume}{511}},
  \bibinfo{pages}{451} (\bibinfo{year}{1998}), \eprint{hep-ph/9707342}.

\bibitem[{\citenamefont{Aarts and Berges}(2002)}]{Aarts:2001yn}
\bibinfo{author}{\bibfnamefont{G.}~\bibnamefont{Aarts}} \bibnamefont{and}
  \bibinfo{author}{\bibfnamefont{J.}~\bibnamefont{Berges}},
  \bibinfo{journal}{Phys. Rev. Lett.} \textbf{\bibinfo{volume}{88}},
  \bibinfo{pages}{041603} (\bibinfo{year}{2002}), \eprint{hep-ph/0107129}.

\bibitem[{\citenamefont{Epelbaum et~al.}(2014)\citenamefont{Epelbaum, Gelis,
  and Wu}}]{Epelbaum:2014yja}
\bibinfo{author}{\bibfnamefont{T.}~\bibnamefont{Epelbaum}},
  \bibinfo{author}{\bibfnamefont{F.}~\bibnamefont{Gelis}}, \bibnamefont{and}
  \bibinfo{author}{\bibfnamefont{B.}~\bibnamefont{Wu}}, \bibinfo{journal}{Phys.
  Rev. D} \textbf{\bibinfo{volume}{90}}, \bibinfo{pages}{065029}
  (\bibinfo{year}{2014}), \eprint{1402.0115}.

\bibitem[{\citenamefont{Wetterich}(1993)}]{Wetterich:1992yh}
\bibinfo{author}{\bibfnamefont{C.}~\bibnamefont{Wetterich}},
  \bibinfo{journal}{Phys. Lett. B} \textbf{\bibinfo{volume}{301}},
  \bibinfo{pages}{90} (\bibinfo{year}{1993}), \eprint{1710.05815}.

\bibitem[{\citenamefont{Mesterházy et~al.}(2015)\citenamefont{Mesterházy,
  Stockemer, and Tanizaki}}]{Mesterhazy:2015uja}
\bibinfo{author}{\bibfnamefont{D.}~\bibnamefont{Mesterházy}},
  \bibinfo{author}{\bibfnamefont{J.~H.} \bibnamefont{Stockemer}},
  \bibnamefont{and} \bibinfo{author}{\bibfnamefont{Y.}~\bibnamefont{Tanizaki}},
  \bibinfo{journal}{Phys. Rev. D} \textbf{\bibinfo{volume}{92}},
  \bibinfo{pages}{076001} (\bibinfo{year}{2015}), \eprint{1504.07268}.

\bibitem[{\citenamefont{Litim}(2001)}]{Litim:2001up}
\bibinfo{author}{\bibfnamefont{D.~F.} \bibnamefont{Litim}},
  \bibinfo{journal}{Phys. Rev. D} \textbf{\bibinfo{volume}{64}},
  \bibinfo{pages}{105007} (\bibinfo{year}{2001}), \eprint{hep-th/0103195}.

\bibitem[{\citenamefont{Wang and Heinz}(1996)}]{Wang:1995qf}
\bibinfo{author}{\bibfnamefont{E.-k.} \bibnamefont{Wang}} \bibnamefont{and}
  \bibinfo{author}{\bibfnamefont{U.~W.} \bibnamefont{Heinz}},
  \bibinfo{journal}{Phys. Rev. D} \textbf{\bibinfo{volume}{53}},
  \bibinfo{pages}{899} (\bibinfo{year}{1996}), \eprint{hep-ph/9509333}.

\bibitem[{\citenamefont{Gasenzer et~al.}(2010)\citenamefont{Gasenzer, Kessler,
  and Pawlowski}}]{Gasenzer:2010rq}
\bibinfo{author}{\bibfnamefont{T.}~\bibnamefont{Gasenzer}},
  \bibinfo{author}{\bibfnamefont{S.}~\bibnamefont{Kessler}}, \bibnamefont{and}
  \bibinfo{author}{\bibfnamefont{J.~M.} \bibnamefont{Pawlowski}},
  \bibinfo{journal}{Eur. Phys. J. C} \textbf{\bibinfo{volume}{70}},
  \bibinfo{pages}{423} (\bibinfo{year}{2010}), \eprint{1003.4163}.

\bibitem[{\citenamefont{Corell et~al.}(2019)\citenamefont{Corell, Cyrol,
  Heller, and Pawlowski}}]{Corell:2019jxh}
\bibinfo{author}{\bibfnamefont{L.}~\bibnamefont{Corell}},
  \bibinfo{author}{\bibfnamefont{A.~K.} \bibnamefont{Cyrol}},
  \bibinfo{author}{\bibfnamefont{M.}~\bibnamefont{Heller}}, \bibnamefont{and}
  \bibinfo{author}{\bibfnamefont{J.~M.} \bibnamefont{Pawlowski}}
  (\bibinfo{year}{2019}), \eprint{1910.09369}.

\bibitem[{\citenamefont{Aarts et~al.}(2002)\citenamefont{Aarts, Ahrensmeier,
  Baier, Berges, and Serreau}}]{Aarts:2002dj}
\bibinfo{author}{\bibfnamefont{G.}~\bibnamefont{Aarts}},
  \bibinfo{author}{\bibfnamefont{D.}~\bibnamefont{Ahrensmeier}},
  \bibinfo{author}{\bibfnamefont{R.}~\bibnamefont{Baier}},
  \bibinfo{author}{\bibfnamefont{J.}~\bibnamefont{Berges}}, \bibnamefont{and}
  \bibinfo{author}{\bibfnamefont{J.}~\bibnamefont{Serreau}},
  \bibinfo{journal}{Phys. Rev. D} \textbf{\bibinfo{volume}{66}},
  \bibinfo{pages}{045008} (\bibinfo{year}{2002}), \eprint{hep-ph/0201308}.

\bibitem[{\citenamefont{Blaizot et~al.}(2011)\citenamefont{Blaizot, Pawlowski,
  and Reinosa}}]{Blaizot:2010zx}
\bibinfo{author}{\bibfnamefont{J.-P.} \bibnamefont{Blaizot}},
  \bibinfo{author}{\bibfnamefont{J.~M.} \bibnamefont{Pawlowski}},
  \bibnamefont{and} \bibinfo{author}{\bibfnamefont{U.}~\bibnamefont{Reinosa}},
  \bibinfo{journal}{Phys. Lett. B} \textbf{\bibinfo{volume}{696}},
  \bibinfo{pages}{523} (\bibinfo{year}{2011}), \eprint{1009.6048}.

\bibitem[{\citenamefont{Carrington et~al.}(2018)\citenamefont{Carrington,
  Friesen, Meggison, Phillips, Pickering, and Sohrabi}}]{Carrington:2017lry}
\bibinfo{author}{\bibfnamefont{M.}~\bibnamefont{Carrington}},
  \bibinfo{author}{\bibfnamefont{S.}~\bibnamefont{Friesen}},
  \bibinfo{author}{\bibfnamefont{B.}~\bibnamefont{Meggison}},
  \bibinfo{author}{\bibfnamefont{C.}~\bibnamefont{Phillips}},
  \bibinfo{author}{\bibfnamefont{D.}~\bibnamefont{Pickering}},
  \bibnamefont{and} \bibinfo{author}{\bibfnamefont{K.}~\bibnamefont{Sohrabi}},
  \bibinfo{journal}{Phys. Rev. D} \textbf{\bibinfo{volume}{97}},
  \bibinfo{pages}{036005} (\bibinfo{year}{2018}), \eprint{1711.09135}.

\bibitem[{\citenamefont{Berges et~al.}(2005)\citenamefont{Berges, Borsanyi,
  Reinosa, and Serreau}}]{Berges:2005hc}
\bibinfo{author}{\bibfnamefont{J.}~\bibnamefont{Berges}},
  \bibinfo{author}{\bibfnamefont{S.}~\bibnamefont{Borsanyi}},
  \bibinfo{author}{\bibfnamefont{U.}~\bibnamefont{Reinosa}}, \bibnamefont{and}
  \bibinfo{author}{\bibfnamefont{J.}~\bibnamefont{Serreau}},
  \bibinfo{journal}{Annals Phys.} \textbf{\bibinfo{volume}{320}},
  \bibinfo{pages}{344} (\bibinfo{year}{2005}), \eprint{hep-ph/0503240}.

\bibitem[{\citenamefont{Prüfer et~al.}(2019)\citenamefont{Prüfer, Zache,
  Kunkel, Lannig, Bonnin, Strobel, Berges, and Oberthaler}}]{Prufer:2019kak}
\bibinfo{author}{\bibfnamefont{M.}~\bibnamefont{Prüfer}},
  \bibinfo{author}{\bibfnamefont{T.~V.} \bibnamefont{Zache}},
  \bibinfo{author}{\bibfnamefont{P.}~\bibnamefont{Kunkel}},
  \bibinfo{author}{\bibfnamefont{S.}~\bibnamefont{Lannig}},
  \bibinfo{author}{\bibfnamefont{A.}~\bibnamefont{Bonnin}},
  \bibinfo{author}{\bibfnamefont{H.}~\bibnamefont{Strobel}},
  \bibinfo{author}{\bibfnamefont{J.}~\bibnamefont{Berges}}, \bibnamefont{and}
  \bibinfo{author}{\bibfnamefont{M.~K.} \bibnamefont{Oberthaler}}
  (\bibinfo{year}{2019}), \eprint{1909.05120}.

\end{thebibliography}

\end{document}